\documentclass{cernrep} 
\usepackage{texnames}
\usepackage[T1]{fontenc}

\def\ga{\mathrel{\raise.3ex\hbox{$>$\kern-.75em\lower1ex\hbox{$\sim$}}}}
\def\la{\mathrel{\raise.3ex\hbox{$<$\kern-.75em\lower1ex\hbox{$\sim$}}}}

\usepackage{varwidth}
\usepackage{xcolor}

\title{Quantum Field Theory and the Electroweak Standard Model}
\author{E. Boos}
\institute{M.~V.~Lomonosov Moscow State University, Skobeltsyn Institute of Nuclear Physics (SINP MSU), Moscow 119991, Russia}

\begin{document}
\maketitle
\setlength{\unitlength}{1mm}

\begin{abstract}
The Standard Model is one of the main intellectual achievements for about the
last 50 years, a result of many theoretical and experimental studies. In this lecture a brief introduction
to the electroweak part of the Standard Model is given. Since the Standard Model is a
quantum field theory, some aspects for understanding of quantization of abelian and non-abelian gauge
theories are also briefly discussed. It is demonstrated how well the electroweak Standard Model works in
describing a large variety of precise experimental measurements at lepton and hadron colliders.
\end{abstract}



\section{Introduction}
The Standard Model (SM) of strong and electroweak (EW) interactions is the basis for understanding of nature at extremely
small distances. In high-energy physics usually the relativistic system of units is used in which
the Planck constant $\hbar$ and the speed of light $c$ are equal to unity, $\hbar=c=1$. Taking into account well-known
values for $\hbar = 1.055 \cdot 10^{27}$ erg s, $c= 3\cdot 10^{10}$ cm$/$s and the positron electric charge $e = 1.6\cdot10^{-19}$ C and using the
relation between the electronvolt and erg (1 eV = $e \cdot$  1 V = 1 V $\cdot 1.6\cdot 10^{-19}$ C = $1.6\cdot10^{-12}$ erg), one easily gets the
following very useful relation between length and energy units:  $1/$GeV = $2\cdot 10^{-14} $ cm. Due to the Heisenberg uncertainty principle,
 $\Delta x \Delta p \ge 1/2 $, the above relation allows us to understand which energies (momentum transfers) are needed approximately to probe certain distances:\\
100 GeV $\rightarrow$ $10^{-16}$ cm,\\
1 TeV $\rightarrow$ $10^{-17}$ cm,\\
10 TeV $\rightarrow$ $10^{-18}$ cm.\\
Therefore, at the LHC one can study the structure of matter at distances of $10^{-18}$--$10^{-17}$ cm. 
For small distances of the order of $10^{-16}$ cm
or correspondingly  100 GeV energies the SM  works very well, as follows from many studies and measurements.

\noindent The SM is a quantum field theory; it is based on a few principles and requirements:
\begin{itemize}
\item gauge invariance with lowest dimension (dimension four) operators;
 SM gauge group: $SU(3)_C \times SU(2)_L \times U(1)_Y$;
\item correct electromagnetic neutral currents and correct charge currents with (V--A)
structure as follows from four fermion Fermi interations (\ref{Fermi}) 
\begin{equation}
\frac{G_{\rm F}}{\sqrt 2}\left [ \bar \nu _{\mu}\cdot \gamma_{\alpha}(1-\gamma_5)\cdot \mu \ \right] \left [ \bar e \cdot \gamma_{\alpha}(1-\gamma_5)\cdot \nu_\mathrm{e} \ \right]+\mathrm{h.c.};
\label{Fermi}
\end{equation}
\item three generations without chiral anomalies;
\item Higgs mechanism of spontaneous symmetry breaking.
\end{itemize}

\begin{figure}
\centering
\includegraphics{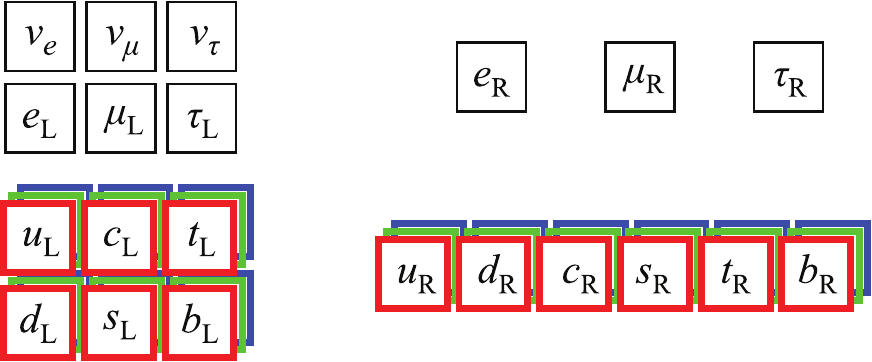}
\caption{Fermion generation}
\label{pic:gen}
\end{figure}

\noindent Fermions are combined into three generations forming left doublets and right 
singlets with respect to the weak isospin (see Fig.\ref{pic:gen}).

\begin{equation*}
f_{\mathrm{L, R}}=\frac12 (1\mp \gamma_5)f,
\end{equation*}
\begin{equation*}
I_{f}^{3\mathrm{L},3\mathrm{R}}=\pm\frac 12, 0: \begin{array}{c}
L_1=\begin{pmatrix}\nu_{\rm e} \\ {e}^-\end{pmatrix}_{\rm L}, e_{\mathrm{R}_1}={e}^-_{\rm R}, Q_1 = \begin{pmatrix}{u} \\ {d} \end{pmatrix}_{\rm L}, u_{\mathrm{R}_1}=u_{\rm R},  d_{\mathrm{R}_1}=d_{\rm R},\\
L_2=\begin{pmatrix}\nu_{\mu} \\ \mu^-\end{pmatrix}_{\rm L}, e_{\mathrm{R}_2}=\mu^-_{\rm R}, Q_2 = \begin{pmatrix}{c} \\ {s} \end{pmatrix}_{\rm L}, u_{\mathrm{R}_1}=c_{\rm R},  d_{\mathrm{R}_1}=s_{\rm R},\\
L_3=\begin{pmatrix}\nu_{\tau} \\ \tau^-\end{pmatrix}_{\rm L}, e_{R_3}=\tau^-_{\rm R}, Q_3 = \begin{pmatrix}{t} \\ {b} \end{pmatrix}_{\rm L}, u_{\mathrm{R}_1}=t_{\rm R},  d_{\mathrm{R}_1}=b_{\rm R}.
\end{array}
\end{equation*}

\begin{figure}

\centering

\includegraphics[width=50mm,height=50mm]{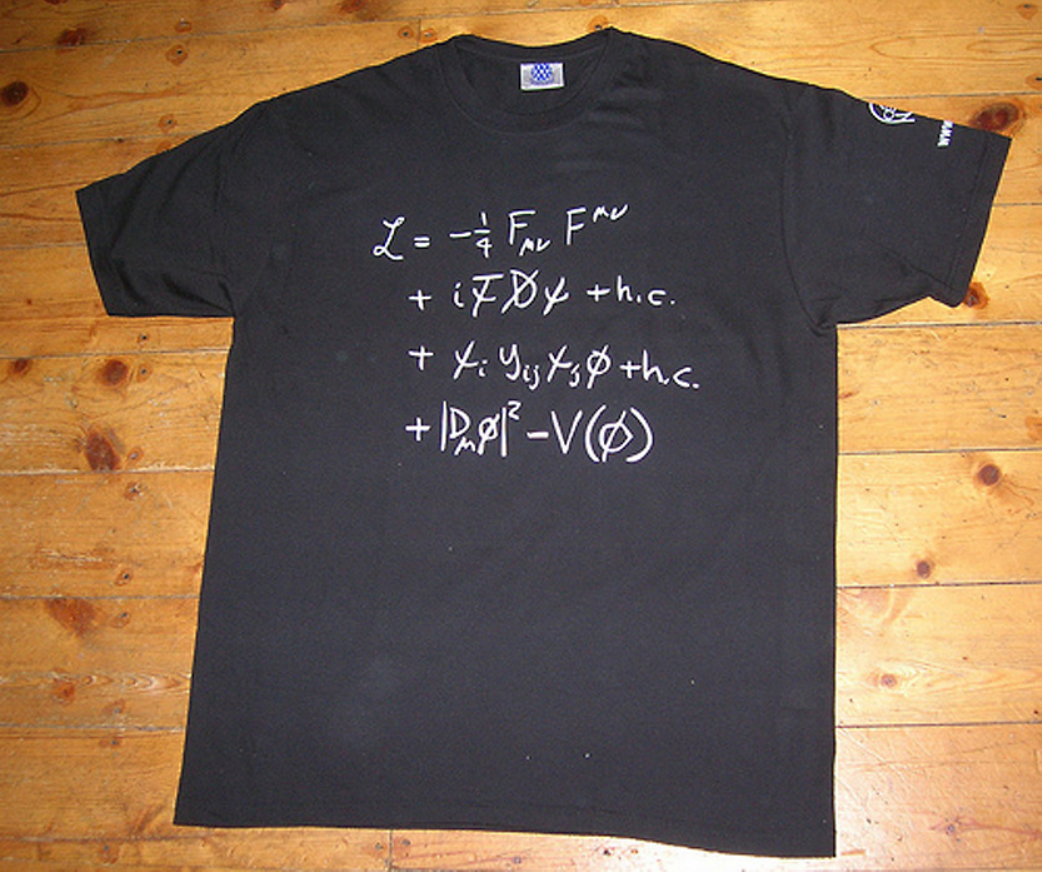}

\caption{CERN T-shirt with SM Lagrangian}

\label{pic:T-sh}

\end{figure}

\noindent The SM Lagrangian written in accord with the mentioned requirements looks very simple:

 \begin {equation*}
\begin{array}{rcl}
L&=& -\frac{1}{4} W_{\mu\nu}^{i}(W^{\mu\nu})^i-\frac{1}{4} B_{\mu\nu}B^{\mu\nu}-\frac{1}{4} G_{\mu\nu}^a(G^{\mu\nu})^a\\[5pt]
&&+\sum_{f=\ell,q}\bar\Psi^f_{\rm L}(\mathrm{i}D^{\rm L}_{\mu}\gamma^{\mu})\Psi_{\rm L}^{\dagger}+\sum_{f=\ell,q}\bar\Psi^f_{\rm R}(\mathrm{i}D^{\rm R}_{\mu}\gamma^{\mu})\Psi_{\rm R}^{\dagger}+L_{\rm H},
\end{array}
\label{int1}
\end{equation*}
\begin{equation*}
L_{\rm H} = L_{\Phi}+L_{\mathrm{Yukawa}},
\label{int1a}
\end{equation*}
\begin{equation*}
L _{\Phi}= D_{\mu}\Phi^{\dag}D^{\mu}\Phi - \mu^2\Phi^{\dag}\Phi-\lambda (\Phi^{\dag}\Phi)^4,
\label{int1b}
\end{equation*}
 \begin{equation*}
 L_{\mathrm{Yukawa}} = -\Gamma_d^{ij}\bar {Q'_{\rm L}}^i\Phi {d'_{\rm R}}^j + \mathrm{h.c.} -\Gamma_u^{ij}\bar {Q'_{\rm L}}^i\Phi^C {u'_{\rm R}}^j + \mathrm{h.c.} -\Gamma_{\rm e}^{ij}\bar {L'_{\rm L}}^i\Phi {e'_{\rm R}}^j + \mathrm{h.c.}
 \label{int1c}
 \end{equation*}
The field strength tensors and covariant derivatives have very familiar forms:
\begin {equation*}
W^i_{\mu\nu} = \partial_{\mu}W_{\nu}^i-\partial_{\nu}W_{\mu}^i+g_2\varepsilon^{ijk}W_{\mu}^jW_{\nu}^k,
\label{int2}
\end{equation*}
 \begin {equation*}
 B_{\mu\nu}=\partial_{\mu}B_{\nu}-\partial_{\nu}B_{\mu},
\label{int3}
\end{equation*}
 \begin {equation*}
 G^a_{\mu\nu} = \partial_{\mu}A_{\nu}^a-\partial_{\nu}A_{\mu}^a+g_Sf^{abc}A_{\mu}^b A_{\nu}^c,
\label{int4}
\end{equation*}
 \begin {equation*}
 \begin{array}{c}
 D^\mathrm{L}_{\mu} = \partial_{\mu}-{\rm i}g_2W_{\mu}^i\tau^i - ig_1 B_{\mu}\left( \frac{Y^f_{\rm L}}{2}\right)-{\rm i}g_SA^a_{\mu}t^a,\\
D^\mathrm{R}_{\mu} = \partial_{\mu}- ig_1 B_{\mu}\left( \frac{Y^f_{\rm R}}{2}\right)-{\rm i}g_SA^a_{\mu}t^a,
\end{array}
\label{int5}
\end{equation*}
where $i=1,2,3,$ $a=1,\dots,8$; $W^i_{\mu}$ are gauge fields for the weak isospin group, $B_{\mu}$ are gauge fields for the weak hypercharge group and $A_{\mu}$ are gluon gauge fields for the strong $SU_C(3)$ colour group.
 \begin {equation*}
Y_f=2Q_f-2I^3_f \Rightarrow Y_{L_i} = -1, Y_{e_{R_i}}=-2, Y_{Q_i}=\frac 13, Y_{u_{R_i}}=\frac43, Y_{d_{R_i}}=-\frac 23.
 \end{equation*}
The Lagrangian is so compact that its main part can be presented on the CERN T-shirt (see Fig.~\ref{pic:T-sh}).

It is hard to imagine that such a simple
Lagrangian allows one to describe basically all the phenomena of the microworld. But the SM Lagrangian, being expressed in terms of physics components, is not that simple, leading after quantization to many interaction vertices between particles or quanta of corresponding quantum fields.

This lecture is organized as follows. In the next section some aspects  of quantum field theory are briefly discussed.
After a motivation as to why do we need a quantum field theory,  we consider scalar fields and introduce the Feynman propagator and functional integral approach as a quantization method. The functional integral given in the holomorphic representation allows us to clarify boundary conditions and show the connection between
the Green functions and S-matrix elements. Feynman diagrams are introduced.
The formalism is extended to the fermion and gauge fields stressing peculiarities in the quantization procedure and Feynman rule derivation.
In the next section a construction of the EW SM Lagrangian is presented.
We discuss experimental facts and theory principles based on which the EW part of the SM Lagrangian for fermion and gauge fields is constructed. We show explicitly which conditions on weak hypercharges allow us to get correctly electromagnetic and charge current (CC)
interactions and predict additional neutral currents (NCs). We demonstrate how potentially dangerous chiral anomalies cancelled out.  Then spontaneous symmetry breaking, the Goldstone theorem and the appearance of Nambu--Goldstone bosons are briefly discussed.
The Brout--Englert--Higgs--Hagen--Guralnik--Kibble mechanism of spontaneous symmetry breaking is introduced leading to non-zero masses of the gauge fields and appearance of the Higgs boson. Very briefly we discuss in addition to the unitary gauge the covariant gauge, propagators of Goldstone bosons and ghosts. At the end of the section it is shown how the spontaneous symmetry breaking mechanism leads to non-zero masses for the fermions in the SM and how very naturally the Cabibbo--Kobayashi--Maskawa mixing matrix appears.
In the next section we concentrate on some phenomenological aspects of the EW SM such as
connections between the Fermi constant $G_{\rm F}$, the Higgs vacuum expectation value $v$, consistency  of low-energy measurements and W, Z mass measurements, W-, Z-boson decay widths and branching ratios, number of light neutrinos, two-fermion processes in ${\rm e}^+{\rm e}^-$ collisions, tests of the gauge boson self-interactions, top-quark decays and the EW top production (single top).
Briefly we discuss the EW SM beyond the leading order, renormalization and running coupling in quantum electrodynamics (QED), as a simplest example,
running masses and running parameters in the SM, precision EW data and global parameter fits.
 Concluding remarks are given in the next section. The quantum chromodynamics (QCD) part of the SM and the phenomenology of the Higgs boson are not discussed
in these lectures as they are addressed in other lectures of the School.

For a deeper understanding of the topics discussed, one can recommend  a number of 
very good textbooks and reviews \cite{bib:bjorkin65, bib:abers73, bib:itzykson80,
bib:halzen84, bib:cheng84, bib:faddeev91, bib:peskin95,bib:weinberg96, bib:zee03}
and lectures given at previous schools and specialized reviews
\cite{bib:buchmuller05, bib:altarelli08, bib:rubakov09, bib:hollik09, bib:alterelli13}, which have been used in preparation of this lecture.

\section{Introductory words to quantum theory}
In classical mechanics a system evolution follows from the principle of least action:
$$\delta S = \delta \int \limits_{t_i}^{t_f} {\rm d}t L(q(t),\dot q(t) )=0; \,\,\,\,\,\int \limits_{t_i}^{t_f}
\left [  \frac{\partial L}{\partial q}\delta q + \frac{\partial L}{\partial \dot q}\delta (\dot q)\right ]=0; \,\,\,\,\,
\delta (\dot q)=\frac {\rm d}{{\rm d}t}\delta q.$$
For an arbitrarily small variation $\delta q$, one gets the well-known Lagrange equation of motion
\begin{equation}
\frac{\partial L}{\partial q}=\frac {\rm d}{{\rm d}t}\left(\frac{\partial L}{\partial \dot q}
\right).
\label{Lagr_eq}
\end{equation}
For a non-relativistic system described by the Lagrangian
$ L=\frac{m \dot q^2}{2}-V(q)$, the second Newton law follows from Eq. (\ref{Lagr_eq}),
$$m\ddot q = -\frac{\partial V}{\partial q} = F.$$

The Hamiltonian of the system is related to the Lagrangian in the following well-known way:
$$H(p,q)=p \dot q - L(q,\dot q),$$
where $\dot q$ is a solution of the equation $p=\frac{\partial L}{\partial \dot q}$.

In quantum mechanics the coordinate and momentum are replaced by corresponding operators $p,q \to \hat p, \hat q$
with postulated commutator relation $[\hat p(0), \hat q(0)] = -{\rm i}\hbar$.
In the Heisenberg picture the system evolution is described by the Heisenberg equation with time-dependent operators; for example, the equation
for the  coordinate operator has the following form:

\begin{equation}
 \frac{\partial \hat q}{\partial t} = \frac{\rm i}{\hbar}[\hat H, \hat q ]
\label{Heisenberg_eq}
\end{equation}
with a formal solution
$$\hat q(t) = {\rm e}^{\frac{\rm i}{\hbar}\hat H t}\hat q(0) {\rm e}^{-\frac{\rm i}{\hbar}\hat H t}.$$
This easily follows from the equalities
$$ \frac{\partial \hat q}{\partial t} =\frac{\rm i}{\hbar}\hat H {\rm e}^{\frac{\rm i}{\hbar}\hat H t}\hat q(0)
{\rm e}^{-\frac{\rm i}{\hbar}\hat H t}
 + {\rm e}^{\frac{\rm i}{\hbar}\hat H t}\hat q(0)
{\rm e}^{-\frac{\rm i}{\hbar}\hat H t} \hat H = \frac{\rm i}{\hbar}\hat H\hat q(t) -\frac{\rm i}{\hbar}\hat q(t)\hat H = \frac{\rm i}{\hbar}[\hat H,\hat q].
$$
For the coordinate and momentum operators, one can prove the following inequality:
$$\triangle q\cdot \triangle p \ge 1/2,$$ which is called the Heisenberg uncertainty principle.

Let us recall a simple proof of the Heisenberg uncertainty principle.
The mid value of any operator $\hat A$  and its dispersion
are given by the relations
$$\langle \psi|\hat A|\psi \rangle = \overline{A}, \,\,\,
\langle \psi|(\hat A-\overline{A})^2|\psi \rangle.$$

Let us take the following operator constructed from the momentum and coordinate operators ($\left[\hat p,\hat q\right]=-{\rm i}\hbar$)
with arbitrary constant $\gamma$:
$$\hat A=\hat p+{\rm i} \gamma \hat q - (\overline{p}+{\rm i}\gamma \overline{q}).$$
Then the conjugated operator has the form
$$\hat A^{\dag}=\hat p-{\rm i} \gamma \hat q - (\overline{p}-{\rm i}\gamma \overline{q}).$$
For any state  $|\psi \rangle$ 
we have $$\langle \psi|\hat A^{\dag}A|\psi \rangle\ge 0$$,
$$\langle \psi|\left[(\hat p-\overline{p})-{\rm i}\gamma (\hat q-\overline{q})\right]
\left[(\hat p-\overline{p})+{\rm i}\gamma (\hat q-\overline{q})\right]|\psi \rangle$$
$$=(\Delta p)^2+\gamma^2\Delta q^2-{\rm i}\gamma(\hat q \hat p - \hat p \hat q) =( \Delta p)^2+\gamma^2\Delta q^2+\gamma\hbar \ge 0.$$
This is true for any value of $\gamma$ and therefore the determinant is not positive:
$$\frac{\hbar^2}{4}- \Delta q^2\Delta p^2\le 0.$$
Thus, we immediately arrive at the uncertainty principle:
$$\Delta q\Delta p \ge  \frac{1}{2}\hbar.$$

Let us consider the simplest system, the harmonic oscillator, described by the Hamiltonian (we put $\hbar = 1$)
$$\hat H = \frac{1}{2}(\hat p^2+\omega^2\hat q^2),$$
and construct two operators $\hat a$ and $\hat a^{\dag}$ which are called the annihilation and creation operators:
$$\hat q = \frac{1}{\sqrt{2\omega}}(\hat a+\hat a^{\dag});\,\,\,\,\,\,\hat p = -{\rm i}\sqrt {\frac{\omega}{2}}(\hat a-\hat a^{\dag}).$$

\begin{equation}
\Longrightarrow \hat a = \sqrt {\frac{\omega}{2}}\hat q+{\rm i} \frac{1}{\sqrt{2\omega}}\hat p;\,\,\,\,\,\,\hat a^{\dag} = \sqrt {\frac{\omega}{2}}\hat q-{\rm i} \frac{1}{\sqrt{2\omega}}\hat p.
\label{OP}
\end{equation}
From the equation $[\hat p, \hat q] = -{\rm i}$, one gets $[\hat a, \hat a^{\dag}] = 1$ and the Hamiltonian takes the form
$$\hat H = \frac{\omega}{2}(\hat a\hat a^{\dag}+\hat a^{\dag}\hat a).$$
It is easy to show the following commutation relations with the Hamiltonian:
$$[\hat H,\hat a] = -\omega \hat a\,\,\,\,\mbox{and}\,\,\,\,[\hat H,\hat a^{\dag}] = \omega \hat a^{\dag}.$$
From (\ref{Heisenberg_eq}),
$$\frac{\mathrm{d}\hat a}{{\rm d}t} = \mathrm{i}[\hat H \hat a] = -{\rm i} \omega \hat a\,\,\,\,\Longrightarrow\,\,\,\, \hat a(t) = \hat a (0) {\rm e}^{-{\rm i} \omega t}; \,\,\,\,  \hat a^{\dag}(t) = \hat a^{\dag} (0) {\rm e}^{\rm i \omega t}.$$
Let us consider states with definite energy:
$$\hat H|E\rangle = E|E\rangle.$$
Then the state $\hat a|E\rangle$ ($\hat a^{\dag}|E\rangle$) corresponds to the energy $(E-\omega)$ ($(E+\omega)$). Indeed,
$$\hat H \hat a|E\rangle = \hat a \hat H |E\rangle - \omega \hat a|E\rangle = (E-\omega) \hat a|E\rangle,$$
$$\hat H \hat a^{\dag} |E\rangle =  (E+\omega) \hat a^{\dag} |E\rangle.$$

Let us construct states (Hilbert space of states) starting from the `vacuum' state $|0\rangle$:
$$\hat a|0\rangle=0.$$
What is the energy of the vacuum state? This is
$$\hat H|0\rangle=\frac{\omega}{2}(\hat a\hat a^{\dag}+\hat a^{\dag}\hat a)|0\rangle=\frac{\omega}{2}|0\rangle.$$
The state $|n\rangle$ we introduce as $|n\rangle=(\hat a^{\dag})^n|0\rangle$; its energy is given by
$$\hat H(\hat a^{\dag})^n|0\rangle=\omega(n+1/2)|n\rangle.$$
As we know, such a construction is very successful in describing non-relativistic quantum phenomena (spectra of atoms, molecules, nuclei etc).

But there are  well-known problems:
\begin{itemize}
\item In experiments we have not only particle creation and annihilation but also production of new particles and antiparticles.
\item Relativity and causality might be in conflict with quantum principles.\\
If in four-dimensional areas $X$ and $Y$ points are separated by the space-like interval $(x-y)^2 < 0$, events in the points $x$ and $y$
are causally independent. But in QM we have the uncertainty principle, $\Delta p \Delta x \ge 1$, and
in some regions of the order of $\Delta x ~ 1/m$ close to indistinct boundaries of the areas $X$ and $Y$ the causality might be violated.

\end{itemize}

Quantum field theory allows us to resolve both of these problems simultaneously.

To describe a particle and an antiparticle with mass $m$, momentum $\vec{k}$ and energy $\omega_k = k^0 = \sqrt{\vec{k}^2+m^2}$,
let us consider two sets of creation and annihilation operators for each momentum point $\vec{k}$
 $\hat a,\hat a^{\dag}$ and $\hat b,\hat b^{\dag}$.
 Vacuum is defined by the requirements $\hat a|0\rangle = \hat b|0\rangle =0$.
The Hamiltonian for every momentum point $\vec{k}$ obviously has the following form:
$$H_k = \frac{\omega_k}{2}(\hat a(\vec{k}) \hat a^{\dag}(\vec{k})+ \hat a^{\dag}(\vec{k})\hat a(\vec{k})
+\hat b (\vec{k})\hat b^{\dag}(\vec{k})+ \hat b^{\dag}(\vec{k})\hat b(\vec{k})),$$
where $\omega_k = k^0 = \sqrt{\vec{k}^2+m^2}$. Please note that the mass parameter $m$ is the same for all oscillators with different $\vec{k}$.

One can use a different normalization of an integral measure in order to get the commutation relations:
$$\int \overline{\mathrm{d}k}[\hat a(\vec{k})\hat a^{\dag}(\vec{k})]=1;\,\,\,\,\int \overline{\mathrm{d}k}[\hat b(\vec{k})\hat b^{\dag}(\vec{k})]=1.$$
We are using
$$\overline{\mathrm{d}k}=\frac{{\rm d}^3\vec{k}}{(2\pi)^3 2\omega_k}\Rightarrow[\hat a(\vec{k})\hat
a^{\dag}(\vec{k'})]=(2\pi)^2 2\omega_k\delta(\vec{k}-\vec{k'}).$$

Total momentum and charge operators taken in so-called `normal ordering' have the following form:
$$\widehat{P}^{\mu} = \int \overline{\mathrm{d}k}k^{\mu}\left[\hat a^{\dag}
(\vec{k})\hat a(\vec{k})+ \hat b^{\dag}(\vec{k})\hat b(\vec{k})\right],$$
$$\widehat{Q} = \int \overline{\mathrm{d}k}\left[\hat a^{\dag}
(\vec{k})\hat a(\vec{k}) - \hat b^{\dag}(\vec{k})\hat b(\vec{k})\right].$$
One can prove the following commutator relations, which clarify the meaning of the operators:
$$\left[\widehat{P}^{\mu},\hat a^{\dag}(\vec{k})\right] = k^{\mu}\hat a^{\dag}(\vec{k}); \,\,\,\,
\left[\widehat{P}^{\mu},\hat a(\vec{k})\right] = -k^{\mu}\hat a(\vec{k}),$$
$$\left[\widehat{Q},\hat a^{\dag}(\vec{k})\right] = \hat a^{\dag}(\vec{k}); \,\,\,\,
\left[\widehat{Q},\hat b^{\dag}(\vec{k})\right] = -\hat b^{\dag}(\vec{k}).$$
Now one can construct the field operator:
$$\widehat{\Phi}(x) = \int\overline{\mathrm{d}k}\left[{\rm e}^{-{\rm i}kx}\hat a(\vec{k})+{\rm e}^{{\rm i}kx}\hat b^{\dag}(\vec{k})\right].$$
The momentum operator acts on the field operator leading to a coordinate translation:
$$ {\rm e}^{\mathrm{i}\widehat P_{\mu}y^{\mu}}\widehat{\Phi}(x) {\rm e}^{-{\rm i}\widehat P_{\mu}y^{\mu}}=\widehat{\Phi}(x+y).$$
Indeed, one may prove this in a very simple way. Let us introduce an operator $\widehat A (\alpha)$ depending on some numerical parameter
$\alpha$:
$$ {\rm e}^{{\rm i}\alpha y^{\mu}\widehat P^{\mu}}\hat a(k) {\rm e}^{-{\rm i}\alpha y^{\mu}\widehat P^{\mu}}=\widehat A (\alpha).$$
The operator obeys the following equation:
$$\frac{\mathrm{d}A}{\mathrm{d} \alpha}=\mathrm{i} y^{\mu}\widehat P^{\mu}\widehat A-{\rm i} y^{\mu}\widehat P^{\mu}\widehat A=-{\rm i} y^{\mu}\left[\widehat P^{\mu}\widehat A\right].$$
One finds a solution in the form $\widehat A = \hat a(\vec{k})f(\alpha)$, from which one can immediately see the needed translation relation:
$$\Longrightarrow  \hat a(k)\frac{\mathrm{d}f}{\mathrm{d} \alpha} = -{\rm i}y^{\mu}f(\alpha)k^{\mu}\hat a(k)\,\,\,\,\Longrightarrow \frac{\mathrm{d}f}{\mathrm{d} \alpha} = -{\rm i}y^{\mu}k^{\mu}f(\alpha)\,\,\,\,\Longrightarrow $$
$${\rm e}^{\mathrm{i}\alpha y^{\mu}\widehat P^{\mu}}\hat a(k) {\rm e}^{-{\rm i}\alpha y^{\mu}\widehat P^{\mu}}=\hat a(k){\rm e}^{-{\rm i}k^{\mu}y_{\mu}}\,\,\,\,\Longrightarrow \widehat{\Phi}(x)=\widehat{\Phi}(x+y).$$

It is very important to note that such a field operator obeys the Klein--Gordon equation:

$$[\Box^2+m^2]\widehat{\Phi}(x) = \int \overline{\mathrm{d}k}\left[ {\rm e}^{-{\rm i}kx}
(-k^2+m^2)\hat a(k)+ {\rm e}^{\mathrm{i}kx}(-k^2+m^2)\hat b(k)\right] = 0$$
because of 
$$-k^2+m^2 = -k_0^2+\vec k^2+m^2 = -\omega^2_k+\omega^2_k = 0$$

What charge has the state created by the operator $\widehat{\Phi}(x)$? Let us act on the vacuum state by the $\widehat{\Phi}(x)$ operator: $\widehat{\Phi}(x)|0\rangle$. This state has the following charge:

$$ \widehat Q\widehat{\Phi}(x)|0\rangle = -\widehat{\Phi}(x)|0\rangle.$$
In the same way, one gets
$$\widehat Q\widehat{\Phi}^{\dag} (x)|0\rangle = \widehat{\Phi}^{\dag} (x)|0\rangle.$$

This means that the field operator  $\widehat{\Phi}(x)$ acting on vacuum produces the state with the negative charge (--1) and the field operator $\widehat{\Phi}^{\dag}(x)$ produces the state with the positive charge (+1).

Now let us consider two space--time points $x_1$ and $x_2$:\\
\begin{center}
\begin{picture}(110,25)
\put(15,20){\circle*{2}}
\put(95,20){\circle*{2}}
\put(6,7){$t_1,\vec x_1$}
\put(86,7){$t_2,\vec x_2$}
\end{picture}
\end{center}
and two-point correlation functions---products of field operators between vacuum states.
If $t_1 < t_2$,  the operator $\widehat{\Phi} (x_1)|0\rangle$ in the correlator
$$\langle0|\widehat{\Phi}^{\dag} (x_2)\widehat{\Phi} (x_1)|0\rangle$$
creates the charge --1 at $t_1$, and the operator $\widehat{\Phi}^{\dag} (x_2)\widehat{\Phi} (x_1)|0\rangle$ annihilates this charge at $t_2$. So, charge --1 propagates from the point $x_1$ to $x_2$ and $t_2>t_1$.
If $t_2 < t_1$, the operator $\widehat{\Phi}^{\dag} (x_2)|0\rangle$ in the correlator
$$
\langle0|\widehat{\Phi} (x_1)\widehat{\Phi}^{\dag} (x_2)|0\rangle
$$
creates the charge +1 at $t_2$, and $\widehat{\Phi} (x_1)\widehat{\Phi}^{\dag} (x_2)|0\rangle$ annihilates this charge at $t_1$. So, charge +1 propagates from the point $x_2$ to $x_1$ and $t_2<t_1$.

Since both these actions do not change the vacuum, we should take both correlators into account to see
the causal relation of events in points $x_1$ and $x_2$:
$$\langle0|T\{ \widehat{\Phi}^{\dag} (x_2)\widehat{\Phi} (x_1)\}|0\rangle$$
$$=\langle0|\widehat{\Phi}^{\dag} (x_2)\widehat{\Phi} (x_1)|0\rangle\Theta(t_2-t_1)+\langle0|\widehat{\Phi} (x_1)\widehat{\Phi}^{\dag} (x_2)|0\rangle\Theta(t_1-t_2)$$
$$={\rm i}\int \frac{{\rm d}^4k}{(2\pi)^4}\frac{{\rm e}^{-{\rm i}k(x_2-x_1)}}{k^2-m^2+{\rm i}0} = D_{\rm c}(x_2-x_1).$$
The function 
\begin{equation}
D_{\rm c}(x)=\mathrm{i} \int  \frac{{\rm d}^4k}{(2\pi)^4}\frac{{\rm e}^{-{\rm i}kx}}{k^2-m^2+{\rm i}0}
\label{Feynman_prop}
\end{equation}
is called the Feynman propagator. Obviously, the Feynman propagator is a Green function of the Klein--Gordon equation.
One can check that all the commutators between the field operators in the points $x$ and $y$ separated by the space-like interval
$(x-y)^2 < 0$ are equal to zero. So, the causality takes place. Also, one can construct multiparticle states by acting of the creation operators (operators have different quantum numbers corresponding to different kinds of particles)
on the vacuum state
$ \left |\vec k_1,\dots, \vec k_n \right>=\prod_{i=1}^n\hat a^{\dag}(\vec k_i)\left |0 \right>$.
The energy and momentum of the states are then obtained by acting of the operators
$$\hat P^0\left |\vec k_1,\dots, \vec k_n \right> = \hat H\left |\vec k_1,\dots, \vec k_n \right>=\left(\sum^n_{i=1}k^0_i \right) \left |\vec k_1,\dots, \vec k_n \right>,$$
$$\widehat {\vec P}\left |\vec k_1,\dots, \vec k_n \right>=\left(\sum^n_{i=1}\vec k_i \right) \left |\vec k_1,\dots, \vec k_n \right>.$$

Of course, one can get the same results using the usual canonical quantization with the correspondence
$$q(t),\,\,\dot q(t),\,\,L(q,\dot q),\,\,S=\int {\rm d}tL(q,\dot q) \Longleftrightarrow$$
$$ \varphi(x),\,\,\partial_{\mu} \varphi(x),\,\,L(\varphi,\partial_{\mu} \varphi),\,\,S=\int {\rm d}^4xL(\varphi,\partial_{\mu} \varphi).$$
The field momentum is then
$$\pi(x) = \frac{\partial L}{\partial(\partial_{0} \varphi(x))};\,\,\,\,\partial_{0} \varphi(x)=\dot \varphi.$$
The Lagrangian of the complex scalar field has the form
$$L = \partial_{\mu} \varphi^{\dag}\partial^{\mu}\varphi-m^2\varphi^{\dag}\varphi,$$
$$\pi(x) = \dot \varphi ^{\dag}(x);\,\,\,\pi^{\dag}(x) = \dot \varphi (x).$$
In the same way, as in classical mechanics, the equation of motion comes from the principle of least action:
$$\frac{\partial L}{\partial \varphi} = \partial _{\mu}\frac{\partial L}{\partial_{\mu} \varphi}\longrightarrow\,\,\,(\Box - m^2)\varphi = 0.$$
The Feynman propagator $D_{\rm c}$ introduced above is a Green function of the equation of motion.
For the field and momentum operators, one naturally assumes the equal-time commutation relation
$$\left.[\hat {\pi}(\vec x, t),\hat \varphi (\vec x', t')]\right|_{t=t'} = -{\rm i}\delta (\vec x - \vec x').$$
The Lagrangian is invariant under a global phase shift:
$$\varphi(x) \rightarrow {\rm e}^{\mathrm{i}\alpha \varphi(x)},\,\,\,\,\alpha\equiv \mathrm{Const}.$$
Such a combination, called a current, is conserved (this is a simple example of the first Noether theorem):
$$j^{\mu}(x) = {\rm i}\partial^{\mu}\varphi^{\dag}\varphi -  i\varphi^{\dag}\partial^{\mu}\varphi, $$
$$\partial_{\mu}j^{\mu} = {\rm i} \Box\varphi^{\dag}\varphi+{\rm i}\partial^{\mu}\varphi^{\dag}\partial_{\mu}\varphi-{\rm i}\partial_{\mu}\varphi^{\dag}\partial^{\mu}\varphi- \mathrm{i}\varphi^{\dag}\Box\varphi=0.$$
Conservation of the current leads to the conservation of the charge:
$$Q=\int {\rm d}^3\vec x j^{0} = \int \mathrm{d}\vec x(\mathrm{i}\dot \varphi^{\dag} \varphi - \mathrm{i}\varphi^{\dag}\dot \varphi),$$
$$\frac{\mathrm{d}Q}{{\rm d}t} = \int \mathrm{d}\vec x \partial_{0}j^{0}
 = \int \mathrm{d}\vec x (\partial_{i}j^{i}) = \int_{\Omega}\mathrm{d}\vec n \vec j =0$$
for falling-off fields.

\section{Functional integral in quantum mechanics}
However, for our further consideration, the functional integral approach to quantum field theory is more useful. In particular,
it allows us to quantize non-abelian gauge field theories, to clarify better boundary conditions and renormalization procedure and to get a reduction formula (connection between S-matrix elements and the Green functions).

Once more we begin with the quantum mechanics as a simple example.
$$L(q_i,\dot q_i) \,\,\,\, \rightarrow \,\,\,\, p_i=\frac{\partial L}{\partial q_i}, \,\,\,\,
H(q_i, p_i) = \left.\dot q_i p_i-L(q_i,\dot q_i)\right|_{\dot q = f(t)}$$
$$\left[\hat q^i(t),\hat p^j (t) \right] = \mathrm{i}\hbar \delta^{ij}\hat1.$$

Let us consider a simple system described by the non-relativistic Hamiltonian
$$H(\hat p, \hat q) = \frac {\hat p^2}{2m} +V(\hat q).$$
In the Schr\"odinger picture, the evolution of a quantum system follows from  the Schr\"odinger equation
$$\mathrm{i}\frac{\partial}{\partial t}\left|\Psi \right> = \hat H \left|\Psi \right>.$$
The formal solution of the Schr\"odinger equation is
$$\left|\Psi (t)\right> = {\rm e}^{-{\rm i}\hat Ht}\left|\Psi (0)\right>.$$
One can define states:
$$\left|q,t \right>:\,\,\,\, \hat q\left|q,t \right> =  q\left|q,t \right>.$$
Then the wave function coordinate representation is
$$\Psi(q,t)= \left<q|\Psi(t)\right>=\left<q\left|{\rm e}^{-{\rm i}\hat Ht }\right| \Psi (0)\right>\mbox{  and  }\left<q|q'\right> = \delta(q-q').$$
If we introduce a complete set of states $\left| q_0\right>$ such that
$$\hat 1 = \int \mathrm{d}q_0\left |q_0 \right>\left< q_0 \right |,$$
we can write
$$\Psi(q,t) = \int \mathrm{d}q_0\left<q\right| {\rm e}^{-{\rm i}\hat Ht}\left | q_o \right >\left< q_0|\Psi(0)\right> = \int \mathrm{d}q_0K(q, q_0, t)\left< q_0|\Psi(0)\right>,$$
where $K$ is the so-called kernel of the Schr\"odinger equation.

Obviously, $[\hat H, \hat H] = 0$ and therefore
$${\rm e}^{-{\rm i} \hat H T} =  {\rm e}^{-{\rm i} \hat H (t_{n+1} - t_n)}\cdot {\rm e}^{-{\rm i} \hat H (t_{n} - t_{n-1})}\,...\,{\rm e}^{-{\rm i} \hat H (t_{1} - t_{0})}.$$

\begin{center}
\begin{picture}(110,25)
\put(15,20){\circle*{2}}
\put(30,20){\circle*{2}}
\put(45,20){\circle*{2}}
\put(95,20){\circle*{2}}
\put(15,20)%
{\line(1,0){80}}
\put(12,7){$t_0$}
\put(28,7){$t_1$}
\put(42,7){$t_2$}
\put(63,7){...}
\put(86,7){$t_{n+1} = T$}
\end{picture}
\end{center}

At each time moment $t_i$ we can introduce a unity operator
$$\hat 1 = \int \mathrm{d}q_i \left|q_i\right>\left< q_i \right |.$$

Then, for the kernel $K$,  we obtain the following formula:
$$K(q, q_0, T-t_0) = \int \lim_{n \rightarrow \inf} \prod ^n_{i=1}\mathrm{d}q_i \left| q_{n+1-i}\right>\left < q_i\left| {\rm e}^{-{\rm i}\hat H \delta t}\right | q_{n-i}\right> \dots  \left < q_1\left| {\rm e}^{-{\rm i}\hat H \delta t}\right | q_{0}\right>,$$
$${\rm e}^{\varepsilon(\hat A+\hat B)} = {\rm e}^{\varepsilon\hat A}+{\rm e}^{\varepsilon\hat B}(1+o(\varepsilon ^2)),$$
$$ \left < q_{i+1}\left| {\rm e}^{-{\rm i}\hat H \delta t}\right | q_{i}\right> =  \left < q_{i+1}\left| {\rm e}^{-{\rm i}\frac {\hat p^2}{2m} \delta t}\cdot {\rm e}^{-{\rm i} \hat V( q) \delta t}\right | q_{i}\right>.$$
For very small $(\delta t)$, ${\rm e}^{-{\rm i} \hat V(q) \delta t}$ could be factorized  out, and therefore
$$\left < q_{i+1}\left| {\rm e}^{-{\rm i}\hat H \delta t}\right | q_{i}\right> \approx  {\rm e}^{-{\rm i} \hat V(q_i) \delta t} \cdot \left < q_{i+1}\left| {\rm e}^{-{\rm i}\frac {\hat p^2}{2m} \delta t}\right | q_{i}\right>.$$
The last term can be expressed in the following way:
$$ \left < q_{i+1}\left| {\rm e}^{-{\rm i}\frac {\hat p^2}{2m} \delta t}\right | q_{i}\right> = \int \frac{\mathrm{d}p}{2\pi}\left <q_{i+1}\left | {\rm e}^{-{\rm i}\frac {\hat p^2}{2m} \delta t}\right |p \right>\left<p|q_i\right > =
\int \frac{\mathrm{d}p}{2\pi}{\rm e}^{-{\rm i}p(q_{i+1}-q_i)\delta t}{\rm e}^{-{\rm i}\frac {\hat p^2}{2m} \delta t},
$$
where $\left< q|p\right> = {\rm e}^{\mathrm{i}pq}$, $\left< p|q\right> = {\rm e}^{-{\rm i}pq}$ and $\hat 1 = \int \frac{\mathrm{d}p}{2\pi}\left | p\right> \left< p \right|$.

One could make the following substitution of the integration variable:
$$p' = \left[\frac {p}{\sqrt{2m}} - \sqrt{2m}(q_{i+1}-q_i)\frac{1}{2}\right]  \sim {\rm e}^{\mathrm{i}\frac{m}{2}(q_{i+1} - q_i)^2}\frac{1}{ \delta t}.$$
Then, for the kernel $K$, one gets
$$K(q,q_0;t)=N\int \mathrm{d}q_i \prod_i{\rm e}^{\mathrm{i}\left[\frac{m}{2}\left(\frac{q_{i+1}-q_i}{\delta t^2}\right)^2 - V(q_i)\right] \delta t}
= N\int_{Dq}{\rm e}^{\mathrm{i} \int_o^t {\rm d}t [\frac{mv^2}{2} - V(q)]} = \int {\rm e}^{\mathrm{i}S}D(q).$$
For our consideration it is not needed, but if one takes the integral of ${\rm d}p$, one gets the following representation
for the functional integral measure:
$$D(q) = \lim_{n \rightarrow \inf,
\delta t = \frac {t}{n}\rightarrow 0}\cdot \sqrt{\frac{m}{2\pi {\rm i} \delta t}}\prod_{T=1}^n
\left( \sqrt{\frac{m}{2\pi {\rm i}\delta t}}\mathrm{d}q_{T} \right).$$
We do not discuss mathematical aspects of how well such a construction is determined.

The formula for the kernel, being written as
\begin{equation}
K(q,q_0;t)= \int D(q){\rm e}^{\mathrm{i}S} = \int D(q){\rm e}^{\mathrm{i}\int_0^t {\rm d}t L(q,\dot q,t)},
\label {kernel}
\end{equation}
can be generalized to the case of quantum field theory. For that we need to recall a few simple but important formulas
for multidimensional Gaussian integrals.

\section{Gaussian integrals}
 Let us consider $n$ coordinates $y_1,\dots, y_n$ as a formal vector $$y=\left(
\begin{array}{c} y_1\\ \vdots \\ y_n
\end {array} \right).$$
Obviously, $y^{\rm T} = (y_1 \dots y_n)$ with a formal definition of `scalar' product
$$y^{\rm T} \cdot x = \sum_i y_i x_i.$$
The well-known answer for the Gaussian integral has the form
\begin{equation}
Z = \int {\rm d}x_1\dots {\rm d}x_n \cdot {\rm e}^{-\frac{1}{2}x^\mathrm{T} A x} = \frac{(2\pi)^{n/2}}{\sqrt {\det A}},
\label {Z}
\end{equation}
where $A$ is a positive-definite $n \times n$ matrix.

\emph{Problem}. Take the integral and obtain the above formula.

\emph{Reminder}: $\det A = \prod _{i=1}^n
\lambda_i$, where $\lambda_i$ is the eigenvalue of the matrix $A$.

The integral (\ref{Z}) is an analogue of the integral in a field theory
without external sources, as we shall see in the next part. An analogue of the functional integral with a source is as follows:
\begin{equation}
Z [J]= \int {\rm d}x_1\dots {\rm d}x_n \cdot {\rm e}^{-\frac{1}{2}x^\mathrm{T} A x+J^{\rm T}x},
\label {ZS}
\end{equation}
where $J$ is some vector $J^{\rm T} = (J_1 \dots J_n)$. If we make a substitution of integration variable
$$ x' = x-(A^{-1})J,$$
the integral (\ref{ZS}) will take the form
 $$ Z [J]= {\rm e}^{\frac{1}{2}J^{\rm T} A^{-1} J}\cdot \int {\rm d}x'_1\dots {\rm d}x'_n \cdot {\rm e}^{-\frac{1}{2}x'^{\mathrm{T}} A x'}.$$
 So, the integral $Z[J]$ is given by
 $$Z[J] = {\rm e}^ {\frac{1}{2}J^{\rm T} A^{-1} J }\cdot Z={\rm e}^ {\frac{1}{2}J^{\rm T} A^{-1} J }\cdot  \frac{(2\pi)^{n/2}}{\sqrt {\det A}}.$$
Generalization of the above formula for the case of complex variables of integration is straightforward.

Let $z=x+{\rm i}y$ and $z^*=x-{\rm i}y$. We need to compute the integral
$$Z_C = \int \prod ^n_{k=1}\mathrm{d}z_k^*\mathrm{d}z_k {\rm e}^{-z^{\dag}Bz},$$
 where $B$ is a Hermitian matrix and $z^{\dag} = (z^*)^{\rm T}$. One can diagonalize the
 quadratic form $z^{*T}Bz$ by applying a unitary transformation of variables $z'=U\cdot z$ such that the matrix $UBU^{\dag}$ becomes diagonal:
$$UBU^{\dag} = \left(
\begin{array}{ccc}
\lambda _1 & & 0\\
& \ddots & \\
0 & & \lambda_n
\end{array}
 \right).$$
 The integral $Z_C$ then takes the form
\begin {equation}
Z_C = \int \prod ^n_{k=1}\mathrm{d}z'^*_k \mathrm{d}z'_k {\rm e}^{-\lambda_k | z'_k|^2} = \int \prod_{k=1}^n {\rm d}x_k \mathrm{d}y_k {\rm e}^{-\lambda_k(x_k^2+y_k^2)}
 = \frac{\pi^n}{\det B}.
\label{interg_numbers}
\end{equation}
 If we add the external complex `source' $J$,
 $$Z_C[J] = \int \prod ^n_{k=1}\mathrm{d}z^*_k \mathrm{d}z_k {\rm e}^{-z^{\dag}Bz+J^{\dag}z+z^{\dag}J},$$
 with the shift of variables of integration we get
 \begin {equation}
 Z_{C}[J]={\rm e}^{J^{\dag}B^{-1}J}\cdot Z_C = {\rm e}^{J^{\dag}B^{-1}J}\cdot \frac{\pi^n}{\det B}.
 \label{Zcs}
 \end {equation}
 In the field theory we have to consider interacting fields. So, we need to consider more complicated integrals involving source `interactions':
 $$Z_{\rm int}[J] = \frac{1}{Z_{\rm int}}\int \prod {\rm d}x {\rm e}^{-\frac{1}{2}xAx + Jx - V(x)},$$
 where $Z_{\rm int} = Z_{\rm int}[0]$.

 If we expand the exponent ${\rm e}^{-V(x)}$ in `perturbation' theory we can easily get the following form for the integral $Z_{\rm int}[J]$:
 $$Z_{\rm int}[J] = \frac{1}{Z[0]}{\rm e}^{-V[\frac{\partial}{\partial J}]}\cdot Z[J]={\rm e}^{-V[\frac{\partial}{\partial J}]}\cdot {\rm e}^{\frac{1}{2}JA^{-1}J}.$$

 \section{Functional integral in quantum field theory}

As we have seen already, a transition from mechanics to a field theory could be done by means
of a formal correspondence between the coordinate
and its derivative and the field $\varphi(x)$ and  its derivative:
$$ q(t) \rightarrow \varphi (x);\,\,\,\,\,\dot q(t) \rightarrow \partial_{\mu}\varphi(x).$$
With this analogy one can immediately write down the following formula for the evolution kernel in the case of quantum field theory:
\begin{equation}
Z[J]=\int D(\varphi){\rm e}^{\mathrm{i}\int {\rm d}^4 x L(\varphi, \partial_{\mu} \varphi)+{\rm i}\int {\rm d}^4 x J(x) \varphi(x)},
\end{equation}
where the measure $ D(\varphi) = \prod _x {\rm d} \varphi (x)$ corresponds to the integration
 over all possible trajectories (field configurations).

Now all the formulas we derived for Gaussian integrals in the previous section can be applied here using
the functional derivative instead of the usual one. For example,
$$\frac{\delta J(y)}{\delta J(x)}=\delta^{(4)}(x-y).$$
If we consider the Lagrangian for the free scalar field
\begin{equation}
L = \frac{1}{2}\partial_{\mu}\varphi\partial^{\mu}\varphi- \frac{1}{2}m^2\varphi^2 = -\frac{1}{2}\varphi \left( \Box^2+m^2\right)\varphi \equiv -\frac{1}{2}\varphi D^{-1}_\mathrm{c}\varphi,
\label{lagr}
\end{equation}
we can get for $Z[J]$,
\begin{equation}
Z[J]=\exp\left({\frac{1}{2}\int {\rm d}^4 x {\rm d}^4 y J(x)D_{\rm c}(x-y)J(y)}\right),
\label{ZJ}
\end{equation}
where the normalization of the measure is taken such that
$$Z[0]=1.$$
Here the function $D_{\rm c}$ is the Green function of the equation of motion:
$$D^{-1}_\mathrm{c} \cdot D_{\rm c} = 1,$$
which more accurately means that
$${\rm i}\left( \Box^2+m^2\right)_xD_{\rm c}(x - y) = \delta^{(4)}(x-y).$$
In the momentum representation, by taking the Fourier transform of both sides of this equation, one gets
$$ {\rm i}\left(-p^2 +m^2\right )D(p) = 1.$$
The formal solution of the equation is
$$D(p) = \frac{\rm i}{p^2-m^2}.$$
But we need to fix how to deal with the pole. The only possible choice is to add $+{\rm i}\varepsilon$. In this case
 the function $D_{\rm c}(p)$ has the familiar form of the Feynman propagator:
$$D_{\rm c}(p) = \frac{\rm i}{p^2-m^2+{\rm i}\varepsilon}.$$
Indeed, such fixing of the denominator leads to the fact that in the expression for the functional integral
\begin{equation}
Z[J]=\int D(\varphi)\exp\left({\mathrm{i}\int {\rm d}^4 x \left[\frac{1}{2}\partial_{\mu}\varphi
 \partial^{\mu}
\varphi-\frac{1}{2}m^2\varphi ^2+J \varphi\right]}\right),
\end{equation}
$m^2$ gets a shift $m^2 \rightarrow m^2 - \mathrm{i}\varepsilon$ and the term

$$\int D(\varphi)\exp\left({-\varepsilon \int {\rm d}x \varphi^2(x)}\right)$$
ensures the convergence of the integral.

On the other hand, as we have discussed already, such a form of the Feynman propagator leads to Feynman boundary conditions, namely if $x^0<y^0$ the particle propagates from $\vec x$ to $\vec y$, and if $x^0>y^0$  the corresponding antiparticle propagates from  $\vec x$ to $\vec y$, as follows from the expression
$$D_{\rm c}(x-y) = -\int \frac{{\rm d} \vec p}{(2\pi)^32 \omega_p}{\rm e}^{{\rm i} \vec p (\vec x - \vec y)}
\left[\Theta (t) {\rm e}^{-{\rm i} \omega_p t}+\Theta (-t) {\rm e}^{{\rm i} \omega_p t}\right].$$
So, the Feynman propagator is a Green function and, in other words, the inverse quadratic
form in the action (\ref{lagr}).

In the case of an interacting potential $V(\varphi)$ we get from (\ref{ZJ}) the following general expression for the generating functional:
\begin{equation}
Z_V[J] = \exp{\left(-{\rm i}\int {\rm d}^4x V\left(\frac{\delta}{{\rm i}\delta J(x)}\right)\right)}\cdot \exp{\left(\frac{1}{2}\int {\rm d}y {\rm d}z J(y)D_{\rm c}(y-z)J(z)\right).}
\label{ZVJ}
\end{equation}
As will be discussed, one can get the Green functions by taking the needed number of functional derivatives with respect to the source.
 However, this is not enough, since we do not
know how exactly the functional integral, called the generalized functional integral, and the Green functions
are related to the S-matrix elements needed to compute physics observables.

\section{Functional integral in holomorphic representation}

We start with  the harmonic oscillator as we did before:
$$H(p,q) = \frac{\hat p^2}{2}+\frac{\omega^2 \hat q^2}{2}.$$
The creation and annihilation operators have the form (\ref{OP}) with the commutator
\begin{equation}
[\hat a, \hat a^{\dag}]=1.
\label{61}
\end{equation}
The Hamiltonian of the system taken in a normal form (all creation operators are on the right-hand side) is
$$H=\omega \hat a^{\dag} \hat a.$$
The commutator relation (\ref{61}) has a very nice representation in terms
of holomorphic functions, which are introduced by means of the following scalar product:
$$\left<f_1|f_2 \right > = \int \left(f_1(a^*)\right)^*f_2(a^*){\rm e}^{-a^*a}\frac{{\rm d}a^*{\rm d}a}{2 \pi {\rm i}}.$$
With such a definition of the scalar product, the set of functions $\Psi_n(a^*) = \frac{(a^*)^n}{\sqrt{n!}}$, $n\ge 0$,
forms an orthonormal basis
\begin{equation}
\left<\Psi_n|\Psi_m \right > =\frac{1}{\sqrt{n!m!}} \int a^n (a^*)^m {\rm e}^{-a^*a}\frac{{\rm d}a^*{\rm d}a}{2 \pi {\rm i}}=\delta_{nm}.
\label{62}
\end{equation}
One can easily prove that $\sum_n|\Psi_n \left> \right <|\Psi_n| = 1$.

\emph{Problem.} Prove the relation (\ref{62}).

The operators $\hat a^{\dag}$ and $\hat a$ act according to the following rules:
\begin{equation}
\hat a ^{\dag}\cdot f(a^*) = a^* f(a^*), \,\,\,\,\,\,\hat af(a^*) = \frac{\rm d}{{\rm d}a^*}f(a^*).
\label{63}
\end{equation}
By direct substitution, one can prove the following relation:
\begin{equation}
\left<f_1|\hat a^{\dag}f_2 \right > =\left<\hat af_1|f_2 \right >,
\label{64}
\end{equation}
which means that the operators  $\hat a^{\dag}$ and $\hat a$ are conjugate to each other.

Now we will show a few simple formulas for the holomorphic representation given above, which are useful for a construction of the S-matrix.

Let us take some operator $\hat A$ with matrix element in our basis
\begin{equation}
A_{nm} =\left<\Psi_n|\hat A|\Psi_m \right >.
\label{65}
\end{equation}
The function
\begin{equation}
A(a^*,a) =\sum_{nm}A_{nm}\frac{(a^*)^n}{\sqrt{n!}}\frac{a^m}{\sqrt{m!}}=\sum_{nm}|n \left> \right < m|
\label{66}
\end{equation}
is called the kernel of the operator $\hat A$.
The kernel of a product of two operators is given by the convolution of kernels:
$$A_1A_2(a^*a) = \int A_1(a^* \alpha)A_2(\alpha ^* a){\rm e}^{-\alpha^* \alpha}\frac{{\rm d}\alpha^* {\rm d}\alpha}{2 \pi {\rm i}}.$$
The operator $\hat A$ can be decomposed into a formal series  of normal ordered creation and annihilation operators:
\begin{equation}
\hat A =\sum_{nm}K_{nm}{(\hat a^{\dag})^n}{(\hat a)^m}.
\label{67}
\end{equation}
The following function is called the normal symbol of the operator $\hat A$:
\begin{equation}
K(a^*,a) =\sum_{nm}K_{nm}(a^*)^n a^m.
\label{68}
\end{equation}

\emph{Problem.} Prove the relation between the kernel and the normal symbol of the operator $\hat A$:
\begin{equation}
A(a^*,a)={\rm e}^{a^* a}K(a^*,a).
\label{69}
\end{equation}
(Check the equality (\ref{69}) for the particular case $\hat A = \hat a^{\dag n}\hat a^l$.)

Now we use the relations (\ref{67}), (\ref{68}) and (\ref{69}) to construct the functional integral in the holomorphic representation.

Let the Hamiltonian of some system be $\hat H (\hat a^{\dag}, a)$. The evolution operator has the form
$$\hat U = {\rm e}^{-{\rm i}\hat H \cdot \triangle t}.$$
From (\ref{67}) and (\ref{69}), one can get the following formula for the kernel of the evolution operator:
\begin{equation}
U(a^*,a)={\rm e}^{[{a^* a}-{\rm i}h({a^* a})] \triangle t}
\label{610}
\end{equation}
for a small time interval $\triangle t$.

In the case of a finite interval, we can split it into small pieces $t'' - t' = N\cdot \triangle t$
and using our orthonormal bases (\ref{62}) we get the following form for the normal symbol of
the evolution operator, which is a convolution of products of the evolution operators:
$$U(a^*,a;t'',t')= \int \exp \left([a^*\alpha_{N-1} -\alpha_{N-1}^*\alpha_{N-1}+\cdots-\alpha_1^*\alpha_1+\alpha_1^*\alpha_0]\right. $$
$$-\left. {\rm i}\triangle t[h(a^*, \alpha_{N-1})+\cdots+h(\alpha^*,\alpha_0)] \right)\cdot \prod_{k=1}^{N-1} \frac{\mathrm{d}\alpha^*_kd \alpha_k}{2\pi \mathrm{i}}.$$
In the limit $\triangle t \rightarrow 0$, $N \rightarrow \infty$, $\triangle N = t'' - t'$, one gets
\begin{equation}
U(a^*,a;t'',t')=\int {\rm e}^{a^*\alpha(t'') }\cdot \exp \left(\int_{t'}^{t''}[-\alpha^* \alpha - {\rm i}h(\alpha^*,\alpha)]{\rm d}t \right) \cdot \prod_t \frac{{\rm d}\alpha ^* {\rm d}\alpha}{2 \pi {\rm i}},
\label{611}
\end{equation}
where the boundary conditions are $\alpha^*(t'') = a^*$, $\alpha(t') = a$.

In our case for the  harmonic oscillator $h(a^*,a)=\omega a^* a$ the integral (\ref{611}) can be easily computed.
To do this, one should take a variation
$$\delta \left[ a^* \alpha(t'')+\int_{t'}^{t''}[-\alpha^* \alpha - \mathrm{i}h(\alpha^*,\alpha)]{\rm d}t\right]$$
$$=a^* \delta \alpha (t'') +\int_{t'}^{t''}{\rm d}t[-\delta \alpha^* \dot \alpha - \alpha^*\delta \dot \alpha - {\rm i}\omega \alpha^* \delta \alpha- {\rm i}\omega  \delta \alpha^* \alpha]$$
$$=a^* \delta \alpha (t'') -\alpha^*(t'')\delta \alpha(t'')+\int^{t''}_{t'}{\rm d}t[-\delta\alpha^*(\dot \alpha+{\rm i}\omega \alpha)+\delta \alpha(\alpha^*-{\rm i} \omega \alpha^*)].$$
The extremum condition gives us the answer. Note that the first two terms cancel each other because
of the boundary condition $\alpha^*(t'') = a^*$. Extremum conditions can be simply solved:
\begin{equation}
a(t) = {\rm e}^{-{\rm i}\omega(t-t')}a(t'),\,\,\,\,\,a^*(t) = {\rm e}^{-{\rm i}\omega(t''-t)}a^*(t).
\label{612}
\end{equation}
Then, for the normal symbol of the evolution operator, we get
\begin{equation}
U(a^*,a;t'',t')= \exp (a^*a {\rm e}^{-{\rm i}\omega(t-t')}),
\label{613}
\end{equation}
substituting the solution (\ref{612}) into the exponent in (\ref{611}) where a non-zero contribution comes only from the first term in the exponent.

An important consequence of this is that for the operator $\hat A$ with the kernel $A(a^*,a)$ the kernel of the operator ${\rm e}^{\mathrm{i} \hat H_0 t'' }\hat A {\rm e}^{-{\rm i} \hat H_0 t' }$ is given by the convolution of corresponding kernels and is equal to
\begin{equation}
A\left( a^* {\rm e}^{\mathrm{i}\omega t''}, a {\rm e}^{-{\rm i}\omega t'} \right).
\label{614}
\end{equation}
This relation demonstrates the power of the holomorphic representation in which the evolution is simply reduced to the substitution of arguments:
$$a\rightarrow a {\rm e}^{-{\rm i} \omega t}.$$

Now we come back to the field theory. The field operator is given by
$$\hat \Phi (x) = \int \frac{\mathrm{d}\vec k}{(2\pi)^3 2k^0}[{\rm e}^{-{\rm i}kx}\hat a +  {\rm e}^{\mathrm{i}kx}\hat a^{\dag}(k)]$$
and the corresponding 4-momentum operator is
$$\hat P^{\mu}=\int \frac{\mathrm{d}\vec k}{(2\pi)^3 2k^0}k^{\mu}a^{\dag}(k)a(k).$$
Vacuum $\left| 0 \right>$ is the state $\hat a (k)\left |0 \right>=0$
and the one-particle state is $\left |\vec k \right>=\hat a^{\dag} (k)\left |0 \right>$.
The operator $\hat P^{\mu}$ acts on the one-particle state as
$$\left( \hat P^0 = \hat H \right)\left |\vec k \right> = k^0 \left |\vec k
\right>, \,\,\,\,\, \widehat {\vec P}\left |\vec k \right> = \vec k \left |\vec k \right>.$$
A multiparticle state is constructed  as $ \left |\vec k_1,\dots, \vec k_n \right>=\prod_{i=1}^n\hat a^{\dag}(\vec k_i)\left |0 \right>$.
Obviously,
$$\hat H\left |\vec k_1,\dots, \vec k_n \right>=\left(\sum^n_{i=1}k^0_i \right) \left |\vec k_1,\dots, \vec k_n \right>,$$
$$\widehat {\vec P}\left |\vec k_1,\dots, \vec k_n \right>=\left(\sum^n_{i=1}\vec k_i \right) \left |\vec k_1,\dots, \vec k_n \right>.$$
If $\hat H$ is the Hamiltonian of a system and $\hat H^0$ is a free Hamiltonian,
then the S-matrix is determined as the following limit of the evolution operator:
\begin{equation}
\hat S = \lim _{
\begin{array}{c}
\scriptstyle t'\to -\infty\\
\scriptstyle t''\to +\infty
\end{array}}{\rm e}^{\mathrm{i}\hat H_0t''}\hat U(t'',t'){\rm e}^{-{\rm i}\hat H_0t'}.
\label{615}
\end{equation}
With such a definition, it is clear that $\hat S = 1$ if $\hat H = \hat H^0$.

From the formulas (\ref{611}) and (\ref{614}), we can obtain the following representation for the kernel of the S-matrix:
\begin{equation}
S(a^*, a) = \lim _{
\begin{array}{c}
\scriptstyle t'\to -\infty\\
\scriptstyle t''\to +\infty
\end{array}} {\rm e}^{\int \frac{\mathrm{d}\vec k}{(2\pi)^3 2k^0}\left(\alpha^*(\vec k,t")\alpha (\vec k, t'')
 -\int_{t'}^{t''} [\alpha^*(\vec k,t)\dot{\alpha}(\vec k,t)+h(\alpha^*,\alpha)]{\rm d}t \right)}\cdot  \prod_{t,\vec k}\frac{\mathrm{d}\alpha^*\mathrm{d}\alpha}{2 \pi \mathrm{i}},
\label{616}
\end{equation}
where $\alpha ^*(\vec k, t'') = a^*(\vec k){\rm e}^{\mathrm{i}\omega t''},\;\;\alpha (\vec k, t') = a(\vec k){\rm e}^{-{\rm i}\omega t'}$.

Now we consider the system with an external source $J(x)$:
$$L=\frac{1}{2}\partial_{\mu}\varphi\partial^{\mu}\varphi-\frac{1}{2}m^2\varphi^2 +J(x)\varphi.$$
The kernel of the interaction operator $V=-J(x)\hat \varphi(x)$ is given by
$$V(a^*,a)=\int \frac{\mathrm{d}\vec k}{(2\pi)^3 2k^0}\left[ \gamma(\vec k,t)a^*(\vec k)+\gamma^*(\vec k,t)a(\vec k)\right],$$
where $\gamma(\vec k,t) = -\int J(\vec x,t){\rm e}^{-{\rm i}kx}\mathrm{d}\vec x$.
In order to take the integral (\ref{616}) once more, we should solve the extremum conditions
\begin{equation}
\begin{array}{c}
\dot \alpha(\vec k,t)+{\rm i}\omega (\vec k)\alpha(\vec k,t)+{\rm i}\gamma(\vec k,t)=0,\\
\dot \alpha^*(\vec k,t)-{\rm i}\omega (\vec k)\alpha^*(\vec k,t)-{\rm i}\gamma^*(\vec k,t)=0
\end{array}
\end{equation}
with the boundary conditions
\begin{equation}
\alpha ^*(\vec k, t'') = a^*(\vec k){\rm e}^{\mathrm{i}\omega t''},\;\;\alpha (\vec k, t') = a(\vec k){\rm e}^{-{\rm i}\omega t'}.
\end{equation}
The solution of these equations has the following form:
$$ \alpha (\vec k, t) = a(\vec k){\rm e}^{-{\rm i} \omega t} - \mathrm{i}{\rm e}^{-{\rm i} \omega t}\int_{t'}^{t''} {\rm e}^{\mathrm{i} \omega s}\gamma(\vec k, s)\mathrm{d}s,$$
$$ \alpha^* (\vec k, t) = a^*(\vec k){\rm e}^{\mathrm{i} \omega t} + \mathrm{i}{\rm e}^{\mathrm{i} \omega t}\int_{t'}^{t''} {\rm e}^{-{\rm i} \omega s}\gamma^*(\vec k, s){\rm d}s.$$
If one substitutes the obtained solutions into the integral (\ref{616}), one will get the following formula for the kernel of the S-matrix:
 \begin{eqnarray}
S_{J}(a^*,a) = \exp \left(\int \frac{\mathrm{d}\vec k}{(2\pi)^3 2k^0}\right .\left[ a^*(\vec k)a(\vec k) \right .\nonumber\\
+\int {\rm d}t \int \mathrm{d}\vec x  J (\vec x, t)\left(a^*(\vec k){\rm e}^{\mathrm{i} \omega t - i\vec k \vec x}+a(\vec k){\rm e}^{-{\rm i} \omega t + i\vec k \vec x}\right)/(2\omega)\\
\left .\left . -\frac{1}{2}\int_{-\infty}^{\infty}{\rm d}t\int_{-\infty}^{\infty}\mathrm{d}s\int \mathrm{d}\vec x d\vec y J (\vec x, t)J (\vec x, t)/(2 \omega ) {\rm e}^{-{\rm i}\omega |t-s|}{\rm e}^{\mathrm{i}\vec k \vec x - i\vec k \vec y}\right]\right). \nonumber
\end{eqnarray}
A transition from the kernel to the normal symbol for the S-matrix corresponds to omitting the first term in the exponent.

\begin {figure}[hb] \begin{center}
\includegraphics{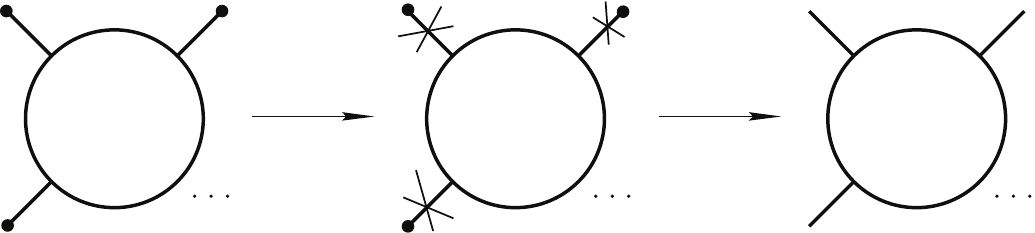}
\caption{Transition from the Green functions to the S-matrix elements}
\label{ss'}
\end{center}
\end {figure}

Now let us recall the solution of the free Klein--Gordon equation
$$\varphi _0(x) = \int \frac{\mathrm{d}\vec k}{(2\pi)^3 2k^0}\left[a(\vec k){\rm e}^{\mathrm{i} \omega t} + a^*(\vec k){\rm e}^{-{\rm i} \omega t} \right]$$
and the Green function (the propagator)
$$D_{\rm c} (x) = {\rm i}\int \frac{\mathrm{d}\vec k}{(2\pi)^3 2\omega}{\rm e}^{\mathrm{i} \vec k \vec x}{\rm e}^{-{\rm i} \omega_k |t|}=\mathrm{i}\int \frac{{\rm d}^4 k {\rm e}^{-{\rm i}kx}}{k^2-m^2+{\rm i}0}.$$
In terms of $\varphi_0(x)$ and $D_{\rm c}(x)$, one can rewrite the normal symbol for the S-matrix in the following way:
\begin{equation}
S_{J}(\varphi_0) = \exp\left( \mathrm{i}\int {\rm d}^4x J(x) \varphi_o(x) + \frac{1}{2}\int {\rm d}^4x{\rm d}^4yJ(x)D_{\rm c}(x-y)J (y) \right).
\label{619}
\end{equation}
Generalization of (\ref{619}) to the case of the integration potential $V(\varphi)$ is rather straightforward.
We can explore the relation we already faced with
\begin{equation}
\exp\left({-{\rm i}\int {\rm d}^4x V(\varphi)}\right) = \exp\left({- \mathrm{i}\int V\left(\frac{\delta}{\mathrm{i} \delta J}\right){\rm d}^4x}\right)\left.\exp\left({\mathrm{i}\int \varphi J {\rm d}^4y}\right)\right|_{J =0}.
\label{620}
\end{equation}
Then the expression for the normal symbol of the S-matrix takes a simple and elegant form:
\begin{eqnarray}
S_{V}(\varphi_0)  =\exp \left( {- \mathrm{i}\int V\left(\frac{\delta}{i \delta J}\right){\rm d}^4x}\right)\nonumber\\
\left.\cdot \exp \left({\mathrm{i}\int J(y) \varphi_0(y) {\rm d}^4y+\frac{1}{2}\int {\rm d}^4z{\rm d}^4yJ(y)D_{\rm c}(y-z)J (z)}\right)\right|_{J =0}.
\label{621}
\end{eqnarray}
If we do not put the source $J \to 0$ after taking functional derivatives, we have the normal symbol of the S-matrix in the presence of an external source:
\begin{eqnarray}
S_{V}(\varphi_0, J)  =\exp \left( {- \mathrm{i}\int V\left(\frac{\delta}{i \delta J}\right){\rm d}^4x}\right)\nonumber\\
\cdot \exp \left({\mathrm{i}\int J(y) \varphi_0(y) {\rm d}^4y+\frac{1}{2}\int {\rm d}^4z{\rm d}^4yJ(y)D_{\rm c}(y-z)J (z)}\right).
\label{622}
\end{eqnarray}
One can see that if we put $\varphi_0=0$ we get the same formula we as already derived for the generating functional
\begin{eqnarray}
Z_{V}(J)  =\exp \left( {- \mathrm{i}\int V\left( \frac{\delta}{i \delta J}\right) {\rm d}^4x}\right)
\cdot \exp \left({\frac{1}{2}\int {\rm d}^4z{\rm d}^4yJ(y)D_{\rm c}(y-z)J (z)}\right).
\label{623}
\end{eqnarray}
This observation allows us to get a very important relation, which is called the Lehmann--Symanzik--Zimmermann reduction formula.
To do this, we introduce in (\ref{622}) some arbitrary external field $\varphi (x)$ instead of  $\varphi_0 (x)$.
Then, by direct computation of functional derivatives with respect to $\varphi (x)$ and $J(x)$, one can check the following relation:
\begin{eqnarray}
\int \prod_i {\rm d}x_i\varphi_0(x_i)\left[\frac{1}{\mathrm{i}}\frac{\delta}{\delta \varphi_1}\dots\frac{\delta}{\delta \varphi_n}\left . S_{V}(\varphi_0, J)\right|_{\varphi,J =0}\right. \nonumber\\
\left.- \frac{1}{\mathrm{i}}\frac{\delta}{\delta\tilde J_1(x_1)}\dots\frac{\delta}{\delta \tilde J_n(x_n)}\left . Z(\tilde J)\right|_{J =0}\right] = 0,
\label{624}
\end{eqnarray}
where $\tilde J (x) = \int D_{\rm c}(x-y) J(y)$.
If we keep in mind that the expression of the S-matrix
$$S_{V}(\varphi_0)  =\sum_n\frac{1}{n}\int {\rm d}x_1 \dots {\rm d}x_n \varphi_0(x_1) \dots \varphi_0(x_n) S_n(x_1\dots x_n)$$
gives us the coefficient functions $S_n(x_1,\dots, x_n)$ of S-matrix scattering elements,
 and on the other hand the expansion of the generating functional
$$Z_V(J) = \sum_n \frac{1}{n!} \int {\rm d}x_1 \dots {\rm d}x_n J(x_1) \dots J(x_n) \frac {\delta Z}{\mathrm{i}\delta J(x_1)\dots \mathrm{i}\delta J(x_n) }$$
gives us the Green functions
$$G_n(x_1 \dots x_n)=\frac {\delta Z}{\mathrm{i}\delta J(x_1)\dots \mathrm{i}\delta J(x_n) },$$
we observe a simple correspondence. The reduction formula (\ref{624}) tells us how to compute
S-matrix elements by computing corresponding Green functions:
\begin{enumerate}
\item one should compute the Green function;
\item multiply all legs to the inverse propagator or, in other words, apply the operator
$$\prod_{i=1}^n \left(\Box_i + m^2 \right)$$ to each of the legs;
\item multiply the result by the product of the corresponding free fields:
$$\frac{1}{n}\prod_i \varphi_0(x_i).$$
\end{enumerate}
Schematically, the procedure is shown in Fig. \ref{ss'}.

This rule is very general and could be applied for all types of fields, not only for the scalar fields.

\section{Generating functional for Green functions and perturbation theory}

Now we  return to the generating functional written in the form
\begin{equation}
Z[J] = \int D(\varphi)\cdot \exp \left(\mathrm{i} \int {\rm d}^4x L(\varphi, \partial_{\mu}\varphi)+{\rm i} \int {\rm d}^4x J(x)\varphi(x) \right),
\end{equation}
where $L = \frac{1}{2}\partial_{\mu}\varphi\cdot \partial^{\mu}\varphi-\frac{1}{2}m^2\varphi^2 - V(\varphi)$.
If we take the second derivative,
$$
\left.\frac{\delta^{(2)}Z}{\mathrm{i}\delta J(x_1)\mathrm{i}\delta J(x_2)}\right|_{J=0} = \left< \varphi_1(x_1)\varphi_2(x_2)\right>,
$$
we obtain the Feynman propagator $D_{\rm c}(x_1-x_2)$, as can be easily seen from the form for $Z[J]$:
\begin{equation}
Z[J] =\exp\left( -{\rm i} \int {\rm d}x V\left( \frac{\delta}{\delta J}\right)\right)\cdot \exp\left( \frac{1}{2}\int \mathrm{d}y \mathrm{d}z J(y)D_{\rm c}(y-z)J(z)\right).
\end{equation}
We obtain the same function as we have obtained from the time-ordered product of field operators (see (\ref{Feynman_prop}))
$$
D_{\rm c}(x_1 - x_2)=\left< 0\right|T\{ \hat \Phi(x_1) \hat \Phi(x_2)\}\left| 0\right> =  \left< \varphi_1(x_1)\varphi_2(x_2)\right>.
$$
This is always the case. Derivatives  of the generating functional automatically give $T$-products of the corresponding field operators:
$$
\frac{\delta^{(2)}Z}{\mathrm{i}\delta J(x_1)\dots \mathrm{i}\delta J(x_n)}
\equiv \left< \varphi_1(x_1)\dots \varphi_n(x_n)\right>=\left< 0\right|T\{ \hat \Phi(x_1) \dots \hat \Phi(x_n)\}\left| 0\right>.
$$
At this point let us consider as an example a theory with $V(\varphi) = \frac{\lambda}{4!}\varphi^4 $.
In such a theory if we take two derivatives and expand the exponent on $\lambda$, we get
\begin{equation}\begin{array}{rcl}
\left< \varphi_1(x_1)\varphi_2(x_2)\right>&=&D_{\rm c}(x_1 - x_2)+ \lambda\left[D_{\rm c}(x_1 - x_2) D^2_\mathrm{c}(0)+\cdots\right]
\nonumber\\
&&+\frac{1}{2}\left(\frac{\lambda}{4!}\right)^2\left[ 72 D^2_\mathrm{c}(0)D^2_\mathrm{c}(x_1 - x_2)+24 D^4_\mathrm{c}(x_1-x_2)+\cdots\right] +\cdots.
\end{array}
\label{73}
\end{equation}

\begin{figure}[htb] \begin{center}
\includegraphics{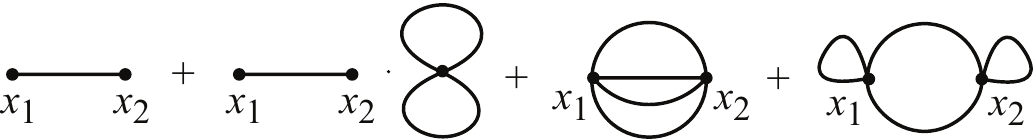}
\caption{Illustration of (\ref{73})}
\label{DDD}
\end{center}
\end{figure}

It is very useful to introduce the Feynman rules and Feynman diagrams. The closed line corresponds to the propagator $$\frac{1}{\mathrm{i}}D_{\rm c}(x-y).$$
Each interaction vertex corresponds to $$-{\rm i}\lambda =\left. -{\rm i} \frac{{\rm d}^4V(\varphi)}{{\rm d}\varphi ^4}\right|_{\varphi =0}.$$
In terms of Feynman rules, the corrections to the two-point correlation function (\ref{73}) are given by Feynman diagrams shown in Fig. \ref{DDD}.

One should add a symmetry factor, which corresponds to possible permutations of equivalent lines.
If one takes the Fourier transformation, one can formulate the rules in momentum space, which are usually used in practical computations.
In momentum space the integral $$\int\frac{{\rm d}p}{(2\pi)^4}$$ corresponds to each loop
and in each vertex the momentum conservation law takes place.

We will be more specific and precise later in the formulation of Feynman rules for the  case of the SM.

But now we need to consider a few more properties of the generating functional. As we have seen already from the two-point
correlator the perturbative expansion contains disconnected diagrams, which are not really needed in computations.
The way out of this problem is to consider the logarithmic function of the generating functional:
$$ {\rm i}W[J] = \ln Z[J],\qquad Z[J]={\rm e}^{{\rm i}W[J]}.$$
Then, for the functional derivatives, one gets
$$\frac{\delta W}{\delta J} = \frac{1}{Z}\frac{\delta Z}{\mathrm{i}\delta J},\qquad \frac{\delta^2 W}{\delta J_1 \delta J_2} = {\rm i} \frac{1}{Z}\frac{\delta^2 Z}{\mathrm{i}\delta J_1\mathrm{i}\delta J_2} - \frac{1}{Z^2}\frac{\delta Z}{\mathrm{i}\delta J_1}\frac{\delta Z}{\mathrm{i}\delta J_2}\dots,$$
where additional terms exactly cancel out disconnected pieces in the Green functions.
The property that the logarithmic function leads to connected diagrams is a particular example of a more general theorem
in graph theory.

The functional
\begin{equation}
W[J] = \frac{1}{\mathrm{i}} \ln Z[J]
\label{functional_W}
\end{equation}
is called the generating functional for connected Green functions.

\begin {figure}[ht] \begin{center}
\includegraphics{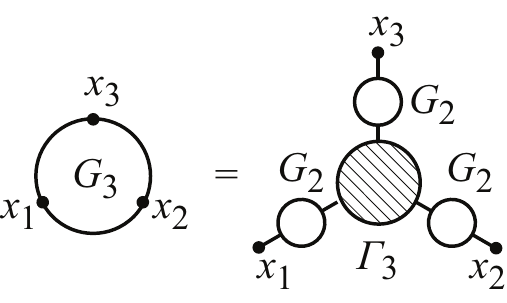}
\caption{Connected three-point Green function is equal to one-particle irreducible three-point function convoluted with three propagators.}
\label{7_2}
\end{center}
\end {figure}

The next important observation is related to the functional Legendre transformation:
\begin{equation}
\Gamma[\varphi_{\rm cl}] = W[J] - \int {\rm d}^4x J(x) \varphi_{\rm cl}(x),
\label{functional_G}
\end{equation}
where $\varphi_{\rm cl} = \delta W/\delta J$ with a formal solution $J=J(\varphi_{\rm cl})$.
If we take functional derivatives of the functional $\Gamma[\varphi_{\rm cl}]$, we get the so-called one-particle irreducible Green functions
(Feynman diagrams corresponding to such functions cannot be split up into disconnected pieces by cutting only one internal line):
\begin{equation}
\frac{\delta \Gamma}{\delta \varphi_{\rm cl}} = \frac{\delta W}{\delta J}\frac{\delta J}{\delta \varphi_{\rm cl}}
 - J(x) -\varphi_{\rm cl}\frac{\delta J}{\delta \varphi_{\rm cl}}.
\label{76}
\end{equation}
The functional $\Gamma$ is called the  effective action. Two terms in (\ref{76}) are cancelled out and we get
\begin{equation}
\frac{\delta \Gamma}{\delta \varphi_{\rm cl}{(x)}} = - J(x).
\label{77}
\end{equation}
Now one can take functional derivatives from both sides of (\ref{77}) and get the following relations:
\begin{equation}
\frac{\delta^2 \Gamma}{\delta \varphi_{\rm cl}(x_1)\delta \varphi_{\rm cl}(x_2)} = -\frac{\delta J(x_1)}{\delta \varphi_{\rm cl}(x_2)}  =-\left[ \frac{\delta W}{\delta J(x_1)\delta J(x_2)}\right]^{-1}.
\label{78}
\end{equation}
If we introduce notation for the connected Green function:
$$G_n(x_1,\dots, x_n) =\left. -{\rm i}\frac{\delta W[J]}{\mathrm{i}\delta J(x_1)\dots \mathrm{i}\delta J(x_2)}\right|_{J=0}$$
and for the one-particle irreducible Green function:
$$\Gamma_n(x_1, \dots, x_n) = \left.-{\rm i}\frac{\delta^{(n)}\Gamma [\varphi_{\rm cl}]}{\delta \varphi_{\rm cl}(x_1)\dots\delta \varphi_{\rm cl}(x_n)}\right|_{\varphi_{\rm cl}=0},$$
then the formula (\ref{78}) can be written in the following form:
$$\Gamma_2 = G_2^{-1},$$
which means that the irreducible two-point Green function is nothing but the inverse propagator.
If we take more derivatives on both sides of (\ref{78}) we obtain relations between connected and one-particle irreducible Green functions.
For the three-point Green function, it is presented schematically in Fig.~\ref{7_2}.
If we restore the Planck constant $\hbar$, the generating functional has the form
$$
Z[J] = \int D(\varphi) \exp\left( \frac{\mathrm{i}}{\hbar}S[\varphi] +{\rm i}\int {\rm d}xJ(x)\varphi(x)\right),
$$
where $S[\varphi]$ is the action.
In the quasiclassical limit $(\hbar \rightarrow 0)$, the functional integral is dominated by the stationary trajectory:
$$\left.\frac{\delta S[\varphi]}{\delta \varphi(x)}\right|_{\varphi = \varphi_{\rm cl}}+J = 0$$
Therefore $$Z[J]\sim \exp\left( \frac{\mathrm{i}}{\hbar}S[\varphi_{\rm cl}] +{\rm i}\int {\rm d}xJ(x)\varphi_{\rm cl}(x)\right)$$
and one can see from (\ref{functional_W}) that
$$W[J] = S[\varphi_{\rm cl}]+\int {\rm d}x J(x) \varphi_{\rm cl}$$
and, by comparing with (\ref{functional_G}), we obtain
$$\Gamma [\varphi_{\rm cl}] = S[\varphi_{\rm cl}].$$
So, we can conclude that the irreducible Green functions are $\Gamma _n(x_1,\dots, x_n) $, the effective vertices of the theory:
\begin{equation}
\Gamma _n(x_1,\dots, x_n) =(-{\rm i})\frac{\delta ^{(n)}S[\varphi_{\rm cl}]}{\delta \varphi_{\rm cl}(x_1)\dots\delta \varphi_{\rm cl}(x_n)}.
\label{effective_vertex}
\end{equation}
We obtain this formula for the case of scalar fields as an example, but it remains true for any theory with corresponding obvious changes.
Note that the formula (\ref{effective_vertex}) is very useful to get Feynman rules for complicated vertices in the interaction Lagrangian.
For example, three- and four-gluon vertices are obtained with all needed symmetry properties.

Up to now we have considered basic ingredients of the quantum field theory for the case of scalar fields.
 However, there are many other fields and corresponding particles which have different spin properties. In the SM
there are leptons and quarks, being fermions with the spin $1/2$, and the boson fields with the spin 1. We begin our brief consideration with spin-$1/2$ fermion  fields.

\section{Fermion fields}

Spin-$1/2$ particles with mass $m$ are described by the four-component field $\Psi$. The Lagrangian for the field has the well-known form
\begin{equation}
L = \bar\Psi {\rm i}\partial_{\mu}\gamma^{\mu}\Psi - m \bar \Psi \Psi.
\end{equation}
The least-action principle leads to the famous Dirac equation of motion:
\begin{equation}
\left( \mathrm{i}\gamma^{\mu} \partial_{\mu} - m\right) \Psi = 0,
\end{equation}
where $\gamma^{\mu}$ ($\gamma^{0}$,  $\gamma^{1}$,  $\gamma^{2}$,  $\gamma^3$) are the Dirac $(4\times 4)$ matrices.
The matrices $\gamma_{\mu}$ obey the anticommutation relation
\begin{equation}
\left\{ \gamma_{\mu} \gamma_{\nu}\right\} = 2\eta_{\mu \nu}.
\end{equation}
There are several representations for $\gamma$-matrices. In the SM chiral or Weyl spinors
are of particular  importance. Therefore, we use the Weyl representation of $\gamma$-matrices
\begin{equation}
 \gamma_{\mu} =\left(
\begin{array}{cc}
0 & \sigma^{\mu}\\
\bar \sigma^{\mu}& 0
\end{array}
\right).
\end{equation}
where $\sigma^0 = I$, $\sigma^i = \tau^{i}$, $\bar \sigma^0 = I$, $\bar\sigma^i = -\tau^i$ and $\tau^i$ are the $(2\times2)$ Pauli matrices.

In this representation the $\gamma^5$-matrix has the following form:
 \begin{equation}
 \gamma^5 =\mathrm{i}\gamma^1\gamma^2\gamma^3\gamma^4=\left(
\begin{array}{cc}
-I &0\\
0& I
\end{array}
\right).
\end{equation}
Chiral spinors
 \begin{equation}
 \Psi_{\mathrm{L,R}} = \frac{1 \mp \gamma_5}{2}\Psi
\end{equation}
in this notation
$$\Psi_{\rm L}  =\left(
\begin{array}{c}
\Psi_1\\
\Psi_2\\
0\\0
\end{array}
\right),\,\,\,\,\,\Psi_{\rm R}  =\left(
\begin{array}{c}
0\\0\\
\Psi_3\\
\Psi_4

\end{array}
\right)
$$
are in fact two-component objects.

In momentum space the function $\Psi (x)$ is decomposed into positive- and negative-energy parts
\begin{equation}
u_{\lambda}(p){\rm e}^{-{\rm i}px} \mbox{  and  } v_{\lambda}(p){\rm e}^{\mathrm{i}px}
\end{equation}
and the spinors $u(p)$ and $v(p)$ obey the Dirac equation in the following form:
\begin{equation}
\begin{array}{c}
\left( p_{\mu}\gamma^{\mu}- m\right)u_{\lambda}(p)=0,\\
\left( p_{\mu}\gamma^{\mu}+ m\right)v_{\lambda}(p)=0.
\end{array}
\end{equation}
The concrete form of spinors is different in different parametrizations  of $\gamma$-matrices, and in the Weyl representation the spinors are
\begin{equation}
u_{\lambda}=\left(
\begin{array}{c}
\sqrt{p^0+\vec p\vec\sigma}\xi_{\lambda}\\
\sqrt{p_0-\vec p\vec\sigma}\xi_{\lambda}
\end{array}\right),
\end{equation}

\begin{equation}
v_{\lambda}=\left(
\begin{array}{c}
\sqrt{p^0+\vec p\vec\sigma}\eta_{\lambda}\\
-\sqrt{p_0-\vec p\vec\sigma}\eta_{\lambda}
\end{array}\right),
\end{equation}
where $\xi$ and $\eta$ are two-component spinors determined by fixing some quantization axis. Left and right chiral spinors are then
$$u_{\mathrm{L,R}} = \frac{1 \mp \gamma_5}{2}u_{\lambda},$$
$$v_{\mathrm{L,R}} = \frac{1 \mp \gamma_5}{2}v_{\lambda}.$$
Normalization conditions and summation over indices are as follows:
$$\bar u_{\lambda} u_{\lambda'}=2m\delta_{\lambda \lambda'},\,\,\,\,\,\bar v_{\lambda}v_{\lambda'}=-2m\delta_{\lambda \lambda'},$$
$$\sum_{\lambda} u_{\lambda} \bar u_{\lambda}=p_{\mu}\gamma^{\mu}+m,\,\,\,\,\,\sum_{\lambda} v_{\lambda} \bar v_{\lambda}=p^{\mu}\gamma_{\mu}-m.$$

Quantization of the Dirac field is similar to the scalar case considered above with a very important difference.
 In order to have correct Fermi statistics and obey the Pauli principle, the commutation relations
 in the scalar case should be replaced by corresponding anticommutation relations:
\begin{equation}
\left\{\hat \pi_{\alpha}(t,\vec x), \Psi_{\beta}(t,\vec x')\right\} = -{\rm i}\delta_{\alpha \beta}\delta(\vec x - \vec x'),
\label{910}
\end{equation}
where $\alpha = 1,2,3,4$ and the field momentum is $$\pi_{\alpha}(t,\vec x)=\frac{\partial L}{\partial \dot \Psi_{\alpha}}=\mathrm{i}\Psi^{\dagger}_{\alpha}.$$
The fermionic field operator may be constructed with the help of spinors obeying the Dirac equation
$$\Psi(x) = \int\frac{\mathrm{d} \vec p}{(2\pi)^3 {p^0}}\sum_{\lambda=1,2}\left[\hat b_{\lambda}(p)u_{\lambda}(p){\rm e}^{-{\rm i}px}+  \hat d^{\dagger}_{\lambda}(p)v_{\lambda}(p){\rm e}^{\mathrm{i}px}\right],$$
$$\bar\Psi(x) = \Psi^{\dagger}\gamma^0=\int\frac{\mathrm{d} \vec p}{(2\pi)^3 {p^0}}\sum_{\lambda=1,2}\left[\hat b^{\dagger}_{\lambda}(p)\bar u_{\lambda}(p){\rm e}^{\mathrm{i}px}+  \hat d_{\lambda}(p)\bar v_{\lambda}(p){\rm e}^{-{\rm i}px}\right].$$
It is easy to check that from the anticommutators (\ref{910}) the creation and annihilation operators satisfy the anticommutation
relations in the following form:
\begin{equation}
\left\{\hat b_{\lambda}(\vec p), \hat b_{\lambda'}(\vec p')\right\}
= \left\{\hat d_{\lambda}(\vec p), \hat d_{\lambda'}(\vec p')\right\}=(2\pi)^3 2p^0 \delta(\vec p - \vec p')\delta_{\lambda \lambda'}.
\end{equation}
Then one-particle and one-antiparticle states are obtained from the vacuum state $\left|0\right>$ by acting of the creation operators:
$$\hat b^{\dagger}_{\lambda}(\vec p)\left|0\right>\mbox{  and  }\hat d^{\dagger}_{\lambda}(\vec p)\left|0\right>.$$
In the same way, by acting of creation operators for particle and antiparticle on the vacuum state,
one gets two-, three-, $\dots$ particle states. Because of zero anticommutators, for any creation operator one gets nicely the Pauli principle:
\begin{equation}
\left\{\hat b^{\dagger}_{\lambda}(\vec p), \hat b^{\dagger}_{\lambda}(\vec p)\right\} = 0\,\,\Rightarrow \,\, \hat b^{\dagger}_{\lambda}(\vec p), \hat b^{\dagger}_{\lambda}(\vec p)\left|X\right>\equiv 0
\end{equation}
for any state $\left|X\right>.$
From the field operators one can get the Feynman propagator (T-ordered correlator) in a similar way as was done for the scalar case:
\begin{equation}
\left< 0\right| T \left( \bar \Psi(x_1)\Psi(x_2)\right)\left|0\right>=
\frac{-1}{\rm i}\int \frac{{\rm d}p}{(2\pi)^4}\frac{p_{\mu}\gamma^{\mu}+m}{p^2-m^2+{\rm i}0}.
\end{equation}

Considering the path-integral method for the fermion field, one can construct the holomorphic representation similar
to the scalar case. However, now we have to deal with anticommuting numbers called Grassmann numbers, which form the Grassmann algebra:
$$\left(a_{\alpha} \right)^* = a_{\alpha}^*,\,\,\,\,\left(a_{\alpha}^* \right)^* = a_{\alpha},$$
$$\left\{a_{\alpha} a_{\beta}\right\} = \left\{a_{\alpha}^* a_{\beta}^*\right\} =\left\{a_{\alpha} a_{\beta}^*\right\} =0,$$
$$ca_{\alpha} = a_{\alpha} c;\,\,\,\,\,c a_{\alpha}^* = a_{\alpha}^* c,$$
where $\alpha = 1, \dots, n$, $c$ are the usual numbers.

A function of Grassmann variables has therefore the generic form
\begin{equation}
\begin{array}{rcl}
f(a,a^*)& =& f_{00}+\sum_{\alpha _1}f_{\alpha_1|0}a_{\alpha _1} +\sum_{\alpha _1}f_{ 0|\alpha_1}a_{\alpha _1}^*\\
&& +
\sum_{\alpha _1\alpha _2}f_{\alpha_1|\alpha_2}a_{\alpha _1}  a_{\alpha _2}^*
+\cdots f_{1\dots n|n\dots 1}a_1\dots a_n a_n^* \dots a^*_1.
\end{array}
\label{911}
\end{equation}
The expression (\ref{911}) reminds us of the norm-ordering operator products we have used already.
The operations of differentiation and integration are defined as follows:
$$\frac{\partial}{\partial a_{\alpha}}a_{\beta} = \frac{\partial}{\partial a_{\alpha}^*}a_{\beta}^* = \delta_{\alpha \beta}, \,\,\,\,\,
\frac{\partial}{\partial a_{\alpha}}a_{\beta}^* = \frac{\partial}{\partial a_{\alpha}^*}a_{\beta} = 0,$$
$$\frac{\partial}{\partial a_{\alpha}}f = f_{\alpha};\,\,\,\, \frac{\partial}{\partial a_{\alpha}^*}f=\bar f_{\alpha}, \,\,\,\,\,
\int \mathrm{d} a_{\alpha}f = f_{\alpha};\,\,\,\, \int \mathrm{d} a_{\alpha}^* f=\bar f_{\alpha},$$
where $f_{\alpha}$ does not depend on $a_{\alpha}$ and $\bar f_{\alpha}$ does not depend on $a_{\alpha}^*$. One can check the anticommutation relation for differentials:
$$\left\{\mathrm{d}a_{\alpha} \mathrm{d}a_{\beta}\right\} = \left\{\mathrm{d}a_{\alpha}^* \mathrm{d}a_{\beta}^*\right\} =\left\{\mathrm{d}a_{\alpha} \mathrm{d}a_{\beta}^*\right\} =0.$$
If we denote
$$\mathrm{d}a^*\mathrm{d}a = \mathrm{d}a^*_1 \dots \mathrm{d}a^*_n \mathrm{d}a_n \dots \mathrm{d}a_1$$
and take into account that for the function defined in (\ref{911})
$$\int \mathrm{d}a^* \mathrm{d}a f(a,a^*) = f_{1\dots n|n \dots 1},$$
we can easily prove the following relation  by  expanding in series in $a$ and $a^*$:
\begin{equation}
\int \mathrm{d}a^* \mathrm{d}a \exp\left(  \sum a^*_{\alpha}A_{\alpha \beta}a_{\beta}
+ \sum \eta^*_{\alpha}a_{\alpha}+{\rm i}\sum \eta_{\alpha}a^*_{\alpha}\right)
 = \det A \exp \left(\eta_{\alpha}^*A^{-1}_{\alpha \beta}\eta_{\beta} \right).
\label{grassman_integral}
\end{equation}
One can see from (\ref{grassman_integral}) that in contrast to the integration with the usual complex numbers (see (\ref{interg_numbers}))
the determinant appears in the numerator in the case of anticommuting Grassmann numbers (\ref{grassman_integral}).

Similar to the case of the scalar field, one can get a formula for the S-matrix normal symbol:
$$S_{\eta}(b^{\dagger},d^{\dagger}, b, d)
= \frac{1}{N}\exp \left( -{\rm i} \int \bar\eta S_C \eta + {\rm i}\int (\bar \eta \Psi_0 +\overline\Psi_0\eta) \right),$$
where $\bar \Psi_0 = \Psi^{\dagger}_0\gamma^0$, $\Psi_0$ is a solution of the free Dirac equation
$$\Psi_0(x) = \int \frac{{\rm d} \vec k}{(2\pi)^32k_0}
\left[ \sum_{\lambda =1, 2} b_{\lambda}(\vec k)u_{\lambda}(\vec k){\rm e}^{-{\rm i}kx}+  \sum_{\lambda =1, 2}
 d^*_{\lambda}(\vec k)v_{\lambda}(\vec k){\rm e}^{{\rm i}kx} \right],$$
$\eta (x)$ is the fermion source and
$$S_{\rm C} (x) = \frac{\rm i}{(2\pi)^4}\int {\rm d}k {\rm e}^{-{\rm i}kx}\frac{\hat k +m}{k^2 -m^2 +{\rm i}0}$$
is the Feynman propagator for the fermion field.

Taking into account the following relation for the functional measure:
$$\prod_{t, k, \lambda} {\rm d}b^*_{\lambda}(t, \vec k) {\rm d}b_{\lambda}(t, \vec k){\rm d}d^*_{\lambda}(t, \vec k){\rm d}d_{\lambda}(t, \vec k)
=  {\mbox {Const.}} \prod_{x, \alpha}{\rm d} \bar \Psi_{\alpha}(x){\rm d} \Psi_{\alpha}(x), $$
one gets the answer for the generating functional for the fermion Green functions in compact form similar to the scalar case:
$$Z(\bar \eta, \eta) = N^{-1}\int \exp\left({\rm i}\int {\rm d}^4 x (L(x)+\bar \eta \Psi +\bar \Psi \eta) \right)\prod_x {\rm d}\bar \Psi (x){\rm d}\Psi (x).$$

\section{Quantization of theories with the gauge fields}

The gauge field was first introduced in QED when the Maxwell equations were rewritten in terms
of the 4-vector potential $A_{\mu}(x)$.

The equation in terms of the field $A_{\mu}$,
$$\partial_{\mu}F^{\mu \nu} = 0,$$
where  $F_{\mu\nu}=\partial_{\mu} A_{\nu} - \partial_{\nu}A_{\mu}$, is invariant under the local $U(1)$ transformation:
$$A'_{\mu} = A_{\mu}+\partial_{\mu}\alpha(x).$$
It is easy to check the $U(1)$ invariance of the QED Lagrangian
\begin{equation}
L=-\frac{1}{4}F_{\mu\nu}F^{\mu\nu}+\bar\Psi({\rm i}\hat D - m)\Psi, \,\,\,\,\Psi\rightarrow {\rm e}^{{\rm i}e\alpha}
\Psi, \,\,\,\,A_{\mu}\rightarrow A_{\mu}+\partial_{\mu}\alpha,
\end{equation}
where $D_{\mu} = \partial_{\mu}-{\rm i}eA_{\mu}$ is the covariant derivative.

In the SM we deal not only with the abelian $U(1)$ group but also with non-abelian $SU(N)$ groups,
$SU(2)$ for the EW and $SU(3)$ for the strong forces.

We are not discussing here generic structure and properties of Lie groups but introduce briefly
the $SU(N)$ group. $SU(N)$ is a group of unitary matrices $U$ $(U^{\dag}U = 1)$  with determinant equal to 1 $(\det U=1)$.
 Elements of the group $U(x)$ may depend on the 
space--time point $x^{\mu}$.

If we want the theory to be invariant under $SU(N)$ transformation the covariant derivative in the Lagrangian
\begin{equation}
L=\bar \Psi ({\rm i}\hat D-m)\Psi
\end{equation}
should transform as
$$D_{\mu}\Psi\rightarrow (D_{\mu}\Psi)^{U} = U D_{\mu}\Psi,$$
$$\left( \partial_{\mu}-{\rm i}gA^U_{\mu} \right) U \Psi  = U\left( \partial_{\mu}-{\rm i}gA_{\mu} \right) \Psi.$$
From this, one gets the following transformation form for the potential $A$:
\begin{equation}
A_{\mu}^{U} = U A_{\mu}U^{-1}+\frac{\rm i}{g}U \partial_{\mu}U^{-1}.
\end{equation}
The kinetic term for the non-abelian $A_{\mu}$ field is constructed as the gauge-invariant operator
\begin{equation}
L_{A} =- \frac{1}{2}\rm{Tr}( F_{\mu\nu}F^{\mu\nu}).
\end{equation}
In the SM all gauge fields are taken in the adjoint representation:
$$A_{\mu}(x) = A^a_{\mu}(x)t^a,$$
where $t^a$ ($a=1,\dots, N-1$ for $SU(N)$) are so-called generators of the group.
As do all generators for the Lie group, the generator $t^a$ obeys the following commutation relation:
$$\left[t^a, t^a \right] = f^{abc}t^c,\,\,\,\,\rm{Tr}( t^a) = 0.$$
One may choose normalization conditions as
\begin{equation}
\rm{Tr}(t^a t^b) = \frac{1}{2}\delta ^{ab}.
\end{equation}
Then the Lagrangian for the gauge field takes the form
\begin{equation}
L = -\frac{1}{4}F_{\mu\nu}^a F^{a\mu\nu},
\end{equation}
where the field strength tensor is
\begin{equation}
F_{\mu\nu}^a = \partial_{\mu}A^{a}_{\nu} - \partial_{\nu}A_{\mu}^a+ g f^{abc}A^b_{\mu}A^c_{\nu}.
\end{equation}
One can express the unitary matrix $U(1)$ in the form $U(x)={\rm e}^{{\rm i} g\alpha^a(x)t^a}$.
 Then the transformation for the gauge field $A^a_{\mu}$ takes the form
\begin{equation}
A^a_{\mu}\rightarrow (A^{\alpha})^a_{\mu}=A^a_{\mu}+\partial_{\mu}\alpha^a+gf^{abc}A^b_{\mu}\alpha^c = A^a_{\mu}+D_{\mu}^{ac} \alpha^c,
\end{equation}
where the covariant derivative in components is
$$
D_{\mu}^{ac} = \partial_{\mu}\delta^{ac}+g f^{abc}A^b_{\mu}.
$$

Part of the SM describing strong interactions is based on the $SU_C(3)$ group. It is called QCD
with the Lagrangian
\begin{equation}
L_{\rm QCD} = -\frac{1}{4}G_{\mu\nu}^a  G^{\mu\nu a} +\bar q_i \left( {\rm i}(D_{\mu})_{ij}\gamma^{\mu}_{ij}-m \delta_{ij}\right)q_j,
\end{equation}
where $q=1,2,3$, $(D_{\mu})_{ij}=\partial _{\mu}\delta_{ij}-{\rm i}g(t^a)_{ij}A_{\mu}^a$.

On the classical level abelian and non-abelian theories look very elegant. However,
problems appear when the theories are quantized.
The reason can be seen already in QED, where we know that in a theory described by the field $A_{\mu}$
we have four components but only two of them are physical degrees of freedom corresponding to two polarizations of the physics photon.
 This problem is manifested in the fact that the quadratic form of the Lagrangian
$$A^{\mu}D^{-1}_{\mu\nu}A^{\nu} = A^{\mu}\left( \Box g_{\mu\nu}-\partial_{\mu}\partial_{\nu} \right)A^{\nu}$$
or, in momentum space,
$$\left( k^2g_{\mu\nu}-k_{\mu}k_{\nu} \right)$$
 does not have an inverse form. As we have seen in cases of scalar and fermion fields the inverse of the differential operator,
 the Green function, is the propagator. So, here one cannot get the propagator in such a way.

The way out of this problem is a correct quantization procedure called the quantization of constrained systems.
 The reason that the functional integral
\begin{equation}
\int \prod_{\mu, x}{\rm d}A_{\mu}(x)\exp \left\{ {\rm i}\int {\rm d}x \left( -\frac{1}{4}F^a_{\mu\nu}F^{a \mu\nu}\right) \right\}
\end{equation}
does not give a reasonable result is that there are an infinite number of gauge
configurations $(A^{\alpha})^a_{\mu}$, which differ only by the gauge transformation,
leading to identical physics results since the action is gauge invariant.
So, one should perform the functional integration taking only one representative
from such gauge configuration.

Without going into details, the final recipe is as follows:
\begin{equation}
\int DA \delta \left(F(A)\right) \det \left( \Delta^F_{gh}\right) {\rm e}^{{\rm i}S(A)},
\end{equation}
where $DA = \prod_{\mu,x}\mathrm{d}A_{\mu}(x)$ and the so-called Faddeev--Popov determinant $\det \left( \Delta^F_{gh}\right)$ is introduced to ensure the
gauge invariance of the functional measure. The $\delta$-function fixes the gauge condition of gauge choice $F(A) = 0$.

We recall briefly the main ideas of the method proposed by Faddeev and Popov.
Let us introduce a functional integral that is equal to unity:
\begin{equation}
1=\int D\alpha \cdot \delta\left( F(A^{\alpha})\right)\det \left(\frac{\delta F}{\delta \alpha}\right),
\label{10FP}
\end{equation}
where $D \alpha = \prod _x \mathrm{d}\alpha (x)$.
Substituting (\ref{10FP}) into the integral (\ref{61}) and performing the gauge transformation $(A^{\alpha})^a_{\mu}(x)\rightarrow A^a_{\mu}(x)$, one gets
\begin{equation}
\int D\alpha \int DA {\rm e}^{{\rm i}S[A]}\cdot \delta \left( F(A)\right)\cdot \det \left(\frac{\delta F}{\delta \alpha}\right).
\label{11FP}
\end{equation}
The factor $\det \left(\frac{\delta F}{\delta \alpha}\right)$ does not depend on $\alpha$ and therefore
 the integration over the gauge group is factorized out. This infinite factor is included into the normalization factor
of the functional measure
and therefore could be dropped.

The main idea of quantization and the Faddeev--Popov method could be illustrated in a very simple example with the usual integrals.
Let us consider an integral
\begin{equation}
I=\int\int_{-\infty}^{\infty} {\rm d}x_1 {\rm d}x_2 {\rm e}^{-x^2_1 -x^2_2 +2x_1x_2} = \int {\rm d}x_1 {\rm d}x_2 {\rm e}^{-x_i A_{ij}x_j},
\end{equation}
where
$$A_{ij} = \left( \begin{array}{rr} 1&-1\\-1&1  \end{array}\right) \mbox{  and  }\det A = 0.$$
We can obviously see that by making a substitution of variables we get
$$I = c\int_{-\infty}^{\infty} {\rm d}x\int_{-\infty}^{\infty} {\rm d}y {\rm e}^{-x^2} \mbox{  when  }x=x_1-x_2,\,\,\,\,y=x_1+x_2.$$
The `action' in the integral $I$ is invariant under translation:
$$x_1\rightarrow x_1+a,\,\,\,\,x_2\rightarrow x_2+a.$$
And, the integral $\int_{-\infty}^{\infty} {\rm d}y$ gives simply the infinite volume of the algebra corresponding to the translation group.

Now let us substitute  $\hat 1$ into the integral $I$:
$$ \int_{-\infty}^{\infty} {\rm d}{\omega}\delta \left(F(x_i+\omega)\right) \det \left(\frac{\delta F}{\delta \omega}\right)= 1,$$
$$\int {\rm d} \omega \int {\rm d}x_1 {\rm d}x_2 {\rm e}^{-x_i A_{ij}x_j}\delta\left(F(x_i+\omega)\right) \det \left(\frac{\delta F}{\delta \omega}\right).$$
After the `gauge' substitution $x_i \rightarrow x_i - \omega$, one can drop the infinite group integral $\int {\rm d}\omega$
and get the final integral in a form symmetric with respect to integration variables:
$$I_{\rm G}= \int {\rm d}x_1 \int  {\rm d}x_2 {\rm e}^{-x_i A_{ij}x_j} \delta\left(F(x_i)\right) \left |\frac{\partial F}{\partial \omega}\right |.$$
The determinant $\det \left(\frac{\delta F}{\delta \alpha}\right)$ is the Faddeev--Popov determinant and
$\Delta_{\rm ch} = \frac{\delta F}{\delta \alpha}$ is the so-called ghost operator.

Obviously, in the abelian case the Lorentz gauge condition, $F=\partial_{\mu}A_{\mu}$,
leads to the ghost operator $\Delta_{\rm ch}=\partial_{\mu}\cdot \partial^{\mu}$. Since the operator does
not depend on the field, it is cancelled in the connected Green functions and does not give any contribution.

However, this is not so in the non-abelian case where $\Delta_{\rm ch}$ is a non-trivial operator depending on the gauge field.
Technically it is convenient to express $\det(\Delta_{\rm ch})$  also as a functional integral.
As we know, a determinant in the numerator appears when one integrates over anticommuting fields:
\begin{equation}
\det(\Delta_{\rm ch}) = \int \prod {\rm d}\bar c {\rm d}c {\rm e}^{{\rm i} \int \bar c \Delta _{\rm ch} c},
\end{equation}
where the anticommuting fields $c$ are called Faddeev--Popov ghosts.

As an example, let us consider the gauge condition in the  covariant form:
\begin{equation}
F(A) = \partial_{\mu}A^{\mu}-a(x)
\end{equation}
with an arbitrary function $a(x)$. The functional integral (\ref{11FP}) in this case takes the form

$$
{\rm e}^{\mathrm{i}W[J]}=\int \prod_{\mu, x}\mathrm{d}A_{\mu}^a (x)
 \mathrm{d}\bar c^a(x) \mathrm{d}c^a(x)\cdot\delta\left(\partial_{\mu}A^{a\mu}-a^a(x) \right) \cdot $$
$$\exp\left( \mathrm{i}\int {\rm d}^4 x\left[ -\frac{1}{4}F^a_{\mu\nu}F^{a\mu\nu}+\bar c^a\left( \Box c^a
- f^{abc}\partial_{\mu}(A^c_{\mu}c^b)\right)+J^a_{\mu}A^{a\mu} \right]\right).
\label{FInt_nonabel}
$$
The Green function does not depend on $a(x)$, so one can integrate (\ref{FInt_nonabel}) with the exponent
$$\int \mathrm{d}a \exp\left({-\frac{a^2}{2 \xi}}\right)\delta\left(\partial_{\mu}A^{\mu}-a \right)  = \exp\left({-\frac{(\partial_{\mu}A^{\mu})^2}{2\xi}}\right).$$
As the result, we get the following quadratic part of the action  in (\ref{FInt_nonabel}):
$$S=\int {\rm d}x\left[ -\frac{1}{4}(\partial_{\mu}A^a_{\nu}-\partial_{\nu}A^a_{\mu})(\partial^{\mu}A^{a\nu}
-\partial^{\nu}A^{a\mu})-\frac{1}{2\xi}\partial_{\mu}A^{a\mu}\partial_{\nu}A^{a \nu}+\bar c \Box c\right],$$
where the numerical parameter $\xi$ is the gauge parameter.
Now there is the inverse form,  and we get the following propagators for non-abelian gauge and ghost fields, respectively:
\begin{equation}
D^{ab}_{\mu \nu}(k)=-{\rm i}\frac{\delta^{ab}}{k^2+{\rm i}0}\left[ g_{\mu\nu}-(1-\xi)\frac{k_{\mu}k_{\nu}}{k^2}\right],
\end{equation}
\begin{equation}
D^{ab}_{\rm ch}(k)=\mathrm{i}\frac{\delta^{ab}}{k^2+{\rm i}0}.
\end{equation}
There are a few famous choices of the parameter $\xi$ used in concrete computations:\\
$\xi = 1$---the 't Hooft--Feynman gauge,\\
$\xi = 0$---the Landau gauge,\\
$\xi = 3$---the Frautschi--Yenni gauge.\\
Of course, any computed physics observable such as cross-section or distribution does not depend on $\xi$. As we see,
$$k^{\mu}D^{ab}_{\mu\nu}=-\xi k_{\nu} {\rm i}\frac {\delta^{ab}}{k^2 + \mathrm{i}0}=-\xi k_{\nu} D^{ab}_{\rm ch}(k).$$
So, ghosts are acting in a way to cancel a dependence on $\xi$ in physics quantities.

Now one can use our formula (\ref{effective_vertex}) in momentum space:
\begin{equation}
\Gamma^{a_1 \dots a_n}_{\mu_1 \dots \mu_n}(p_1 \dots p_n)\cdot (2\pi)^4\delta(p_1 +\dots+p_n)
= \mathrm{i}\frac{\delta^{(n)} S}{\delta A^{a_1}_{\mu_1}(p_1)\dots\delta A^{a_n}_{\mu_n}(p_n) }
\label{pbo}
\end{equation}
in order to get Feynman rules in momentum space for all the vertices in abelian and non-abelian theories.
 Note that by taking functional derivatives in (\ref{pbo}) one gets vertex functions with all needed symmetries.
In the formula  (\ref{pbo}), $\mu$ and $a$ indicate proper Lorentz and other group indices identifying the field.

As an example, let us consider the QED Lagrangian
\begin{equation}
L=\bar\Psi\left( {\rm i} D_{\mu}\gamma^{\mu} - m \right)\Psi, \,\,\,\,D_{\mu}=\partial_{\mu}+{\rm i}eQA_{\mu}.
\end{equation}
For the interaction vertex of the fermion field $\Psi$ with the photon field $A_{\mu}$, we obtain
$$
\Gamma_{\mu}(p_1, p_2; p_3)(2\pi)^4\delta(p_1+p_2+ p_3) $$
$$= \mathrm{i}\frac{\delta^{(3)}}{\delta \bar \Psi (p_1) \delta \Psi (p_2) \delta A_{\mu}(p_3)}\cdot \int {\rm d}x
\mathrm{i} \cdot (\mathrm{i}eQ)\bar \Psi(x) \gamma^{\mu} A_{\mu} (x) \Psi(x)$$
$$=-{\rm i} eQ\frac{\delta^{(3)}}{\delta \bar \Psi (p_1) \delta \Psi (p_2) \delta A_{\mu}(p_3)}\cdot$$ $$\cdot \int {\rm d}x\mathrm{d}q_1 \mathrm{d}q_2 \mathrm{d}q_3
\exp \left(-{\rm i}q_1 x - {\rm i}q_2 x - {\rm i}q_3x \right) \bar \Psi(q_1) \gamma^{\mu} \Psi(q_2)A_{\mu} (p_3) $$
$$=-{\rm i}eQ(2\pi)^4 \delta(p_1+p_2+p_3) \gamma_{\mu}$$
and so $\Gamma_{\mu}(p_1, p_2; p_3) = -{\rm i}eQ \gamma_{\mu}$, where $Q=-1$ for the fermion field (say, an electron) and
$+1$ for the antifermion field (say, a positron) etc.

In a similar way, one gets all other interaction vertices in the case of the SM Lagrangian, which will be considered later.
Also, similar to the scalar field theory, the reduction formula allows us to compute S-matrix
elements from the corresponding connected Green functions by cutting out propagators on all legs
and multiplying by corresponding free-particle wave functions.

A well-known visual way for presenting and computing S-matrix elements is given by Feynman rules---lines for different types of propagators and external particles, and points for vertices:
\vspace{3mm}






\begin{tabular}{cl}
\includegraphics{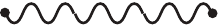}&
$\;\frac{-{\rm i}}{k^2+ i0}\delta^{ab}[g_{\mu\nu}-(1-\xi)\frac{k_{\mu}k_{\nu}}{k^2}]$

for massless gauge field
\vspace{3mm}\\
\includegraphics{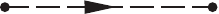}&$\;\frac{i}{k^2+{\rm i}0}\delta^{ab}$ for the ghost field
\vspace{3mm}\\

\includegraphics{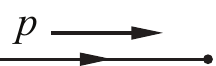}&
$\;u(p)$ for an incoming fermion
\vspace{3mm}\\

\includegraphics{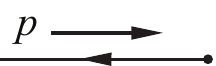}&$\;\bar v(p)$ for an incoming antifermion
\vspace{3mm}\\

\includegraphics{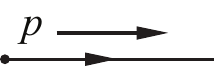}&$\;u(p)$ for an outgoing fermion
\vspace{3mm}\\

\includegraphics{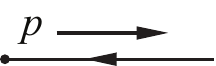}&$\;\bar v(p)$ for an outgoing antifermion
\vspace{3mm}\\

\includegraphics{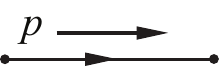}&$\;\mathrm{i}\frac{\displaystyle \hat p +m}{\displaystyle p^2 -m^2 + \mathrm{i}0}$ for a fermion propagator
\vspace{3mm}\\

\includegraphics{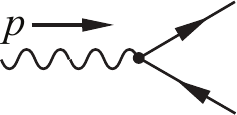}&\begin{tabular}{rl}
&$\mathrm{i}g\gamma_{\mu}(t^a)$ for a fermion--gauge boson vertex\\[12pt]
& \\
\end{tabular}
\\

\includegraphics{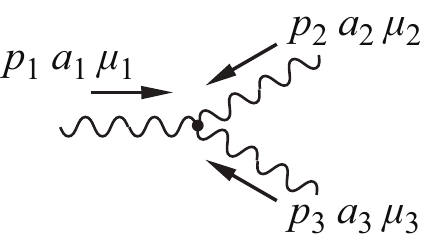}&
\begin{tabular}{rl}
&$gf^{a_1 a_2 a_3}\left[ g_{\mu_1 \mu_2}(p_1-p_2)_{\mu_3}\right.$\\[5pt]
&$\left.+g_{\mu_2 \mu_3}(p_2-p_3)_{\mu_1}+ g_{\mu_3 \mu_1}(p_3-p_1)_{\mu_2}\right]$\\[20pt]
& \\[30pt]
\end{tabular}
\\
\end{tabular}

We do not derive here formulas for cross-sections and decay widths; they are given in many textbooks.
We use notation of the Particle Data Group:
$$\mathrm{d}\sigma _{ab}=\frac{|M|^2}{4\sqrt{(p_a p_b)^2-m_a^2m_b^2}}\mathrm{d}\Phi _n,$$
where
$$\mathrm{d}\Phi_n = (2\pi)^4 \delta(p_i-p_f)\cdot \frac{{\rm d}^3\vec p_1}{(2\pi)^32p^0_1}\dots\frac{{\rm d}^3\vec p_n}{(2\pi)^32p^0_n},$$
$$\mathrm{d}\Gamma = \frac{|M|^2}{2m_a}\mathrm{d}\Phi_n.$$
As we see, one needs to compute the matrix element squared in order to get a scattering cross-section
and a decay width.

Also, one can formulate the Feynman rules for the matrix element squared directly.

\begin{figure}
\centering

\includegraphics{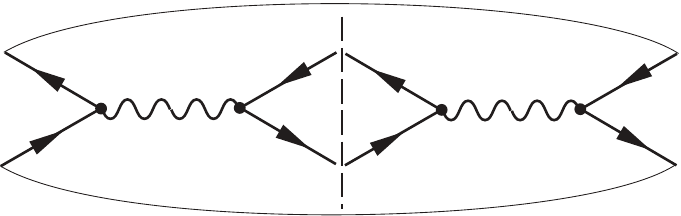}

\caption{$\mu^+\mu^-$ production in ${\rm e}^+{\rm e}^-$ collisions}

\label{10_10}

\end{figure}

In  Fig. \ref{10_10}, the square diagram is shown for $\mu^+\mu^-$-pair production in ${\rm e}^+{\rm e}^-$ collisions in QED.
The crossed lines correspond to external particles summed over polarizations.
Concrete expressions for the sums over polarizations for  particles and antiparticles with different spins
are as follows:
\\

\begin{tabular}{p{0.5\linewidth}p{0.5\linewidth}}
\centerline{1}

&for scalar particles
\end{tabular}
 $$\left\{
\begin{array}{rcr}
\sum_{\lambda}u_{\lambda}(p)\times \bar u_{\lambda}(p) &=&  p_{\mu}\gamma^\mu+m\\
\sum_{\lambda}v_{\lambda}(p)\times \bar v_{\lambda}(p) &= &- p_{\mu}\gamma^\mu+m
\end{array}
\right. \mbox{        for spin-$1/2$ Dirac particles}$$

\begin{tabular}{p{0.5\linewidth}p{0.5\linewidth}}
$\,\,\,\,\,\,\,\,\,\,\,\sum_{\lambda}{e}^{\mu}_{\lambda}(k){e}^{*\nu}_{\lambda}(k) = g_{\mu\nu}-\frac{k_{\mu}k_{\nu}}{k^2}$
& for massless gauge fields \\
&in the Landau gauge
\vspace{2mm}\\
\centerline{$\sum_{\lambda}{e}^{\mu}_{\lambda}(k){e}^{*\nu}_{\lambda}(k) = g_{\mu\nu}$}
& in the 't Hooft--Feynman gauge\\
\centerline{$\sum_{\lambda}{e}^{\mu}_{\lambda}(k){e}^{*\nu}_{\lambda}(k) = g_{\mu\nu}-\frac{k_{\mu}k_{\nu}}{M^2}$}
& for vector fields in the unitary gauge
\end{tabular}
The Feynman rules for propagators and vertices in the case of matrix elements squared are the same as for the case of amplitudes.
Note that the sums over polarizations represent the spin-density matrix  and
coincide with numerators of the propagators of corresponding particles. Note also, in computations using
the Feynman rules for matrix elements squared, that the ghosts
 should be added into initial and final lines together with corresponding gauge-boson lines.
 Each loop with crossed ghost lines should include extra factor (--1)
 with respect to the corresponding gauge-boson loop. This --1 sign reflects the anticommuting property of the ghost fields.

\section{Electroweak interactions in the SM}

As we know from school textbooks, the weak interactions are responsible for decay of
elementary particles. As we shall see, there are also scattering processes due to weak interactions,
 as were predicted by the SM and discovered in experiments.
Studies of weak interactions started from decays, and have a long history, which we do not describe here.
From various experimental studies it was realized that
(1) electron and muon neutrinos are not the same, and the electron neutrino and antineutrino are different.
There are processes $\nu_\mathrm{e} {\rm n} \rightarrow {\rm e}^- {\rm p}$, $\bar\nu_\mathrm{e} {\rm p} \rightarrow {\rm e}^+ {\rm n}$, $\nu_{\mu} {\rm n} \rightarrow \mu^- {\rm p}$,
$\bar\nu_{\mu} {\rm p} \rightarrow {\mu}^+ {\rm n}$,
but there are no processes $\bar\nu_\mathrm{e} {\rm n} \;\lefteqn{\rightarrow}{\,/\,\,} \;{\rm e}^- {\rm p}$, $\nu_\mathrm{e} {\rm p} \;
\lefteqn{\rightarrow}{\,/\,\,} \; {\rm e}^+ {\rm n}$, $\bar\nu_{\mu} {\rm n} \; \lefteqn{\rightarrow}{\,/\,\,}\; {\mu}^- {\rm p}$, $\nu_{\mu} {\rm p} \;
\lefteqn{\rightarrow}{\,/\,\,} \; {\mu}^+ {\rm n}$;

(2) the decays $\mu \rightarrow {\rm eX}$ have never been observed;

(3) only left-handed leptons and right-handed antileptons participate in the process with $|\Delta Q| = 1$ for
leptons from the same generations;

(4) three generations have been observed.

\begin{figure}[h]
\centering

\includegraphics{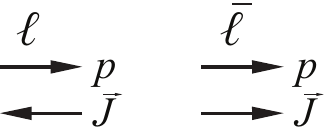}



\end{figure}

\begin{figure}[h]
\centering

\includegraphics{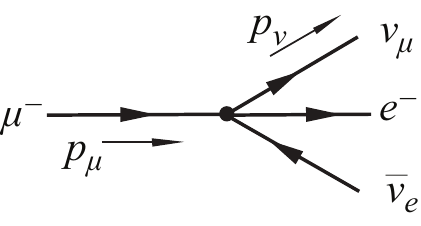}

\caption{Feynman diagram for $\mu^-$ decay due to four fermion interaction}

\label{11}

\end{figure}

The observations lead to the assumption that the lepton interactions with $|ΔQ| = 1$ occur via charge current
in so-called V--A form:
\begin {equation}
J_{\ell} \sim \bar \ell \gamma_{\mu}(1-\gamma_5)\nu_\ell.
\label{11_1}
\end{equation}
The corresponding four-fermion interaction Lagrangian for muon and electron currents is
\begin {equation}
L=\frac{G_{\rm F}}{\sqrt{2}}\bar \mu \gamma_{\sigma}(1-\gamma_5)\nu_{\mu}\bar e \gamma_{\sigma}(1-\gamma_5)\nu_{\rm e}+\mathrm{h.c.},
\label{11_2}
\end{equation}
where the notation $\mu$, $e$, $\nu$ stands for the corresponding fermion fields and $G_{\rm F}$ is the
well-known Fermi constant with dimension  $[m]^{-2}$.
With the help of the Lagrangian (\ref{11_2}), one can easily  compute the decay
width of $\mu^- \rightarrow {\rm e}^- \bar \nu _\mathrm{e} \nu_{\mu}$, the Feynman diagram for which is
shown in Fig. \ref{11}.


From dimensional analysis, one can say without any computations that $\Gamma \sim G_{\rm F}^2 \cdot m_{\mu}^5$.

The formula for the width is
\begin {equation}
\Gamma_{\mu} = \frac{G^2_{\rm F}m^5_{\mu}}{192 \pi^3}\cdot f\left(\frac{m_{e}^2}{m^2_{\mu}}\right),
\end{equation}
where $f=1+0\left(\frac{m_{e}^2}{m^2_{\mu}}\right)$. Today's $G_{\rm F}$ is measured from the muon decay very precisely:
$$G_{\rm F} = (1.166371 \pm 6\cdot 10^{-6})\cdot 10^{-5}\;\mbox{GeV}^{-2}.$$
As we know, any fermion field $\Psi(x)$ may be presented as
$$\Psi = \frac{1-\gamma_5}{2}\Psi+ \frac{1+\gamma_5}{2}\Psi=\Psi_{\rm L}+\Psi_{\rm R}.$$
Therefore, the current (\ref{11_1}) involves the left component of the fermion field only and has the form
\begin {equation}
J_{\ell}=\bar \Psi_{\rm L} \gamma_{\mu}\Psi_{\rm L}.
\label{11_3}
\end{equation}
We want to construct a quantum field theory (the SM) which obeys a few requirements:
\begin{itemize}
\item{correct electromagnetic neutral currents and V--A charge currents (Fermi);}
\item{three generations without chiral anomalies;}
\item{gauge-invariant dimension-four operators.}
\end{itemize}

 Because there are two leptons (charge lepton and corresponding neutrino) in each generation
and the left components interact, a very natural assumption is to choose the gauge group for the
EW part of the SM  to be
\begin{equation}
SU_L(2) \otimes U_Y(1),
\label{11_4}
\end{equation}
where $SU_L(2)$ is called the weak isospin group (the weak isospin is an analogue of the usual isospin introduced
by Heisenberg to describe the proton and the neutron)  and $U_Y(1)$ is the weak hypercharge group.
 The hypercharge group is needed because we need to have somehow the $U(1)$ group in order to describe
the electromagnetic interactions, as we already know. But if one includes  simply the electromagnetic group $U_{\rm em}(1)$
instead of $U_Y(1)$ the construction would not give us interactions correctly for all the fermions.
The EW part $SU_L(2) \otimes U_Y(1)$ and  the $SU_{C}(3)$ group for strong interactions
are combined into the gauge group of the SM.

The fermion fields are taken to be in three generations, and in each
generation the left components are combined into $SU(2)$ doublets and the right components transform as $SU(2)$ singlets:
  \begin {equation}
\begin{array}{ccc}
\displaystyle{ \nu_\mathrm{e} \choose e}_{\rm L}&\displaystyle {\nu_{\mu} \choose \mu}_{\rm L}&\displaystyle {\nu_{\tau} \choose \tau}_{\rm L}\\
e_{\rm R}&\mu_{\rm R}& \tau_{\rm R}.
\end{array}
\label{11_5}
\end{equation}
Right-handed neutrinos are present in the original version of the SM.

For the quarks, a similar structure of representations is assumed but with singlet right-handed components
for up and down types of quarks:
  \begin {equation}
\begin{array}{ccc}
\displaystyle{ u \choose d}_{\rm L}&\displaystyle {c \choose s}_{\rm L}&\displaystyle {t \choose b}_{\rm L}\\
u_{\rm R},d_{\rm R}&c_{\rm R}, s_{\rm R}& t_{\rm R}, b_{\rm R}.
\end{array}
\label{11_6}
\end{equation}

We begin with the construction of the gauge and fermion parts.
Now the requirements of the gauge invariance and lowest possible dimension four of terms fix
uniquely the form of the EW interaction Lagrangian. The strong interactions are described by the $SU_C(3)$ gauge group:
 \begin {equation}
\begin{array}{rcl}
L&=& -\frac{1}{4} W_{\mu\nu}^{i}(W^{\mu\nu})^i-\frac{1}{4} B_{\mu\nu}B^{\mu\nu}-\frac{1}{4} G_{\mu\nu}^a(G^{\mu\nu})^a\\[5pt]
&&+\sum_{f=\ell,q}\bar\Psi^f_{\rm L}(\mathrm{i}D^{\rm L}_{\mu}\gamma^{\mu})\Psi_{\rm L}^{\dagger}+\sum_{f=\ell,q}\bar\Psi^f_{\rm R}(\mathrm{i}D^{\rm R}_{\mu}\gamma^{\mu})\Psi_{\rm R}^{\dagger},
\end{array}
\label{11_7}
\end{equation}
where the field strength tensors and covariant derivatives have forms very familiar to us:
$$W^i_{\mu\nu} = \partial_{\mu}W_{\nu}^i-\partial_{\nu}W_{\mu}^i+g_2\varepsilon^{ijk}W_{\mu}^jW_{\nu}^k$$
 \begin {equation}
 B_{\mu\nu}=\partial_{\mu}B_{\nu}-\partial_{\nu}B_{\mu},
\label{11_8}
\end{equation}
$$G^a_{\mu\nu} = \partial_{\mu}A_{\nu}^a-\partial_{\nu}A_{\mu}^a+g_Sf^{abc}A_{\mu}^b A_{\nu}^c,$$
$$D^{\rm L}_{\mu} = \partial_{\mu}-{\rm i}g_2W_{\mu}^i\tau^i - \mathrm{i}g_1 B_{\mu}\left( \frac{Y^f_{\rm L}}{2}\right)-{\rm i}g_SA^a_{\mu}t^a,$$
 \begin {equation}
D^{\rm R}_{\mu} = \partial_{\mu}- \mathrm{i}g_1 B_{\mu}\left( \frac{Y^f_{\rm R}}{2}\right)-{\rm i}g_SA^a_{\mu}t^a,
\label{11_9}
\end{equation}
where $i=1,2,3$, $a=1,\dots,8$; $W^i_{\mu}$ are gauge fields for the weak isospin group, $B_{\mu}$ are gauge fields for the weak hypercharge group and $A_{\mu}$ are gluon gauge fields for the strong $SU_C(3)$ colour group.
 The gauge fields are taken in the adjoint representations and the lepton and the quark
fields are in the fundamental representation of $SU_{L}(2)$ and $SU_C(3)$ groups. The strongly interacting part
of the SM related to the $SU_C(3)$ colour group is called QCD,
 which is covered in a separate course of lectures.

Often when the SM is described the weak hypercharges $Y^f_{\rm L}$ and $Y^f_{\rm R}$ are chosen
from the beginning
such that the Gell-Mann--Nishijima formula is satisfied for each of the left and right chiral fermions:
\begin {equation}
\begin{array}{ccc}
Q_f&=&(T^f_3)_{\rm L} + \frac{\displaystyle Y^f_{\rm L}}{\displaystyle 2},\\[6pt]
Q_f&=&(T^f_3)_{\rm R} + \frac{\displaystyle Y^f_{\rm R}}{\displaystyle 2},\\
\end{array}
\label{11_55}
\end{equation}
where $(T_3)_{\nu_{\ell}}=1/2$ and $(T_3)_{\ell}=-1/2$ are projections of weak isospin operators,
${+}1/2$ for the up-type and ${-}1/2$ for the down-type fermions.

However, we do not know at this moment why the Gell-Mann--Nishijima formula should work in our case for the
EW part. So, we do not assume from the beginning the Gell-Mann--Nishijima relations for weak hypercharges.
Let us take weak hypercharges as free parameters for a moment, and try to fix them from two physics requirements:
\begin{enumerate}
\item correct electromagnetic interactions;
\item V--A weak charge currents.
\end{enumerate}
Let us consider for simplicity only fermions from the first generation.
 We will consider the case of three generations
later by introducing the quark mixing matrix.

From the covariant derivatives for the left and right chiral fields (\ref{11_9}),
one gets the following Lagrangian for leptons of the first generation:
 \begin {equation}
 L^{\ell} =- \mathrm{i}^2\left( \bar\nu_{e_{\rm L}}\bar e_{\rm L} \right) \gamma_{\mu}
\left(
\begin{array}{cc}
\frac{\displaystyle 1}{\displaystyle 2} g_2W^3_{\mu}+g_1\frac{\displaystyle Y^{\ell}_{\rm L}}{\displaystyle 2}
B_{\mu}& g_2\frac{\displaystyle W^+_{\mu}}{\displaystyle \sqrt{2}}\\
g_2\frac{\displaystyle W^-_{\mu}}{\displaystyle \sqrt{2}} &- \frac{\displaystyle 1}
{\displaystyle 2} g_2 W^3_{\mu}+g_1\frac{\displaystyle Y^{\ell}_{\rm L}}{\displaystyle 2}B_{\mu}
\end{array}
\right) \left(
\begin{array}{c}
\nu_{e_{\rm L}}\\e_{\rm L}
\end{array}
\right)
\label{11_11}
\end{equation}
$$+\bar e_{\rm R} \gamma_{\mu}g_1\frac{Y^{\ell}_{\rm R}}{2}B_{\mu}e_{\rm R}.$$
Here the relation following from the Pauli matrices is used:
 \begin {equation}
 {\tau}^iW^i = \frac{\sigma^i}{2}W^i=\frac{1}{2}\left(
\begin{array}{cc}
W^3_{\mu}&\sqrt{2}W^+_{\mu}\\
\sqrt{2}W^-_{\mu}& -W^3_{\mu}
\end{array}
\right),
\label{11_10}
\end{equation}
where $W_{\mu}^{\pm} = \left( W^1_{\mu}\mp \mathrm{i}W^2_{\mu}\right)/\sqrt{2}$.

Products of non-diagonal elements give us the form of the charge current:
 \begin {equation}
L^{\ell}_\mathrm{CC}=\frac{g_2}{\sqrt{2}}\bar \nu_{e_{\rm L}}\gamma_{\mu}W^+_{\mu}e_{\rm L}+\mathrm{h.c.}=\frac{g_2}
{2\sqrt{2}}\bar \nu_{\rm e}\gamma_{\mu}(1-\gamma_5)W^+_{\mu}e+\mathrm{h.c.}
\label{11_12}
\end{equation}
The interaction Lagrangian (\ref{11_12}) contains  the lepton charge current with the needed V--A structure.

Products of diagonal elements in (\ref{11_11}) lead to the neutral current interaction Lagrangian\footnote{Question for students: why can arbitrary hypercharge not exist in the case of non-abelian gauge symmetry?}:
 \begin {equation}
\begin{array}{rcl}
L^{\ell}_\mathrm{NC}&=&\bar \nu_{e_{\rm L}}\gamma_{\mu}\left( \frac{\displaystyle 1}{\displaystyle 2} g_2W^3_{\mu}
+g_1\frac{\displaystyle Y^{\ell}_{\rm L}}{\displaystyle 2}B_{\mu}\right)\nu_{e_{\rm L}}\\[5pt]
&&+\bar {e_{\rm L}}\gamma_{\mu}\left( -\frac{\displaystyle 1}{\displaystyle 2} g_2W^3_{\mu}
+g_1\frac{\displaystyle Y^{\ell}_{\rm L}}{\displaystyle 2}B_{\mu}\right){e_{\rm L}}\\[5pt]
&&+\bar e_{\rm R} \gamma_{\mu}g_1\frac{\displaystyle Y^{\ell}_{\rm R}}{\displaystyle 2}B_{\mu}e_{\rm R}.
\end{array}
\label{11_13}
\end{equation}

Generically, the neutral component of the $W$ field and the $B$ field can mix with some mixing angle $\theta_\mathrm{W}$:
 \begin {equation}
W^3_{\mu}=Z_{\mu}\cos \theta_\mathrm{W}+A_{\mu}\sin \theta_\mathrm{W},
\label{11_14}
\end{equation}
$$B_{\mu}=-Z_{\mu}\sin \theta_\mathrm{W}+A_{\mu}\cos \theta_\mathrm{W}.$$
The angle $\theta_\mathrm{W}$ is called the Weinberg mixing angle.

One of these fields, say $A$, we try to identify with the photon---it should not interact with the neutrino and should have the well-known
Dirac interaction with the electron field as needed in QED.
These two physics requirements lead to three simple equations:
\begin{equation}
\frac{1}{2}\left(-\frac{g_2}{2}\sin \theta_\mathrm{W}+\frac{g_1}{2} Y^{\ell}_{\rm L}\cos \theta_\mathrm{W}\right)+ \frac{1}{2}\frac{g_1}{2} Y^{\ell}_{\rm R} \cos \theta_\mathrm{W}
=Q_e e\;\;(Q_e = -1),
\label{11_15a}
\end{equation}
\begin{equation}
\frac{1}{2}\left(\frac{g_2}{2}\sin \theta_\mathrm{W}-\frac{g_1}{2} Y^{\ell}_{\rm L}\cos \theta_\mathrm{W}\right)+ \frac{1}{2}\frac{g_1}{2} Y^{\ell}_{\rm R} \cos \theta_\mathrm{W}
=0,
\label{11_15b}
\end{equation}
\begin{equation}
\frac{g_2}{2}\sin \theta_\mathrm{W}+\frac{g_1}{2} Y^{\ell}_{\rm L}\cos \theta_\mathrm{W} =0.
\label{11_15c}
\end{equation}
The first equation (\ref{11_15a}) comes from the coefficient in front of the $\gamma_{\mu}$ structure in the interaction
of the electron with the $A_{\mu}$ field, the second  equation (\ref{11_15b}) follows from the coefficient in front of
the $\gamma_{\mu}\gamma_5$ structure and the third one (\ref{11_15c}) comes from the absence of the interaction of
the neutrino field with $A_{\mu}$.

Therefore,
\begin{equation}
-\frac{g_2}{2}\sin \theta_\mathrm{W}+\frac{g_1}{2} Y^{\ell}_{\rm L}\cos \theta_\mathrm{W} = \frac{g_1}{2} Y^{\ell}_{\rm R} \cos \theta_\mathrm{W}
=Q_e e.
\label{leptonY}
\end{equation}
From the equations (\ref{leptonY}), we get
$$g_1 Y^{\ell}_{\rm L} \cos \theta_\mathrm{W} = -e,$$
$$g_2 \sin \theta_\mathrm{W} = e$$
and
\begin{equation}
Y^{\ell}_{\rm R} = 2 Y^{\ell}_{\rm L}.
\label{lep_hyperch_LR}
\end{equation}
As we can see, the hypercharges of the left and right chiral leptons are proportional but not fully fixed.

In the quark sector there are both left and right chiral components for up and down quarks:
$${ u \choose d}_{\rm L},\;\;u_{\rm R},\;\;d_{\rm R}.$$
Then, from the Lagrangian (\ref{11_7}), the interaction of the quarks with gauge fields is
 \begin {equation}
\begin{array}{c}
 \left( \bar u \bar d \right)_{\rm L} \gamma_{\mu}\left(
\begin{array}{cc}
\frac{\displaystyle 1}{\displaystyle 2} g_2 W^3_{\mu}+g_1\frac{\displaystyle Y_{\rm L}^q}{\displaystyle 2}B_{\mu}&g_2\frac {\displaystyle W^+_{\mu}}{\displaystyle \sqrt 2}\\
g_2\frac {\displaystyle W^-_{\mu}}{\displaystyle \sqrt 2}&-\frac{\displaystyle 1}{\displaystyle 2} g_2 W^3_{\mu}+g_1\frac{\displaystyle Y_{\rm L}^q}{\displaystyle 2}B_{\mu}
\end{array}
\right)\left(\begin{array}{c}
{\displaystyle  u }\\[12pt]
{ \displaystyle d}
\end{array}
\right)_{\rm L}\\[7pt]
+\bar u_{\rm R}\gamma_{\mu}g_1\frac{\displaystyle Y_{\rm R}^u}{\displaystyle 2}B_{\mu}u_{\rm R}+\bar d_{\rm R}\gamma_{\mu}g_1\frac{\displaystyle Y_{\rm R}^d}{\displaystyle 2}B_{\mu}d_{\rm R}.
\end{array}
\label{11_16}
\end{equation}
The charge current has, as expected, the needed V--A form:
\begin{equation}
L^q_\mathrm{CC} = \frac{g_2}{2\sqrt{2}}\bar u\gamma_{\mu}(1-\gamma_5)W^+_{\mu}d+\frac{g_2}{2\sqrt{2}}\bar d\gamma_{\mu}(1-\gamma_5)W^-_{\mu}u.
\label{L_CCq}
\end{equation}
In the same way as was done for the lepton case, one should require correct electromagnetic interactions
 for both u and d quarks. This means that one should have a QED electromagnetic Lagrangian with electric
charges $\frac{2}{3}$ for up-quark and $-\frac{1}{3}$ for down-quark fields.
Substituting $W^3_{\mu}$ and $B_{\mu}$ in terms of $A_{\mu}$ and $Z_{\mu}$ fields (\ref{11_13}) into (\ref{11_16}),
we get the following equalities:
\begin {equation}
\frac{1}{2} g_2\sin \theta_\mathrm{W} + \frac{1}{2} g_1 Y^q_{\rm L} \cos \theta_\mathrm{W} = \frac{1}{2} Y^u_{\rm R}g_1\cos\theta_\mathrm{W} = \frac{2}{3}e,
\label{11_17}
\end{equation}
$$-\frac{1}{2} g_2\sin \theta_\mathrm{W} + \frac{1}{2} g_1 Y^q_{\rm L} \cos \theta_\mathrm{W} = \frac{1}{2} Y^d_{\rm R}g_1\cos\theta_\mathrm{W} = -\frac{1}{3}e.$$
From these four equations one gets the following four equalities:
\begin {equation}
 \left\{
\begin{array}{rcl}
g_2 \sin \theta_\mathrm{W} &=& e,\\[5pt]
g_1 Y^q_{\rm L} \cos \theta_\mathrm{W} &=& \frac{1}{3} e,\\ [5pt]
Y^u_{\rm R} &=& -2Y^d_{\rm R},\\ [5pt]
Y^u_{\rm R}+Y^d_{\rm R} &=& 2Y_{\rm L}^q,
\end{array}\right.
\label{11_18}
\end{equation}
which are consistent with equalities we obtained from the lepton sector:
\begin {equation}
 \left \{
\begin{array}{rcl}
g_2 \sin \theta_\mathrm{W} &=& e,\\[5pt]
g_1 Y^{\ell}_{\rm L} \cos \theta_\mathrm{W} &=& -e,\\ [5pt]
Y^{\ell}_{\rm R} = 2Y_{\rm L}^{\ell}.
\end{array}
\right.
\label{11_19}
\end{equation}
Note that, as follows from (\ref{11_18}) and (\ref{11_19}),
$$Y_{\rm L}^{\ell} = -3Y_{\rm L}^q,$$ which means that there is only one independent hypercharge, say $Y_{\rm L}^{\ell}$, and all the others
may be expressed in terms of it.

Let us recall that up to now we did not assume any additional relations such as $(Q=T_3+Y/2)$, which are usually assumed
from the very beginning.

Now we can write the generic Lagrangian for neutral current interactions with introduced bosons $A_{\mu}$ and $Z_{\mu}$
in the following form:
\begin {equation}
 L_\mathrm{NC}=e\sum_f
Q_f J^{\rm em}_{f \mu}A^{\mu}+ \frac{e}{4 \sin \theta_\mathrm{W} \cos \theta_\mathrm{W}} \cdot \sum_f J^Z_{f \mu}Z^{\mu},
\label{11_20}
\end{equation}
where $J^{\rm em}_{f \mu}=\bar f \gamma_{\mu}f$, $Q_{\nu} = 0$, $Q_e = -1$, $Q_u = 2/3$, $Q_d = -1/3$,
$$J^Z_{f\mu} = \bar f \gamma_{\mu}[v_f - a_f \gamma_5]f, $$
$$v_{\nu} = 1,\;\;a_{\nu} = 1,\;\;v_e = -1+4\sin^2 \theta_\mathrm{W},\;\;a_e = -1;$$

$$v_u=1-\frac{1}{3}\left( 4+\frac{Y^u_{\rm R}}{Y^q_{\rm L}}\right) \sin^2 \theta_\mathrm{W}, \;\;\;
a_u=1-\frac{1}{3}\left( 4-\frac{Y^u_{\rm R}}{Y^q_{\rm L}}\right)\sin^2 \theta_\mathrm{W},$$
$$v_d=-1+\frac{1}{3}\left( 2-\frac{Y^d_{\rm R}}{Y^q_{\rm L}}\right) \sin^2 \theta_\mathrm{W}, \;\;\;
a_d=-1+\frac{1}{3}\left( 2+\frac{Y^d_{\rm R}}{Y^q_{\rm L}}\right) \sin^2 \theta_\mathrm{W}.$$

Since the structure of all three generations is the same, the equality (\ref{11_20}) is the same for all leptons and quarks.
Vector and axial-vector  couplings $v_f$ and $a_f$ may be expressed for all the fermions in a compact common form
via the fermion charge $Q_f$ and a component of the fermion weak isospin $T_3^f$:
$$v_f = 2T_3^f - 4 Q_f \sin^2 \theta_\mathrm{W},  \;\;\; a_f = 2T_3^f.$$

Note that the hypercharge parameters $Y^{\ell}_{\rm L}$ and $Y^{\ell}_{\rm R}$ are not present in (\ref{11_20})
 while in the quark sector one free parameter, which as we saw may be expressed in terms of $Y_{\rm L}^{\ell}$,
remains taking into account (\ref{11_18}),
$$\frac{Y^u_{\rm R}}{Y_{\rm L}^q}+  \frac{Y^d_{\rm R}}{Y_{\rm L}^q} = 2.$$

We have obtained the Lagrangian, the sum of (\ref{11_12}), (\ref{L_CCq}), (\ref{11_20}) and the gauge boson kinetic terms from
(\ref{11_7}),  which contains QED with massless fermions, an additional
vector  particle $Z_{\mu}$ interacting with new neutral currents, two charge massless vector particles
$W^{\pm}_{\mu}$ interacting with V--A charge currents  and self-interactions of the gauge particles.

Since our Lagrangian has a rather non-trivial chiral structure, an important
question arises as to whether or not our construction is free of chiral anomalies, which is absolutely needed
for a theory to be self-consistent.

\section{Anomalies} 
Detailed discussion of anomalies is not a subject of our brief notes
and can be found in a number of textbooks (see for example \cite{bib:weinberg96, bib:faddeev91}).

\begin{figure}[hbt]
\centering
\includegraphics{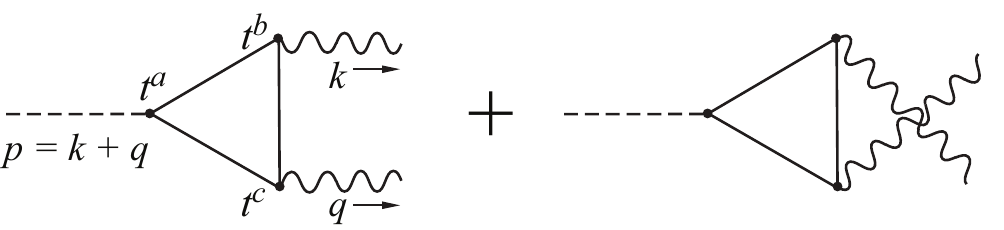}
\caption{Loop corrections}
\label{FA1}
\end{figure}

Generically, anomalies correspond to a situation in the field theory when some
symmetry takes place at the level of a classical Lagrangian but it is violated at
quantum loop level. For us an important anomaly is the chiral anomaly. In short, it means that, for example, the Lagrangian

\begin{equation}
L= \bar \Psi_{\rm L} {\rm i}\gamma^{\mu}\left( \partial_{\mu} - \mathrm{i}gA^a_{\mu}t^a\right)\Psi_{\rm L} =  \bar \Psi {\rm i}\gamma^{\mu}\left( \partial_{\mu} - \mathrm{i}gA^a_{\mu}t^a\right)\frac{1-\gamma_5}{2}\Psi
\label{A1}
\end{equation}
is invariant under the transformation
\begin{equation}
\Psi \rightarrow \exp\left(\mathrm{i}\alpha^a t^a \frac{1-\gamma_5}{2}\right)\Psi,\;\;\;\; A_{\mu} \rightarrow A_{\mu}^a +\frac{1}{g}\partial_{\mu}\alpha^a
+ f^{abc}A_{\mu}^b \alpha^c.
\label{A2}
\end{equation}
This invariance according to the Noether theorem leads to conserving of the current:
\begin{equation}
j^a_{\mu} = \bar \Psi \gamma_{\mu} \frac{1-\gamma_5}{2}t^a \Psi.
\label{A3}
\end{equation}
However, after quantization one finds that the current  (\ref{A3}) cannot be conserved due
to the triangle loop contributions shown in Fig.~\ref{FA1}.

The diagrams in Fig.~\ref{FA1}, after a convolution with the momentum $p_{\mu}$ instead of being equal to zero, are proportional to the factor
\begin{equation}
\frac{g^2}{8 \pi^2} \varepsilon^{\mu \nu \alpha \beta} k^{\alpha}q{\beta}\cdot \mbox{Tr}\left[ t^a\{t^b t^c\} \right].
\label{A4}
\end{equation}

If the anomaly is not vanishing the theory loses  its gauge invariance and therefore cannot be acceptable. (However, in cases of some currents external with respect to the theory, which have nothing to do with symmetries of the theory
and the Noether theorem, anomalies or such currents may take place. This does not lead to any problems.
Moreover, such type of anomalies may have very important physics consequences, as in the case of $\pi^0$ decay to
two photons.)
In the SM there are simultaneously contributions from left and right chiral fermions which
contribute to the anomaly with opposite signs. The anomaly is then proportional to the differences
between traces of group generators coming from fermions with left and right chiralities:
\begin{equation}
\mbox{Anom}\sim  \mbox{Tr}\left[ t^a\{t^b t^c\} \right]_{\rm L} -  \mbox{Tr}\left[ t^a\{t^b t^c\} \right]_{\rm R}.
\label{A5}
\end{equation}
In the theories like QED or QCD there are no $\gamma^5$ matrices involved in the Lagrangians. Therefore, the left and right
 chiral contributions exactly compensate each other, the anomaly is equal to zero and the theories perfectly make sense.

In the EW part of the SM left and right states couple to $U_Y(1)$ gauge bosons  with different hypercharges,
 and only the left components couple to the $SU_2(2)$ gauge bosons.
So, it is not obvious a priori that chiral anomalies vanish.
 In fact, zero anomalies is a request to the SM to be a reasonable quantum theory.

Since the generators of the $SU_{L}(2)$ group are the matrices $t^i = \sigma^i/2$,
then any combination of three $SU_C(2)$ generators gives zero traces in (\ref{A5}) and therefore zero anomalies.
The only potentially dangerous ones are
$$\left( SU_{L}(2)\right)^2\cdot U_Y(1)\mbox{ and }U_Y(1)^3$$
anomalies. For the first type we have to take into account $\{t^i,t^j\} = \frac{1}{2}\delta^{ij}$ and then the only non-zero contribution is
 \begin{equation}
\mbox{Anom}\sim  \mbox{Tr}\left[ Y\{t^i, t^j\} \right]_{\rm L} = \frac{1}{2}\delta^{ij}\mbox{Tr}Y_{\rm L}
= \frac{1}{2} \delta^{ij}\left[N_C \cdot 2Y_{\rm L}^q +2Y_{\rm L}^{\ell}\right].
\label{A6}
\end{equation}

 From the relations (\ref{11_17}) and (\ref{11_18}), as already was mentioned, we have the following relation between hypercharges:
  \begin{equation}
Y^{\ell}_{\rm L} = -3Y_{\rm L}^q.
\label{A7}
\end{equation}
After substitution of (\ref{A7}) into (\ref{A6}), we get
 \begin{equation}
\mbox{Anom}\sim  \frac{1}{2} \delta^{ij}2Y_{\rm L}^q (N_C-3).
\label{A8}
\end{equation}

It is very interesting that the anomaly vanishes only for the number of colours equal to three, as it does in QCD.
However, the value of the hypercharge $Y_{\rm L}^q$ is not fixed.

The second type ($(U_Y(2))^3$) of anomaly for the fermions for each generation is proportional to
\begin{equation}
\begin{array}{rcl}
\mbox{Anom}&\sim& \mbox{Tr}\left( Y_{\rm L}^3 \right) - \mbox{Tr}\left( Y_{\rm R}^3 \right)\\[5pt]
&=&N_C(Y_{\rm L}^q)^3\cdot 2 + (Y_{\rm L}^{\ell})^3\cdot 2 - N_C\left[(Y_{\rm R}^u)^3 + (Y_{\rm R}^d)^3\right] - (Y_{\rm R}^{\ell})^3,
\end{array}
\label{A9}
\end{equation}
where the factor (2) in the left-hand contribution comes from two (u and d) quarks and two (e and $\nu_\mathrm{e}$) leptons.
Taking into account from (\ref{A7}) and (\ref{11_17}) that $Y_{\rm R}^u + Y_{\rm R}^d = 2Y^q_{\rm L}$, one gets from (\ref{A9}) the following:
\begin{equation}
\begin{array}{rcl}
\mbox{Anom}&\sim&  Y_{\rm L}^{\ell}\left[ 2N_C(\frac{1}{3} Y^{\ell}_{\rm L}+Y_{\rm R}^u)^2-6(Y_{\rm L}^{\ell})^2\right] \\[5pt]
&=& Y_{\rm L}^{\ell}\cdot 6 (\frac{1}{3} Y^{\ell}_{\rm L}+Y_{\rm R}^u-Y^{\ell}_{\rm L})(\frac{1}{3} Y^{\ell}_{\rm L}+Y_{\rm R}^u+Y^{\ell}_{\rm L}).
\end{array}
\label{A10}
\end{equation}
In order to get zero for the anomaly,
\begin{equation}
Y_{\rm R}^u=\frac{2}{3}Y_{\rm L}^{\ell}\mbox { or }Y_{\rm R}^u=-\frac{4}{3}Y_{\rm L}^{\ell}.
\label{A11}
\end{equation}
At this point one cannot prefer one of the relations in (\ref{A11}).
This value will be finally fixed only after $SU_{L}(2)\times U_Y(1)$ symmetry breaking.

So, we note once more that we have constructed a theory with the Lagrangian
for massless fermions and gauge bosons, which gives us:
\begin{enumerate}
\item correct V--A charge currents;
\item  correct electromagnetic interactions;
\item no chiral anomalies;
\item predictions of additional neutral currents observed experimentally.
\end{enumerate}

However, obviously such a theory cannot describe nature correctly.
We do not observe massless EW vector particles, except for the photon,
and we do not observe massless fermions, except for, maybe, the neutrinos (or one of the neutrinos).
Massive W and Z bosons, massive leptons and quarks are observed experimentally.

So, we have to introduce masses in the theory. But we cannot do it directly without violation of the basic principle of
 gauge invariance. Indeed, the mass term for the vector field
$m^2_V V^{\mu}V_{\mu}$ is not invariant under the gauge transformation $V_{\mu}\rightarrow V_{\mu}+\partial_{\mu}{\alpha}$,
and  the mass term for fermions $m_{\Psi}\bar \Psi \Psi$,
the Dirac mass, is equal to $m\left( \bar\Psi_{\rm L} \Psi_{\rm R} +\bar \Psi_{\rm R} \Psi_{\rm L} \right)$, and
it is also not gauge invariant. Indeed, the left field  $\Psi_{\rm L}$ is the doublet and
the right field $\Psi_{\rm R}$ is the singlet  with respect  to the group $SU_{L}(2)$.
How to make massive particles without violation of the basic principle of gauge invariance?
There is a way to resolve this problem, which is related to spontaneous symmetry breaking,
 the Nambu--Goldstone theorem and the Brout--Englert--Higgs--Hagen--Guralnik--Kibble mechanism.

\section{Spontaneous symmetry breaking and the\\
Brout--Englert--Higgs--Hagen--Guralnik--Kibble mechanism} 

The situation when the Lagrangian is invariant under some symmetry while the
spectrum of the system is not invariant is very common for spontaneous symmetry breaking
(for example, Ginzburg--Landau theory). But a naive realization of ideas of spontaneous symmetry breaking leads to
a problem manifested in the appearance of so-called Nambu--Goldstone bosons with zero masses.

\begin{figure}[htb]
\centering

\includegraphics{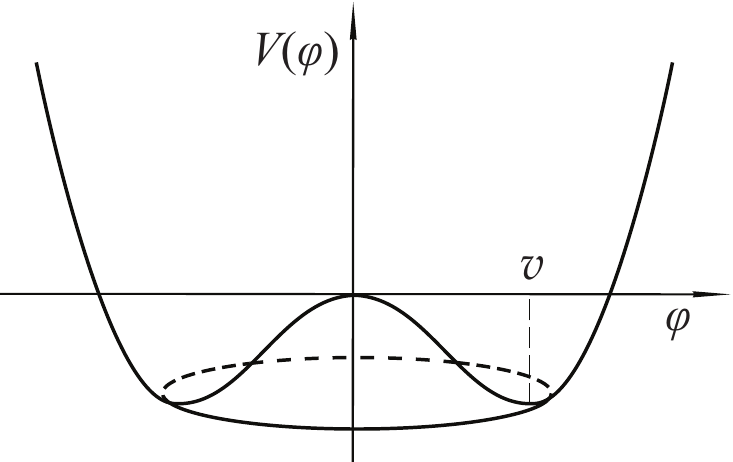}

\caption{Higgs potential}

\label{Higgs}

\end{figure}

To illustrate this, let us consider a very simple scalar model with the Lagrangian
\begin{equation}
L=\partial_{\mu}\varphi^{\dag}\partial^{\mu} \varphi- {\mu}^2\varphi^{\dag}\varphi - \lambda(\varphi^{\dag}\varphi)^2.
\label{12_1}
\end{equation}
The Lagrangian (\ref{12_1}) is invariant under the phase shift 
$\varphi \rightarrow \varphi \mathrm{e} ^{\mathrm{i}\omega}$ with $\omega =$ Const.
The case with $\mu^2>0$ is trivial and not interesting for us. 
In the case $\mu^2 = -|\mu^2|<0$, the potential shown in Fig.\ref{Higgs}
\begin{equation}
V(\varphi)= {\mu}^2\varphi^{\dag}\varphi + \lambda(\varphi^{\dag}\varphi)^2
\label{12_2}
\end{equation}
has a non-trivial minimum:
$$\left.\frac{\mathrm{d}V}{\mathrm{d}\varphi^{\dag}}\right|_{\varphi_0} = -|\mu^2|\varphi_0+2\lambda (\varphi^{\dag}_0 \varphi_0)\varphi_0=0\;\;
\Rightarrow\;\; |\varphi_0|=\sqrt{\frac{|\mu^2|}{2\lambda}}=\frac{v}{\sqrt{2}}>0.$$
The system takes some concrete value for the vacuum solution, say $\varphi_0=+{v}/{\sqrt{2}}$, which violates the phase-shift symmetry.

A complex scalar field can be parametrized by two real fields
\begin{equation}
\varphi= \frac{1}{\sqrt{2}}(v+h(x)){\rm e}^{-{\rm i}\xi (x)/v}.
\label{12_3}
\end{equation}
In terms of the new fields $h(x)$ and $\xi(x)$, the Lagrangian has the following form:
\begin{equation}
\begin{array}{rcl}
L&=& \frac{\displaystyle 1}{\displaystyle 2}\partial _{\mu}h\partial ^{\mu}h-\lambda v^2h^2-\lambda v h^3-\lambda h^4/4 \\[5pt]
&&+\frac{\displaystyle 1}{\displaystyle 2}\partial _{\mu}\xi\partial ^{\mu}\xi+\frac{\displaystyle 2}{\displaystyle v}\partial _{\mu}\xi\partial ^{\mu}\xi h+\frac{\displaystyle 1}{\displaystyle v^2}\partial _{\mu}\xi\partial ^{\mu}\xi h^2+\lambda v^4/4.
\end{array}
\label{12_4}
\end{equation}

The Lagrangian (\ref{12_4}) describes the system of a massive scalar field $h$ with mass
$m^2_h = 2\lambda v^2$ interacting with a massless scalar field $\xi(x)$.
The field $\xi(x)$ is called the Nambu--Goldstone boson field.

This is a particular case of the generic Goldstone theorem.
If the theory is invariant under a global group with $m$ generators
 but the vacuum is invariant under transformations generated only by $\ell$-generators,
then in theory there exist $m-\ell$ massless Nambu--Goldstone bosons.

Consider the system described by the Lagrangian
$$L =  \frac{1}{2} \partial_{\mu}\varphi \partial^{\mu} \varphi - V(\varphi).$$

Let the Lagrangian be invariant under $i=1,\dots, m$ transformations:
$$\varphi \rightarrow \varphi'= \varphi + \delta \varphi, \,\,\,\, \delta \varphi_i = {\rm i}\delta \Theta_A t_{ij}^A \varphi_j.$$
The invariance of the potential means that
\begin{equation}
\delta V = (\partial V/ \partial \varphi_i) \delta \varphi_i = \mathrm{i}\delta \Theta^A (\partial V/ \partial \varphi_i) t_{ij}^A \varphi_j =0.
\label{invar_condition}
\end{equation}
Let the potential have a minimum (vacuum) at some field value $\varphi_i=\varphi_i^0$:
\begin{equation}
(\partial V/ \partial \varphi_i)(\varphi_i=\varphi_i^0)=0.
\label{poten_min}
\end{equation}
We consider the case that the vacuum is invariant under transformations generated only by $\ell$-generators
from all $m$ generators corresponding to the symmetry, which means that
\begin{equation}
t_{ij}^A \varphi_j^0 =0
\label{invar_vac}
\end{equation}
only for $i=1,\dots, \ell$ ($\ell < m $).

The second derivative from the invariance condition (\ref{invar_condition}) at the minimum leads to
$$\frac{\partial^2 V}{\partial \varphi_k \partial \varphi_i}(\varphi_i
=\varphi_i^0)t_{ij}^A \varphi_j^0 + \frac{\partial V}{\partial \varphi_i}(\varphi_i=\varphi_i^0)t_{ij}^A=0.$$
The second term here is equal to zero due to (\ref{poten_min})
and therefore
$$\frac{\partial^2 V}{\partial \varphi_k \partial \varphi_i}(\varphi_i=\varphi_i^0)t_{ij}^A \varphi_j^0 =0.$$
For the first $\ell$, this equality takes place because of (\ref{invar_vac}).
However, for other fields with $i=\ell+1,  \dots, m$, the following equality has to be valid:
\begin{equation}
\frac{\partial^2 V}{\partial \varphi_k \partial \varphi_i}(\varphi_i=\varphi_i^0)=0.
\label{mass_term}
\end{equation}
But the second derivative of the potential in (\ref{mass_term}) is nothing but the mass term for these $i=\ell+1,  \dots, m$ fields. And, the equation (\ref{mass_term}) tells us that the masses of these fields
are equal to zero. So, in such a situation when the vacuum of a system
is not invariant under all the symmetry transformations of the Lagrangian
there are $m-\ell$ massless fields  (Nambu--Goldstone bosons) corresponding to the number of generators
violating the symmetry.

Let us return to the SM gauge group $SU_{L}(2)\times U_Y(1)$ and add to the system
one additional complex scalar field $\Phi(x)$, being an $SU_{L}(2)$ doublet  and a $U_Y(1)$ singlet:
\begin{equation}
L _{\Phi}= D_{\mu}\Phi^{\dag}D^{\mu}\Phi - \mu^2\Phi^{\dag}\Phi-\lambda (\Phi^{\dag}\Phi)^4.
\label{12_5}
\end{equation}

The Lagrangian $L _{\Phi}$ is gauge invariant; the covariant derivative has the form
 \begin{equation}
D_{ \mu} = \partial _{\mu}-{\rm i}g_2 W_{\mu}^i \tau^i-{\rm i}g_1\frac{Y_\mathrm{H}}{2}B_{\mu}.
\label{12_6}
\end{equation}
As in our previous example, let the mass parameter squared be negative, $\mu^2 = -|\mu^2|$, and
therefore the field potential has a non-trivial minimum at $\Phi = v/\sqrt{2}$.

One can parametrize the complex field doublet $\Phi(x)$  by four real fields in the following generic way:
\begin{equation}
\Phi(x)=\exp\left(-{\rm i}\frac{\xi^i(x) t^i}{v}\right)\left(
\begin{array}{c}
0\\
(v+h)/\sqrt{2}
\end{array}
\right),
\label{12_8}
\end{equation}
where four scalar fields $\xi_1,\;\xi_2,\;\xi_3$ and $h$ are introduced.

 The Lagrangian (\ref{12_5}) in invariant under an $SU_{L}(2)$ transformation:
\begin{equation}
\Phi(x)\rightarrow\Phi'(x)=\exp\left(\mathrm{i}g_2\alpha^i t^i\right)\Phi(x),
\label{12_9}
\end{equation}
where $t^i = \sigma^i/2$ are the generators of the $SU_{L}(2)$ gauge group. If we compare (\ref{12_8}) and (\ref{12_9}),
we can choose a special gauge $g_2\alpha^i(x) = \xi^i(x)/v$ such that the unitary factor
 $\exp\left(-{\rm i}\xi^i(x)t^i/v \right)$ disappears from all the formulas.
This gauge is called the unitary gauge. The Higgs field $\Phi(x)$ in this gauge takes therefore the following form:
\begin{equation}
\Phi= \frac{1}{\sqrt 2}\left(
\begin{array}{c}
0\\v+h(x)
\end{array}
\right).
\label{12_10}
\end{equation}
The field $\Phi$ has a non-zero vacuum expectation value and as we know it leads to a violation of  the symmetry of the system.

After such spontaneous symmetry breaking, the substitution of (\ref{12_10}) into the Lagrangian (\ref{12_5})
 with  the covariant derivative (\ref{12_6}) yields the following Lagrangian in terms
of fields $W_{\mu}^{\pm},$ $A_{\mu}$ and $Z_{\mu}$ introduced before:
\begin{equation}
L= \frac{1}{2}(\partial_{\mu}h)^2-\frac{1}{2}(2\lambda v^2)h^2-\lambda v h^3 - \frac{\lambda}{4}h^4\\
\label{12_11}
\end{equation}
$$+M_W^2W_{\mu}^+W^{\mu-} (1+h/v)^2+\frac{1}{2}M_Z^2Z_{\mu}Z^{\mu}(1+h/v)^2,$$
where
\begin{equation}
M_{h}^2 = 2\lambda v^2
\label{mH}
\end{equation}
is the mass of the scalar field $h$  called the Higgs boson and
\begin{equation}
M_W = \frac{1}{2} g_2 v\;\mbox{ and }\;M_Z = \frac{1}{2}\left(g_2 \cos \theta_\mathrm{W}+g_1 Y_H \sin \theta_\mathrm{W}\right)v
\label{mWmZ}
\end{equation}
are masses of the vector fields $W^{\pm}_{\mu}$ and $Z_{\mu}$. The field $A_{\mu}$ is not present in the Lagrangian (\ref{12_11})
 and therefore remains massless only in the case where the corresponding coefficient in front of it is equal to zero:
$$
 -\frac{1}{2}g_2 \sin \theta_\mathrm{W}+g_1 \frac{Y^H}{2} \cos \theta_\mathrm{W}=0.
$$
The condition from the lepton sector (\ref{11_15c})
$$ g_2 \sin \theta_\mathrm{W} = - g_1 Y^{\ell}_{\rm L} \cos \theta_\mathrm{W} $$
tells us that the field $A$ has the correct electromagnetic interactions and has zero mass simultaneously if the charged
lepton and Higgs field hypercharges have equal moduli and opposite signs:

$$ Y_\mathrm{H} = - Y_{\rm L}^l.$$

One should note that if the vacuum is invariant under some group transformation,
then the generator of the group gives zero acting on the vacuum. Indeed, the invariance of the vacuum means that
$${\rm e}^{{\rm i}T_i\Theta_i} \Phi_{\rm vac} = \Phi_{\rm vac};$$
therefore, the generator $T_i$ is given by
$$T_i\Phi_{\rm vac}=0.$$
In our case $$\Phi_{\rm vac} = \frac{1}{\sqrt{2}} {0 \choose v}$$ and the generator of unbroken symmetry should have a generic form:
$$T\Phi_{\rm vac} = \frac{1}{\sqrt{2}}\left(
\begin{array}{cc}
a_{11}&a_{12}\\
a_{21}&a_{22}
\end{array}
\right) {0 \choose v}=0,\;\;\;\Rightarrow\;\;T=\left(
\begin{array}{cc}
a_{11}&0\\
0&0
\end{array}
\right) $$
taking into account the fact that the generator should be Hermitian. For the group $SU_{L}(2)\times U_Y(1)$, such a generator is
$$T_3+\frac{1}{2}Y_\mathrm{H} = \frac{1}{2}\left(
\begin{array}{cc}
1&0\\
0&-1
\end{array}
\right) +\frac{1}{2}Y_\mathrm{H}\left(
\begin{array}{cc}
1&0\\
0&1
\end{array}
\right) = \left(
\begin{array}{cc}
1&0\\
0&0
\end{array}
\right)\;\;\mbox{ only if }\;\;\;Y_\mathrm{H} = +1.$$
This reflects the fact that the vacuum should be neutral, and the remaining group is naturally
the unbroken electromagnetic group $U_{\rm em}(1)$:
$$SU_{L}(2)\times U_Y(1)\rightarrow U_{\rm em}(1),$$
\begin{equation}
T_3+\frac{1}{2}Y_\mathrm{H} =Q_\mathrm{H} = 0, \;\;  Y_\mathrm{H} = +1.
\label{Higgs_hypercharge}
\end{equation}

Because $Y_\mathrm{H}=1$, one gets the following relation:
\begin{equation}
g_2\sin \theta_\mathrm{W} = g_1 \cos \theta_\mathrm{W}.
\label{12_12}
\end{equation}
If one substitutes (\ref{12_12}) into the equality for $M_Z$ (\ref{mWmZ}), one gets the well-known
relation between masses of W and Z bosons:
\begin{equation}
M_W = M_Z\cos \theta_\mathrm{W}.
\end{equation}

The value of the Higgs hypercharge $Y_{\rm H}=1$  fixes the lepton hypercharge  $Y^{\ell}_{\rm L} =-1$.
Now, from the connection between hypercharges (\ref{11_18}, \ref{11_19}) and following from them
$$Y_{\rm L}^{\ell} = -3Y_{\rm L}^q,$$
all the values for hypercharges of leptons and quarks with left and right chiralities are fixed:
$$Y^{\ell}_{\rm R} =-2, Y_{\rm L}^q=Y_{\rm L}^u=Y_l^d=1/3, Y^u_{\rm R} = 4/3, Y^d_{\rm R} = -2/3.$$
This confirms the Gell-Mann--Nishijima relation
$$I_3+\frac{Y}{2} = Q$$
for all three leptons and for all quarks with both chiralities.
It should be so from the relation between $SU_{L}(2)$ and $U_Y(1)$ generators leading to an unbroken $U_{\rm em}(1)$ generator.

Coming back to the Lagrangian (\ref{12_11}) and adding to it
the kinetic terms and self-interactions of the gauge fields $W^{\pm}_{\mu},$ $A_{\mu}$ and $Z_{\mu}$,
which come from the terms of the SM Lagrangian
$$-\frac{1}{4}W^i_{\mu\nu} W^{i\mu\nu} -\frac{1}{4}B_{\mu\nu} B^{\mu\nu},$$
we get the Lagrangian describing the massive Higgs boson $h$,
massive vector fields $W^i_{\mu}$ and $Z_{\mu}$ and massless field $A_{\mu}$.
From the Goldstone theorem we expect $4-1 = 3$ massless Nambu--Goldstone bosons. But they are not present in the Lagrangian.
Three would-be Nambu--Goldstone bosons $\xi_1$, $\xi_2$ and $\xi_3$ are `eaten' by three longitudinal components
of the fields $W^-_{\mu},\; W^+_{\mu}$ and $Z_{\mu}$. One should stress that while the symmetry is spontaneously broken,
the gauge symmetry of the Lagrangian itself remains unbroken.

This is the famous Brout--Englert--Higgs mechanism of
 spontaneous symmetry breaking (Nobel prize for 2012) confirmed by the discovery of the Higgs-like boson
in ATLAS and CMS experiments at the LHC.

Now we consider the fermions, leptons and quarks, of the SM, and  show how the mechanism of
 spontaneous symmetry breaking allows us to get massive fermions without violation of the gauge invariance.

As we know, in the SM the left fermions are the $SU(2)$ doublets and the right fermions are the singlets.
There are only two gauge-invariant dimension-four operators of the Yukawa-type preserving the SM gauge invariance:
\begin{equation}
\bar Q_{\rm L}\Phi d_{\rm R}\; \mbox{ and }\; \bar Q_{\rm L}\Phi^C u_{\rm R},
\label{12_13}
\end{equation}
where $$Q_{\rm L}  = {{u_{\rm L}}\choose{d_{\rm L}}}$$ is the doublet of left fermions and
 $$\Phi=\frac{1}{\sqrt{2}}
 \left(
\begin{array}{c}
0\\v+h
\end{array}
\right)\; \mbox{ and }\;\Phi^C=\mathrm{i}\sigma^2\Phi^{\dag}=\frac{1}{\sqrt{2}}
 \left(
\begin{array}{c}
v+h\\0
\end{array}
\right)$$
are the Higgs and conjugated Higgs $SU_{L}(2)$ doublet fields in the unitary gauge.
Corresponding to (\ref{12_13}), charge conjugated operators have the form
$$\left(\bar Q_{\rm L}\Phi d_{\rm R}\right)^{\dag} = d^{\dag}_{\rm R}\Phi^{\dag}\left(\bar Q_{\rm L}\right)^{\dag}
=d^{\dag}_{\rm R}\gamma^0\gamma^0\Phi^{\dag}\gamma^0Q_{\rm L} = \bar d_{\rm R} \Phi^{\dag}Q_{\rm L}$$ and
$$\left(\bar Q_{\rm L}\Phi^C u_{\rm R}\right)^{\dag} =\bar u_{\rm R}\left(\Phi^C\right)^{\dag}Q_{\rm L}.$$

As one can easily see, the operators of (\ref{12_13}) type lead after spontaneous
symmetry breaking to the needed terms for the fermion masses. Indeed,
$$(\bar u_{\rm L} \bar d_{\rm L}){0 \choose v} d_{\rm R}+\bar d_{\rm R} (0\;\;v)  {{u_{\rm L}}\choose{d_{\rm L}}}
=\bar d_{\rm L} d_{\rm R} + v\bar d_{\rm R} d_{\rm L} = v\left(\bar d_{\rm L} d_{\rm R}+\bar d_{\rm R} d_{\rm L}\right) = v \bar d d, $$
which is the Dirac mass term for the fermion. Similarly, the operator with the $\Phi^C$ field leads to the correct
Dirac mass term for the up-type fermions:
$$v\bar uu.$$

However, most general operators preserving the SM gauge invariance may include mixing of the fermion fields from various generations.
 The most general interaction Lagrangian including operators of (\ref{12_13}) type has the following form:
 \begin{equation}
 L_{\mathrm{Yukawa}} = -\Gamma_d^{ij}\bar {Q'_{\rm L}}^i\Phi {d'_{\rm R}}^j + \mathrm{h.c.} -\Gamma_u^{ij}\bar {Q'_{\rm L}}^i\Phi^C {u'_{\rm R}}^j
 + \mathrm{h.c.} -\Gamma_{\rm e}^{ij}\bar {L'_{\rm L}}^i\Phi {e'_{\rm R}}^j + \mathrm{h.c.},
 \label{12_14}
 \end{equation}
where there are no terms with a right-handed neutrino, and $\Gamma_{u,d,e}$ are
generically possible mixing coefficients. After spontaneous symmetry breaking,  one can rewrite the Lagrangian (\ref{12_14})
in the unitary gauge as follows:
\begin{equation}
L_{\mathrm{Yukawa}} =
- \left[M_d^{ij} \bar{d'_{\rm L}}^i {d'_{\rm R}}^j + M_u^{ij} \bar{u'_{\rm L}}^i{u'_{\rm R}}^j+M_{e}^{ij}
\bar{e'_{\rm L}}^i{e'_{\rm R}}^j + \mathrm{h.c.}\right]\cdot\left(1+\frac{h}{v}\right),
 \label{L_yukawa_unit}
 \end{equation}
where $  M^{ij} = \Gamma^{ij}v/{\sqrt 2}$.

The physics states are the states with definite mass. So, one should diagonalize the matrices in order to get
the physics states for quarks and leptons.
This can be done by unitary transformations for all left- and right-handed fermions:
 $$d'_{\mathrm{L}i} = (U_{\rm L}^d)_{ij}d_{\mathrm{L}j};\,\,\,\,\,d'_{\mathrm{R}i} = (U_{\rm R}^d)_{ij}d_{\mathrm{R}j};\,\,\,\,\,u'_{\mathrm{L}i}
= (U_{\rm L}^u)_{ij}u_{\mathrm{L}j};\,\,\,\,\,u'_{\mathrm{R}i} = (U_{\rm R}^u)_{ij}u_{\mathrm{R}j}$$
 $${\ell}'_{\mathrm{L}} = (U_{\rm L}^{\ell}){\ell}_{\mathrm{L}};\,\,\,\,\,{\ell}'_{\mathrm{R}} = (U_{\rm R}^{\ell}){\ell}_{\mathrm{R}}$$
$$U_{\rm L} U_{\rm L}^{\dag} =1,\;\;\;U_{\rm R} U_{\rm R}^{\dag} =1,\;\;\;U_{\rm L}^{\dag} U_{\rm L} =1.$$
The matrices $U$ are chosen such that
$$(U^u_{\rm L})^{\dag}M_u U_{\rm R}^u=\left(
\begin{array}{ccc}
m_u&0&0\\0&m_c&0\\0&0&m_t
\end{array}
\right);\,\,\,\,\,(U^d_{\rm L})^{\dag}M_d U_{\rm R}^d=\left(
\begin{array}{ccc}
m_d&0&0\\0&m_s&0\\0&0&m_b
\end{array}
\right)$$
$$(U^{\ell}_{\rm L})^{\dag}M_{\ell} U_{\rm R}^{\ell}=\left(
\begin{array}{ccc}
m_e&0&0\\0&m_{\mu}&0\\0&0&m_{\tau}
\end{array}
\right).$$
Therefore, the Yukawa-type Lagrangian (\ref{L_yukawa_unit})  is
$$L_{\mathrm{Yukawa}} = - \left[m_d^{i} \bar{d_{\rm L}}^i {d_{\rm R}}^i +m_d^{*i} \bar{d_{\rm R}}^i {d_{\rm L}}^i + \right.$$
$$\left.+m_u^{i} \bar{u_{\rm L}}^i {u_{\rm R}}^i +m_u^{*i} \bar{u_{\rm R}}^i {u_{\rm L}}^i+ m_{\ell}^{i} \bar{{\ell}_{\rm L}}^i {{\ell}_{\rm R}}^i +m_{\ell}^{*i} \bar{{\ell}_{\rm R}}^i {{\ell}_{\rm L}}^i\right]\cdot\left(1+\frac{h}{v}\right).$$
We consider only real  mass parameters  $m \equiv m^*$. So, the Yukawa Lagrangian after diagonalization
of the mass matrices contains masses of fermions and their interactions with the Higgs boson:
\begin{equation}
\Longrightarrow L_{\mathrm{Yukawa}} = - \left[m_d^{i} \bar{d}^i {d}^i + m_u^{i} \bar{u}^i {u}^i +m_{\ell}^{i} \bar{{\ell}}^i {{\ell}}^i\right]\cdot\left(1+\frac{h}{v}\right).
\label{12_15}
\end{equation}

Now one can easily see what the fermion interactions with the gauge bosons look like in the basis of the fermion physics state with definite masses.

Neutral currents have the same structure (\ref{11_19}) with respect to flavours as the mass terms.
And they, after the unitary rotation $\Psi' \rightarrow U\Psi$, become diagonal simultaneously with the mass terms:
$$\bar\Psi'\hat O_N\Psi' \rightarrow \bar \Psi \hat O \Psi .$$

However, charge currents
$$J_C\sim \bar u_{\rm L} \hat O_{\rm ch}d_{\rm L} +\mathrm{h.c.}$$
contain fermions rotated by different unitary matrices  for the up- and down-type fermions:
$$u' \rightarrow (U^u_{\rm L})u, \, \, \,   d' \rightarrow (U^d_{\rm L})d.$$
Therefore, after the rotation one gets for the charge current
$$J_{\rm C}\sim(U^u_{\rm L})^{\dag} U^d_{\rm L} \bar u_{\rm L} \hat Q d_{\rm L}.$$
The unitary matrix
$$V_{\rm CKM}=(U^u_{\rm L})^{\dag} U^d_{\rm L}$$ is called the Cabibbo--Kobayashi--Maskawa (CKM) mixing matrix:
\begin{equation}
 V_{\rm CKM}= \left(
\begin{array}{ccc}
V_{ud}&V_{us}&V_{ub}\\
V_{cd}&V_{cs}&V_{cb}\\
V_{td}&V_{ts}&V_{tb}
\end{array}
\right).
\label{12_16}
\end{equation}
Concrete values for the elements of the CKM matrix are not predicted in the SM.
One can show that an arbitrary unitary matrix with $N\times N$ complex elements may be
 parametrized by $N(N-1)/2$ real angles and $(N-1)(N-2)/2$ complex phases.
So, the CKM $3\times 3$ matrix contains three real parameters and one complex phase. The presence
of this phase leads to parity and charge (CP) 
violation, 
 which in this sense is a prediction of
the SM with three generations. In this lecture we do not discuss physics of the CKM matrix.
The flavour physics is covered in a special lecture course at the School.

Now we have all parts of the EW part of the SM Lagrangian expressed in terms of physics fields  in unitary gauge:
\begin{equation}
L_{\rm SM}=L_{\rm Gauge}+L_{\rm FG}+L_{\rm H}.
\label{12_17}
\end{equation}
Here, as was mentioned, the self-interactions of the gauge fields $W^{\pm}_{\mu},$ $A_{\mu}$ and $Z_{\mu}$
come from the terms of the SM Lagrangian $-\frac{1}{4}W^i_{\mu\nu} W^{i\mu\nu} -\frac{1}{4}B_{\mu\nu} B^{\mu\nu}$:
\begin{equation}
\begin{array}{rcl}
L_{\rm Gauge}&=&-\frac{\displaystyle 1}{\displaystyle 4}F_{\mu\nu}F^{\mu\nu}-\frac{\displaystyle 1}{\displaystyle 4}Z_{\mu\nu}Z^{\mu\nu}-\frac{\displaystyle 1}{\displaystyle 2}W^+_{\mu\nu}W^{-\mu\nu}\\[5pt]
&&+e\left[W^+_{\mu\nu}W^{-\mu}A^{\nu}+\mathrm{h.c.} + W^+_{\mu}W^{-}_{\nu}F^{\mu\nu}  \right]\\[5pt]
&&+e\frac{\displaystyle c_{\mathrm{W}}}{\displaystyle s_{\mathrm{W}}}\left[W^+_{\mu\nu}W^{-\mu}Z^{\nu}+\mathrm{h.c.} + W^+_{\mu}W^{-}_{\nu}Z^{\mu\nu}  \right]\\[5pt]
&&-{e}^2\frac{\displaystyle 1}{\displaystyle 4s_{\mathrm{W}}^2}\left[(W^-_{\mu}W^+_{\nu}-W^-_{\nu}W^+_{\mu} )W^{-\mu}W^{+\nu}+\mathrm{h.c.} \right]\\[5pt]
&&-\frac{\displaystyle {e}^2}{\displaystyle 4}(W^+_{\mu}A_{\nu}-W^+_{\nu}A_{\mu} )(W^{-\mu}A^{\nu}-W^{-\nu}A^{\mu} )\\[5pt]
&&-\frac{\displaystyle {e}^2}{\displaystyle 4}\frac{\displaystyle c_{\mathrm{W}}^2}{\displaystyle s_{\mathrm{W}}^2}(W^+_{\mu}Z_{\nu}-W^+_{\nu}Z_{\mu} )(W^{-\mu}Z^{\nu}-W^{-\nu}Z^{\mu})\\[5pt]
&&-\frac{\displaystyle {e}^2}{\displaystyle 2}\frac{\displaystyle c_{\mathrm{W}}}{\displaystyle s_{\mathrm{W}}}(W^+_{\mu}A_{\nu}-W^-_{\nu}A_{\mu})(W^{+\mu}Z^{\nu}-W^{-\nu}Z^{\mu} )+\mathrm{h.c.},
\end{array}
\label{12_18}
\end{equation}
where $c_{\mathrm{W}} = \cos \theta_\mathrm{W}$, $s_{\mathrm{W}} = \sin \theta_\mathrm{W}$; the gauge for a photon field may be taken differently, for example $(\partial_{\mu}A^{\mu}) = 0$.
The Lagrangian for the interactions of fermions with the gauge bosons is
\begin{equation}
L_{\rm FG} = \sum_f \bar f(\mathrm{i}\hat \partial)f+L_\mathrm{NC}+L_\mathrm{CC},
\label{12_19}
\end{equation}
 where $L_\mathrm{NC}$  and $L_\mathrm{CC}$ are given by (\ref{11_20}), (\ref{11_13}) and (\ref{L_CCq}).
The Lagrangian for the Higgs boson and its interactions with the gauge and fermion fields is
\begin{equation}
\begin{array}{rcl}
L_{\rm H}& =& \frac{\displaystyle 1}{\displaystyle 2}(\partial^{\mu}h)(\partial_{\mu}h)+\frac{\displaystyle
M_h^2}{\displaystyle 2}h^2-\frac{\displaystyle M_h^2}{\displaystyle 2v}h^3-\frac{\displaystyle M_h^2}{\displaystyle 8v^2}h^4\\[5pt]
&&+\left(M_W^2W_{\mu}^+W^{-\mu}+\frac{1}{2}M_Z^2Z_{\mu}Z^{\mu}\right)
\left(1+\frac{\displaystyle h}{\displaystyle v}\right)^2-\sum_f m_f \bar ff\left(1+\frac{\displaystyle h}{\displaystyle v}\right).
\end{array}
\label{12_21}
\end{equation}
All Feynman rules following from the Lagrangian (\ref{12_17}) can be obtained with the help of the formula (\ref{pbo}).
The kinetic parts of the Lagrangians $L_{\rm Gauge}$ (\ref{12_18}) and $L_{\rm FG}$ (\ref{12_18})
being taken together with the gauge and fermion mass terms from $L_{\rm H}$ (\ref{12_21}) give the propagators
 for fermions, massless photons, massive W$^{\pm}$ and Z bosons, and for the Higgs boson by inverting
the corresponding quadratic forms, as we know already\footnote{The complete list of the Feynman rules for interaction vertices can be found in many books and
listed explicitly, for example, in the model files for the SM used in computer codes like CompHEP, Grace, CalcHEP,
MadGraph, Wizard, Sherpa etc.}.

The propagator for the massive gauge bosons requires special care.
The propagators for massive vector fields follow directly from the Lagrangian
by inverting the quadratic form:
\begin{equation}
V^{\mu}\left( \Box g_{\mu\nu}-\partial _{\mu}\partial^{\nu} +g_{\mu\nu}M_V^2 \right)V^{\nu}.
\label{12_22}
\end{equation}
Then the  propagator for massive vector fields ($V=W,Z$) has the following structure in the unitary gauge:
\begin{equation}
D_{\mu\nu}(p)=\frac{-{\rm i}}{p^2-M^2_V}\left[g_{\mu\nu}-\frac{p _{\mu}p^{\nu}}{M_V^2} \right].
\label{12_23}
\end{equation}
However, the term ${p _{\mu}p^{\nu}}/{M_V^2}$ has a bad ultraviolet behaviour. This leads to the problem of
 proving renormalizability of the SM. To resolve the problem,
 one can use another gauge in which the bad ultraviolet behaviour is absent.
It is convenient to express the Higgs field as follows:
\begin{equation}
  \Phi(x)=\left(
\begin{array}{c}
-{\rm i}w^+_{\mathrm{g}}\\
(v+h+{\rm i}z_{\mathrm{g}})/\sqrt{2}
\end{array}
\right)
\label{12_24}
\end{equation}
and  $\Phi^{\dag}$ contains $w^-_{\mathrm{g}}$, where the notation $w_{\mathrm{g}}^{\pm}$ and $z_{\mathrm{g}}$ for the Goldstone bosons is introduced.
The covariant derivative, being expressed in terms of
the fields $W^{\pm}_{\mu}$, $Z_{\mu}$ and $A_{\mu}$, and constants $e$ and $\sin \theta_\mathrm{W} = s_{\mathrm{W}}$, takes the form
\begin{equation}
D_{\mu}\Phi = \left(
\begin{array}{cc}
\partial_{\mu}-{\rm i}\frac{\displaystyle e(1-2s_{\mathrm{W}}^2)}{\displaystyle 2s_{\mathrm{W}} c_{\mathrm{W}}}Z_{\mu}
-{\rm i}eA_{\mu} &-{\rm i}\frac{\displaystyle e}{\displaystyle \sqrt 2s_{\mathrm{W}}}W^+_{\mu}\\[10pt]
-{\rm i}\frac{\displaystyle e}{\displaystyle \sqrt 2s_{\mathrm{W}}}W^-_{\mu} & \partial_{\mu}+{\rm i}\frac{\displaystyle e}{\displaystyle 2s_{\mathrm{W}}c_{\mathrm{W}}}Z_{\mu}.
\end{array}
\right) \Phi
\label{12_25}
\end{equation}

Simple algebraic manipulation leads to the following Lagrangian for the Higgs-gauge part of the SM:
\begin{equation}
\begin{array}{rcl}
L&=& (D_{\mu}\Phi)^{\dag}(D^{\mu}\Phi)-\lambda\left( \Phi \Phi ^{\dag}-v^2/2 \right)^2\\[7pt]
&=&\frac{\displaystyle 1}{\displaystyle 2}(\partial^{\mu}h)(\partial_{\mu}h)+M_W^2W^+_{\mu}W^{\mu-}(1+h/v)^2
+ \frac{\displaystyle 1}{\displaystyle 2}M_Z^2Z_{\mu}Z^{\mu}(1+h/v)^2\\[5pt]
&&-M_h^2h^2-\lambda v h^3-\frac{\displaystyle \lambda}{\displaystyle 4}h^4\\[5pt]
&&-M_W\partial_{\mu} w_{\mathrm{g}}^+W^{\mu-}-M_W\partial_{\mu} w_{\mathrm{g}}^-W^{\mu+}-M_Z\partial_{\mu}z_{\mathrm{g}} Z^{\mu}\\[5pt]
&&+\partial_{\mu} w_{\mathrm{g}}^+\partial^{\mu} w_{\mathrm{g}}^- +\frac{\displaystyle 1}{\displaystyle 2}\partial_{\mu} z_{\mathrm{g}}\partial^{\mu} z_{\mathrm{g}}\\[5pt]
&&-\lambda h(h+2v)\left(w_{\mathrm{g}}^- w_{\mathrm{g}}^+ + z_{\mathrm{g}}/2\right)-\lambda \left(w_{\mathrm{g}}^- w_{\mathrm{g}}^+ + z_{\mathrm{g}}/2\right)^2\\[5pt]
&&+\mbox{more cubic and quadratic terms involving $w_{\mathrm{g}}^{\pm}$ and $z_{\mathrm{g}}$ fields}.
\end{array}
\label{12_26}
\end{equation}
The first two lines in (\ref{12_26}) are exactly the same as in the SM in the unitary gauge.
 The fourth line involves massless scalar fields, the Goldstone bosons $w_{\mathrm{g}}^{\pm}$ and $z_{\mathrm{g}}$.
There are many terms describing interactions of the Goldstone fields. But we would like to draw attention
to the third  line in (\ref{12_26}) describing the kinetic mixing of the  $w_{\mathrm{g}}^{\pm}$ and $z_{\mathrm{g}}$ fields with
$W^{\pm}$ and $Z$ fields, respectively.
 Such a mixing should be removed from the Lagrangian. This can be achieved by choosing proper gauge conditions.

Indeed, if we add to (\ref{12_26}) the following gauge fixing terms:
\begin{equation}
L_{\rm GF} = -\frac{1}{\xi} \left( \partial_{\mu} W^+_{\mu}-\xi M_W w^+_{\mathrm{g}} \right)\left( \partial_{\nu} W^{\mu-}
-\xi M_W w^-_{\mathrm{g}} \right)-\frac{1}{2\xi} \left( \partial_{\mu} Z^{\mu}-\xi M_Z z_{\mathrm{g}} \right)^2,
\label{12_27}
\end{equation}
the mixing terms are cancelled out. Then the quadratic part of the SM Lagrangian for $W^{\pm}_{\mu}$ and $Z_{\mu}$
fields which gives their propagators is
\begin{equation}
\begin{array}{c}
-\frac{\displaystyle 1}{\displaystyle 4}Z_{\mu\nu}Z^{\mu\nu}+\frac{\displaystyle 1}{\displaystyle 2}M_Z^2Z_{\nu}Z^{\nu}
-\frac{\displaystyle 1}{\displaystyle 2\xi}\left(\partial_{\mu}Z^{\mu}\right)^2\\[10pt]
-\frac{\displaystyle 1}{\displaystyle 2}W_{\mu\nu}^+W^{-\mu\nu}+M_W^2W_{\nu}^+W^{-\nu}
-\frac{\displaystyle 1}{\displaystyle \xi}\left(\partial_{\mu}W^{+\mu}\right)\left(\partial_{\nu}W^{-\nu}\right),
\end{array}
\label{12_28}
\end{equation}
where terms with
$$Z_{\mu\nu} = \partial_{\mu}Z_{\nu}-\partial_{\nu}Z_{\mu},\;\;W^{\pm}_{\mu\nu} = \partial_{\mu}W^{\pm}_{\nu}-\partial_{\nu}W^{\pm}_{\mu}$$
come from the kinetic part of the  SM Lagrangian, as in the unitary case.
Inverting the quadratic form (\ref{12_28}), one gets the propagator of the massive gauge field in the so-called $R_{\xi}$ gauge:
\begin{equation}
D^{\xi}_{\mu\nu}=\frac{-{\rm i}}{k^2-M^2_V}\left[g_{\mu\nu}-(1-\xi)\frac{k_{\mu}k_{\nu}}{k^2-\xi M^2_V}\right],
\label{12_29}
\end{equation}
where $M_V$ is $M_W$ or $M_Z$ and $\xi$ is the gauge parameter.

The unitary gauge is restored by the formal limit $\xi \rightarrow \infty$.
In the Landau gauge $\xi = 0$, we get the transverse structure $\left(g_{\mu\nu}-k_{\mu}k_{\nu}/k^2\right)$,
while in the 't Hooft--Feynman gauge $\xi=1$ the propagator contains only the part with the $g_{\mu\nu}$ tensor.
However, one should stress that in both these gauges as well as in the generic $R_{\xi}$ gauge one should take
into account the appearance of Faddeev--Popov ghost fields. This is done with the help of the Faddeev--Popov method in the functional
integral, which we have described   already. Without going into details, the following ghost
 fields appear: $c_{W}^{\pm}$, $c_Z$ and $c_A$,
corresponding to the gauge fixing terms (\ref{12_27}) and the gauge
 fixing $-(\partial_{\mu}A^{\mu})^2/2\xi$ term
for the photon field. In contrast to pure electrodynamics where the photon ghost fields do not interact
 and can be omitted, in the SM the photon ghost field $c_A$  has non-trivial interactions with
 the ghost $c_{{W}}^{\pm}$ and Goldstone $w_{\mathrm{g}}^{\pm}$ fields.


Propagators of all the ghost fields have the following form:
$$D^c = \frac{\rm i}{p^2-\xi M^2_V},$$
where $M_V^2$ is equal to $M_Z^2$ for $c_Z$, $M_W^2$ for $c_{{W}}$ and 0 for $c_A$ ghost fields.
(The complete set of all Feynman rules for interaction vertices  of Goldstone and ghost fields in
the SM in the $R_\xi$ gauge is rather long and can be found in the mentioned computer codes such as CompHEP.)

So, all the propagators for massive gauge, Goldstone and ghost fields have good ultraviolet behaviour,
and therefore the SM is a renormalizable quantum field theory.

\section{Phenomenology of the SM in lowest order} 

\begin{figure}[h]
\centering

\includegraphics{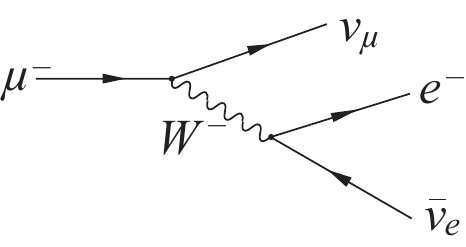}

\caption{$\mu\rightarrow {\rm e}^-\bar\nu_\mathrm{e} \nu_{\mu}$ decay diagram}

\label{mu}

\end{figure}

The Fermi constant $G_{\rm F}$ is measured with high precision from the muon lifetime:
\begin{equation}
G_{\rm F} = 1.166\,378\,7(6)\times10^{-5}\mbox{ GeV}^{-2}.
\label{13_1}
\end{equation}
The decay is described in the SM by the Feynman diagram shown in Fig. \ref{mu}.

Since the muon mass $m_{\mu}\ll M_W$, one can neglect the W-boson momentum in the propagator, and one immediately gets the following relation:
\begin{equation}
\frac{g_2^2}{8M_W^2} = \frac{G_{\rm F}}{\sqrt 2}.
\label{13_2}
\end{equation}
As we have seen, the W-boson mass is obtained in the SM due to the Higgs mechanism and is proportional to the Higgs
field vacuum expectation value $v$:
\begin{equation}
M_W^2 = \frac{1}{4} g^2_2 v^2.
\label{13_3}
\end{equation}
From these two relations, we obtain
\begin{equation}
v=\frac{1}{\sqrt{\sqrt{2}G_{\rm F}} }= 246. 22\mbox{ GeV}.
\label{13_4}
\end{equation}
The Higgs field expectation value $v$ is determined by the Fermi constant $G_{\rm F}$ introduced long before the Higgs mechanism appeared.
At this point one can see the power of the gauge invariance principle; $g_2$ is the
same gauge coupling in the relations (\ref{13_2}) and (\ref{13_3}).

Now from (\ref{13_2}) using the relation (\ref{12_12}), $g_1c_{\mathrm{W}} = g_2s_{\mathrm{W}} = e$ and keeping in mind $M_W = M_Zc_{\mathrm{W}}$, one gets
\begin{equation}
M_W^2\left( 1-\frac{M_W^2}{M_Z^2}\right) = \frac{\pi \alpha_{\rm em}}{\sqrt{2}G_{\rm F}}\equiv A^2_0,
\label{13_5}
\end{equation}
where $\alpha_{\rm em}={\rm e}^2/4\pi$ is the usual electromagnetic fine structure constant. The low-energy
$$\alpha_{\rm em} = \left(137.035\,999\,074(44)\right)^{-1}$$
follows mainly from the electron anomalous magnetic measurements.
One gets $A_0$ very precisely from low-energy experimental results:
\begin{equation}
A_0 = 37.2804\mbox{ GeV}.
\label{13_6}
\end{equation}
On the other hand, one gets $A_0$ from measured values of the masses of W and Z bosons:
\begin{equation}
\begin{array}{rclll}
M_W&=&80.385&\pm 0.015&\mbox{ GeV},\\
M_Z&=& 91.1876&\pm 0.0021&\mbox{ GeV}
\end{array}
\label{13_7}
\end{equation}
by substituting (\ref{13_7}) into the left-hand side of (\ref{13_5}):
\begin{equation}
A_0 = 37.95\mbox{ GeV}.
\label{13_8}
\end{equation}
The values (\ref{13_5}) and (\ref{13_8}) are rather close.
The difference is about 1.5\%. If one takes into account properly the higher-order corrections
to the relation (\ref{13_5}), the agreement between the two numbers will be improved further.

CC and NC interactions of the SM fermions, as has been shown in the previous section,
have the following structure (see (\ref{11_20})):
\begin{equation}
\begin{array}{rcl}
L_\mathrm{CC}&=&\frac{\displaystyle g_2}{\displaystyle 2\sqrt{2}}\sum_{ij}V_{ij}\bar u_i \gamma_{\mu}(1-\gamma_5)d_j=
\frac{\displaystyle e}{\displaystyle 2\sqrt{2}s_{\mathrm{W}}}\sum_{ij}V_{ij}\bar u_i \gamma_{\mu}(1-\gamma_5)d_j,\\[15pt]
L_\mathrm{NC}&=&e\sum_f Q_f\bar f \gamma_{\mu}f A^{\mu}+ \frac{\displaystyle e}{\displaystyle 4s_{\mathrm{W}} c_{\mathrm{W}}}\sum_f \bar f \gamma_{\mu}(v_f-a_f \gamma_5) f Z^{\mu},
\end{array}
\label{13_9}
\end{equation}
where $V_{ij}$ are the CKM matrix elements, $i,\,j = 1,2,3$ the number of the SM fermion generations and
$$\begin{array}{cccc}
v_{u_i} = 1-\frac{8}{3}s_{\mathrm{W}}^2,&a_{u_i}=1;&v_{d_i}=-1+\frac{4}{3}s_{\mathrm{W}}^2,&a_{d_i}=-1;\\[5pt]
v_{\ell}=-1+4s_{\mathrm{W}}^2,& a_{\ell}=-1;&v_{\nu} = 1,&a_{\nu}=1
\end{array}
$$
are the vector and axial-vector coupling constants.

The Feynman rules following from (\ref{13_8}) allow us to get tree level formulas for the W- and Z-boson
partial decay widths, as shown below at tree level:
\begin{eqnarray}
\includegraphics{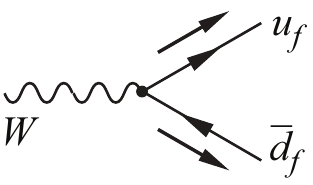} &
\begin{array}{rcl}
\Gamma(W\rightarrow u_f \bar d_f)&=&|V_{ij}|^2 N_c \frac{\displaystyle \alpha}{\displaystyle 12s_{\mathrm{W}}^2} M_W,\\[20pt]
&&
\end{array}
\label{13_10}\\
\includegraphics{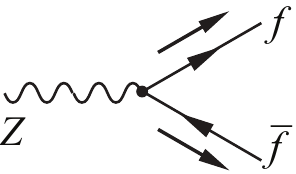} &\begin{array}{rcl}
\Gamma(Z\rightarrow f \bar f)&=&N_c \frac{\displaystyle \alpha M_Z}{\displaystyle 12 \sin^2(2\theta_\mathrm{W})}[v_f^2+a_f^2],\\[20pt]
&&
\end{array}
\label{13_11}
\end{eqnarray}
where the number of colours $N_{\rm c}=3$ for quarks and $N_{\rm c} = 1$ for leptons.

The total W- and Z-boson widths are obtained by summing up all the partial widths (\ref{13_10}) and (\ref{13_11}).
Since CCs for all SM fermions have the same V--A structure, one can very easily obtain branching fractions for W-decay modes:
\begin{equation}
\begin{array}{rclll}
\sum_q \mbox{Br}(W\rightarrow q\bar q)&=&2N_c\cdot\frac{\displaystyle 1}{\displaystyle 9}&=&\frac{\displaystyle 2}{\displaystyle 3},\\[5pt]
\sum_{\ell} \mbox{Br}(W\rightarrow {\ell}\nu)&=&3\cdot\frac{\displaystyle 1}{\displaystyle 9}&=&\frac{\displaystyle 1}{\displaystyle 3}.
\end{array}
\end{equation}
The measured $\mbox{Br}(W\rightarrow {\ell}\nu) = (10.80\pm0.09)\%$ is in reasonable agreement with the simple tree level result
\begin{equation}
\mbox{Br}(W\rightarrow {\ell}\nu) = \frac{1}{9} = (11.11)\%.
\label{13_13}
\end{equation}
QCD corrections to the branching ratio $\mbox{Br}(W\rightarrow q\bar q)$ improve the agreement.

\begin{figure}[thb]
\centering

\includegraphics{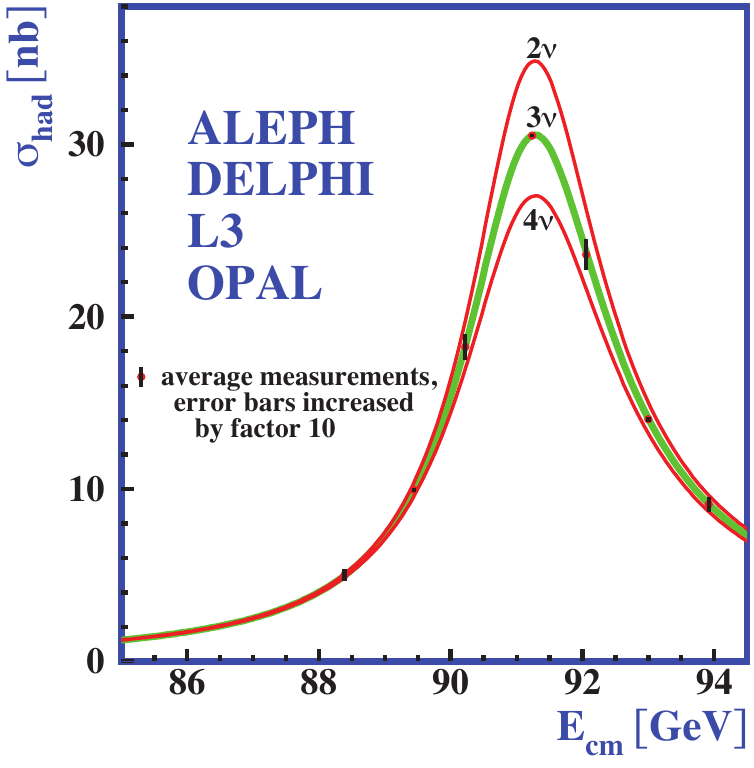}

\caption{$N_{\nu}$ from LEP measurements}

\label{Nnu}

\end{figure}

The decay width of the Z boson to neutrinos, the invisible decay mode, allows us to measure the number
of light $(m_{\nu}<M_Z/2)$ neutrinos by comparing
\begin{equation}
\Gamma^Z_{\rm inv} = \Gamma^Z_{\rm tot}-\Gamma^Z_{\rm had}-\Gamma^Z_{\ell^+\ell^-}
\label{13_14}
\end{equation}
with  the tree level formula obtained from (\ref{13_11}):
\begin{equation}
\Gamma^Z_{\rm inv} =\Gamma^Z_{\nu\bar \nu} = N_{\nu}\cdot \frac{\alpha M_Z}{12\sin^2(2\theta_\mathrm{W})}(1+1).
\label{13_15}
\end{equation}
Experimentally, $\Gamma_{\rm tot}$ is measured  from the shape of the Z-boson resonance according to the well-known Breit--Wigner formula
$$\Gamma^Z_{\rm tot} = 2.4952 \pm 0.0023 \, \mathrm{GeV}.$$
Decay widths to hadrons and charged leptons are measured directly in ${\rm e}^+{\rm e}^-$ collisions (LEP1) to be
$$\Gamma^Z_{\rm had} = 1744.4 \pm  2.0 \, \mathrm{MeV}, $$
$$ \Gamma^Z_{\ell^+\ell^-} = 83.984 \pm 0.086 \, \mathrm{MeV}. $$
As a result, $\Gamma^Z_{\rm inv}$ obtained from (\ref{13_14}) is
$$ \Gamma^Z_{\rm inv} = 0.4990 \pm 0.0015 \, \mathrm{GeV}.$$
 This gives for $N_{\nu}$
$$N_{\nu} = 2.984 \pm 0.008,$$
which is close to the number of known neutrinos. The test is an important
confirmation of three generations of fermions assumed in the SM and observed in experiments as shown in Fig. \ref{Nnu}.

One can make this test by looking at the ratio $\Gamma^Z _{\rm inv}/\Gamma^Z_{{\rm e}^+{\rm e}^-}$.
In the SM, the ratio follows from (\ref{13_11}) and (\ref{13_15}):
\begin{equation}
\frac{\Gamma^Z_{\rm inv}}{\Gamma^Z_{{\rm e}^+{\rm e}^-}} = \frac{2N_{\nu}}{1+(1-4s^2_{\mathrm{W}})^2}.
\label{13_16}
\end{equation}
The measured value (5.942 $\pm$ 0.016) is in agreement with 5.970 coming from
the formula (\ref{13_16}) for $N_{\nu}=3$ and $s_{\mathrm{W}}^2 = 0.2324$.

\begin{figure}[h]
\centering

\includegraphics{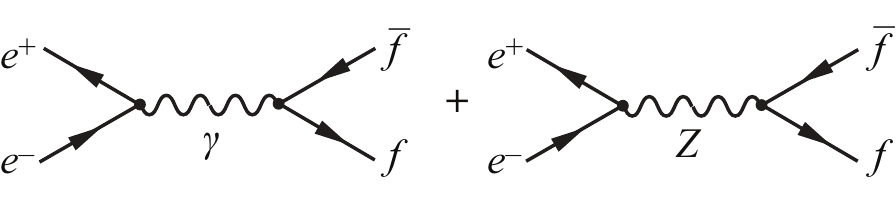}

\caption{${\rm e}^+{\rm e}^-$  diagram}

\label{F13_1}

\end{figure}

\begin{figure}[hbt]
\centering

\includegraphics{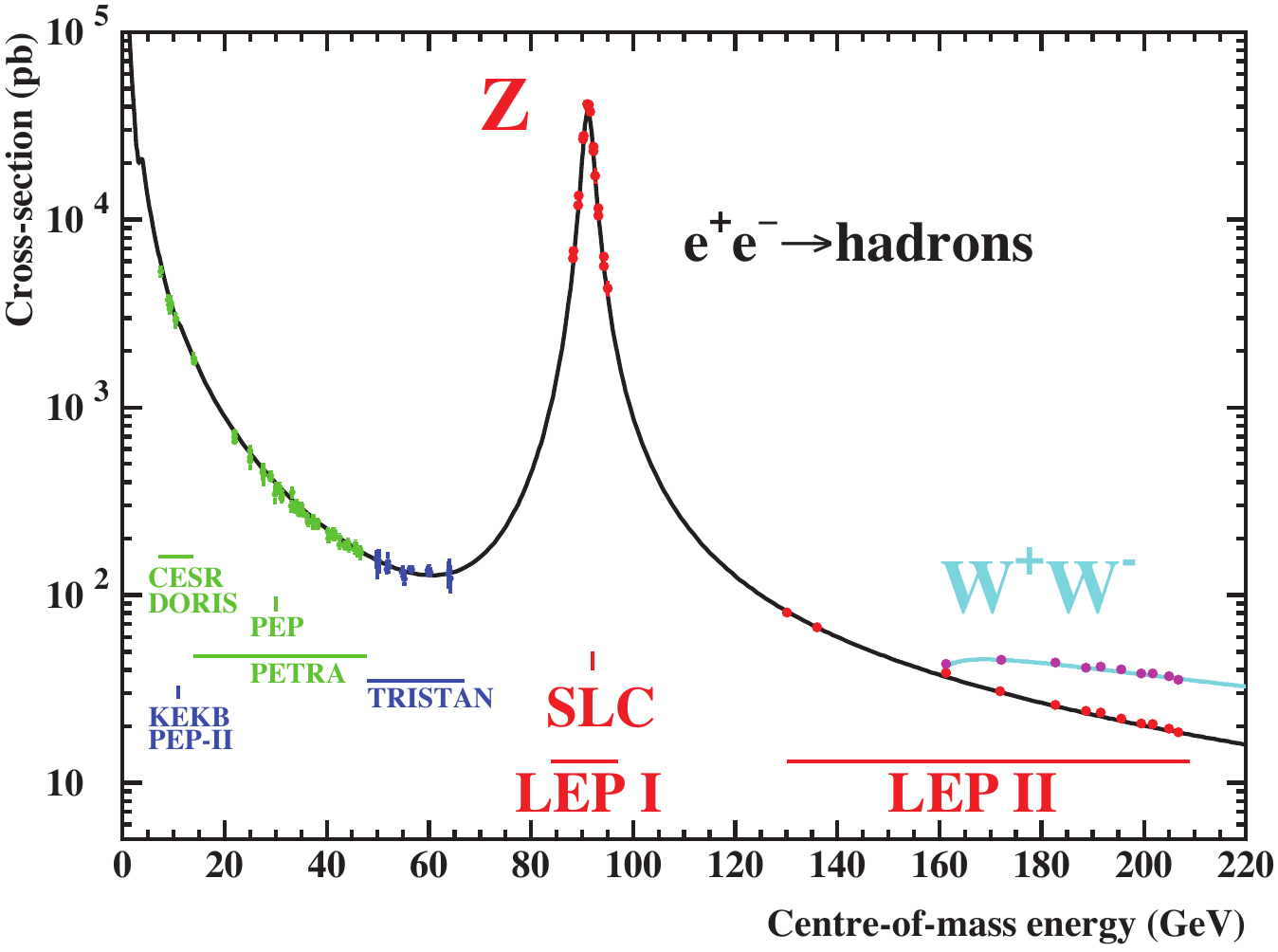}

\caption{The cross section of hadron  production in $e^+e^-$ collisions showing
a good agreement between the SM computation and various experiments 
in the energy range up to 220 GeV including the contribution of the Z-boson resonance.}

\label{eeZ}

\end{figure}

An important part of information about the EW interactions and couplings of the SM fermions comes from ${\rm e}^+{\rm e}^-$ annihilation
to fermion--antifermion pairs. The differential cross-section computed 
from the diagrams shown in Fig. \ref{F13_1}
has a well-known form
neglecting fermion masses compared to the centre of mass energy $\sqrt{s}$:
\begin{equation}
\begin{array}{rcl}
\frac{\displaystyle \mathrm{d}\sigma}{\displaystyle \mathrm{d} \cos \theta}&=&\frac{\displaystyle 2\pi \alpha^2}{\displaystyle 4s}N_C\left\{(1+\cos^2\theta)\cdot \right .\\[10pt]
&&\cdot \left[ Q^2_f-2\chi_1v_ev_fQ_f+\chi_2(a_{e}^2+v_{e}^2)(a_f^2+v_f^2)\right]\\[10pt]
&&\left . +2\cos \theta \left[ -2\chi_1 a_e a_f Q_f+4\chi_2 a_e a_f v_e v_f\right]\right\},
\end{array}
\label{13_17}
\end{equation}
where
$$\chi_1=\frac{1}{16s_{\mathrm{W}}^2c_{\mathrm{W}}^2}\frac{s(s-M_Z^2)}{(s-M_Z^2)^2+M_Z^2\Gamma_Z^2},$$
$$\chi_2=\frac{1}{256s_{\mathrm{W}}^2c_{\mathrm{W}}^2}\frac{s^2}{(s-M_Z^2)^2+M_Z^2\Gamma_Z^2}.$$
The  cross-section obtained from the differential form (\ref{13_17}) 
is in good agreement with the experimental data,
as shown in Fig. \ref{eeZ}.
In the region far below the Z-boson pole, one can neglect the Z-boson exchange diagram and restore the well-known QED formula
\begin{equation}
\frac{\mathrm{d} \sigma}{\mathrm{d}\cos \theta}=\frac{\pi \alpha^2}{2s}Q_f^2 N_C(1+\cos ^2 \theta),\;\;\;\sigma = \frac{4\pi \alpha^2}{3}Q^2 N_C.
\label{13_18}
\end{equation}
From the formula (\ref{13_17}), one can get a number of asymmetries, which have been measured, in particular, at LEP1 and SLC.
In the region close to the  Z-boson pole the photon exchange part is small and can be neglected. Then the forward--backward asymmetry is
$$A_{\rm FB}\equiv \frac{N_{\rm F}-N_{\rm B}}{N_{\rm F}+N_{\rm B}},$$
where $$N_{\rm F} = \int^1_0 \mathrm{d}(\cos \theta)\frac{\mathrm{d} \sigma}{\mathrm{d}\cos \theta},\;\;\;N_{\rm B} = \int^0_{-1} \mathrm{d}(\cos \theta)\frac{\mathrm{d} \sigma}{\mathrm{d}\cos \theta}.$$
Simple integration of (\ref{13_17}) gives the following result:
$$A_{\rm FB} = \frac{3}{2} A_{e}\cdot A_{f},\;\;\;A_{e,f} = \frac{2a_{e,f}v_{e,f}}{a^2_{e,f}+v^2_{e,f}}.$$

\begin{figure}[hbt]
\centering

\includegraphics[width=0.6\textwidth]{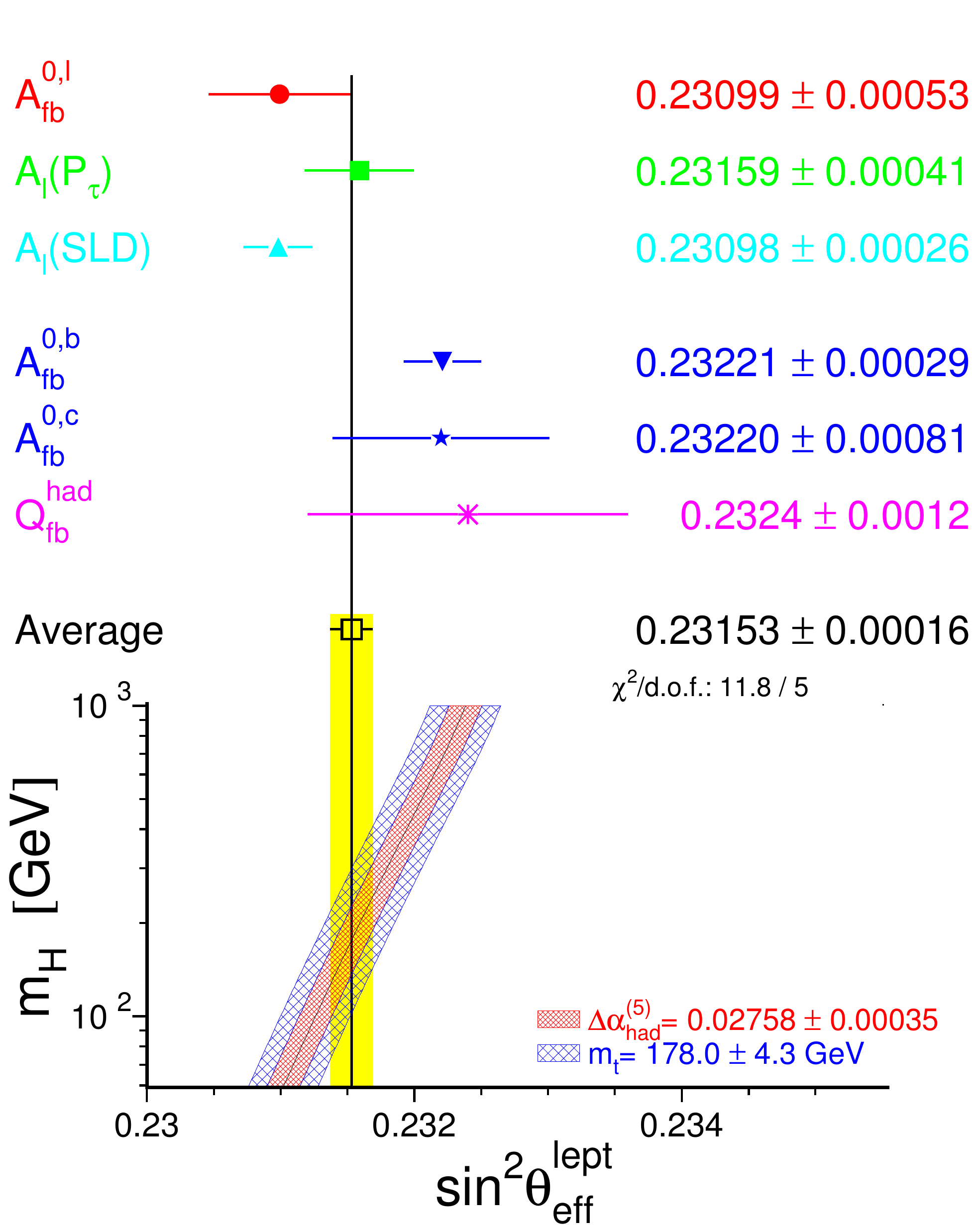}

\caption{Effective electroweak mixing angle from various observables \cite{ALEPH:2005ab}}

\label{effective_sin}

\end{figure}

Measurements of decay widths being proportional to $(a^2_{f}+v^2_{f})$ and asymmetries for different
fermions $f$ allow us to extract the coefficients $a_f$ and $v_f$. Then one can get a precise value for
the Weinberg mixing angle from the relation involving the lepton couplings: 
$$\sin^2\theta_{\rm eff}^{\rm lept} = \frac{1}{4}\left(1 - \frac{v_l}{a_l}\right).$$
Results of the measurements are shown in Fig.\ref{effective_sin} 
 as obtained by the EW working group \cite{ALEPH:2005ab}.

\begin{figure}[hbt]
\centering

\includegraphics{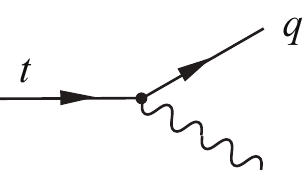}

\caption{Top-quark decay}

\label{F13_2}

\end{figure}

In the SM there are no $1\rightarrow 2$ decays of fermions to the real Z boson due
 to  absence of FCNCs (flavour-changing neutral currents). The top quark is heavy enough to decay to a W boson,
as shown in Fig. \ref{F13_2}.
The decay mode to the b-quark is dominant in the top decays due to the CKM mixing matrix structure, where
$$V_{tb}\sim 1 \gg V_{ts},V_{td}.$$
A direct tree level computation leads to
\begin{equation}
\Gamma_{\rm top}=\frac{G_{\rm F}M_t^3}{8\pi \sqrt{2}}\left(1-\frac{M_W^2}{M_t^2}\right)^2\left(1+2\frac{M_W^2}{M_t^2}\right),
\label{13_19}
\end{equation}
where one neglects the b-quark mass:
$$\Gamma(t \to bW)_{\rm LO}\simeq 1.53\mbox{ GeV},\,\,\,\,\Gamma(t \to bW)_{\rm corr} = 1.42\mbox{ GeV}.$$
The top-quark lifetime $\tau_{\rm top} = 1/\Gamma_{\rm top}$ is about $5\cdot 10^{-25}$~s,
which is much smaller than the typical time of strong bound state formation $\tau_{\rm QCD} \sim 1/\Lambda_{\rm QCD} \sim 3\cdot 10^{-24}$~s.
The top quark decays before hadronization. Therefore, there are no hadrons containing the top quark.

Since the top-quark mass is larger than the W-boson and b-quark masses, one can use the EW equivalence theorem
to get the leading top width up to the term $m_W^2/m_t^2$. According to the EW equivalence theorem, amplitudes with
external W and Z bosons are dominated by the longitudinal polarization of
the bosons $\left({e}^{W,Z}_{L} \sim p^0/M_{W,Z} \right)$. But the longitudinal component in the
SM appears by `eating' the Goldstone bosons $w_{\mathrm{g}},\,\,z_{\mathrm{g}}$. So, one can compute simply the diagram
$$ \includegraphics{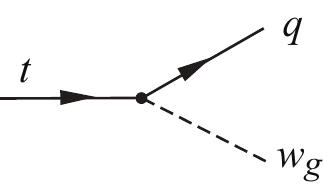}$$
with the Yukawa vertex $M_t/(v\sqrt{2})$. Then one immediately obtains for the top width the following formula:
$$\Gamma = \frac {2}{32\pi}\left( \frac{M_t}{v}\right)^2\cdot M_t = \frac{G_{\rm F} M_t^3}{8 \pi\sqrt{2}},$$
which is exactly equal to the first term in (\ref{13_19}), as expected.

The EW single top quark production is another confirmation of the EW fermion structure of the SM.
There are three mechanisms of single top production at hadron colliders differing by the typical virtuality ($Q_W^2$) of the W boson involved:

\begin{eqnarray}
\includegraphics{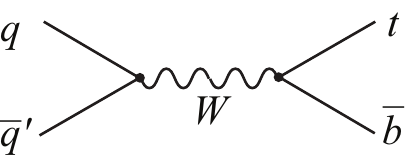} &
\begin{array}{rcl}
s\mbox{-channel,}&\;&Q^2_W>0,\\[20pt]
&&
\end{array}
\nonumber\\
\includegraphics{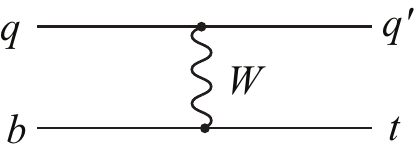} &\begin{array}{rcl}
t\mbox{-channel,}&\;&Q^2_W<0,\\[20pt]
&&
\end{array}
\nonumber\\
\includegraphics{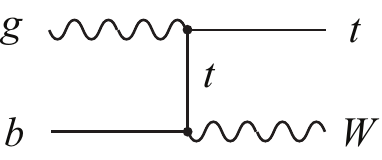} &\begin{array}{rcl}
+W\mbox{-associated,}&\;&Q^2_W\approx M_W^2.\\
&&
\end{array}
\nonumber
\end{eqnarray}quart
 $t$-channel and $s$-channel production mechanisms have been observed at the Tevatron, while $t$-channel and $W$-associated
 production was observed at the LHC.
 There are a number of important QCD next-to-leading-order (NLO) and next-to-next-to-leading-order (NNLO)
corrections which are needed to
be taken into account in order to get SM predictions with needed accuracy to be
compared to experimental results. Up to now a good agreement with SM computations was observed.

\begin{figure}[hbt]
\centering

\includegraphics{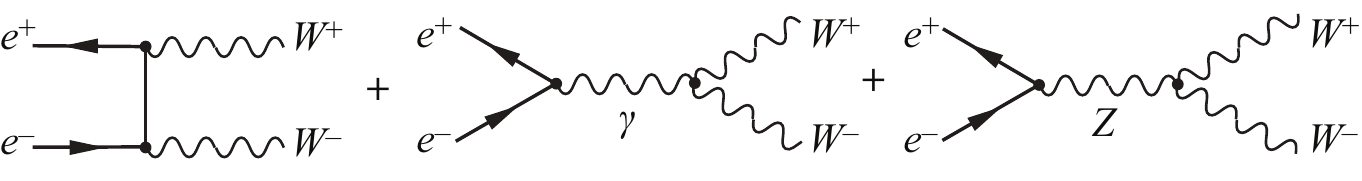}

\caption{${\rm e}^+{\rm e}^-\rightarrow W^+W^-$}

\label{F13_3}

\end{figure}

A well-known example demonstrating correctness of the Yang--Mills interactions of the gauge bosons is
the gauge boson pair production.
 Triple gauge boson vertices WW$\gamma$ and WWZ have been tested at LEP2
(${\rm e}^+{\rm e}^-\rightarrow {\rm W}^+{\rm W}^-$) and at the Tevatron
($q \bar q\rightarrow {\rm W}^+{\rm W}^-$, $q \bar q'\rightarrow {\rm W} \gamma$, $q \bar q'\rightarrow {\rm WZ}$).
The diagrams for the process ${\rm e}^+{\rm e}^-\rightarrow W^+W^-$ form the so-called CC3
set of diagrams, as shown in Fig.~\ref{F13_3}.

\begin{figure}[hbt]
\centering

\includegraphics[width=0.5\textwidth]{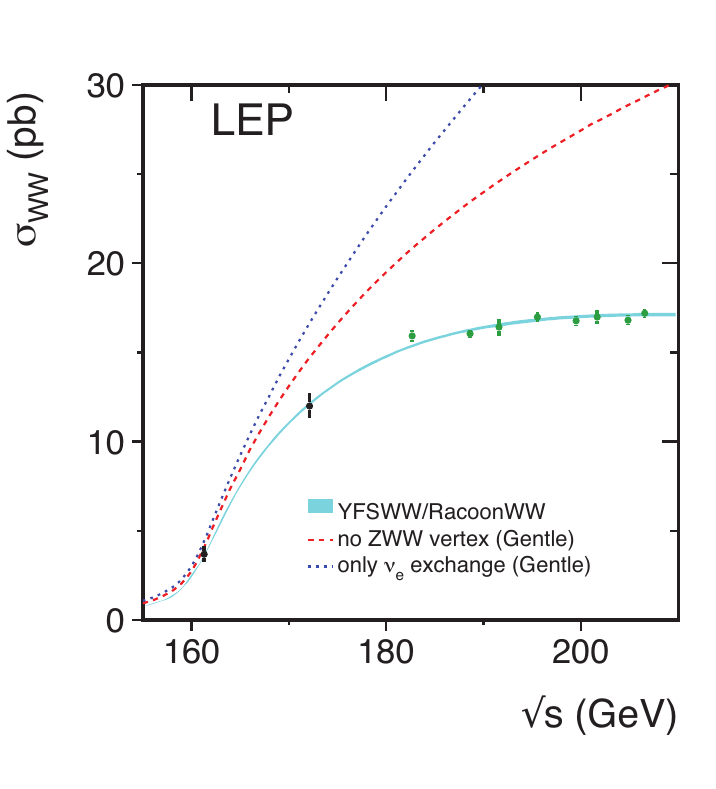}

\caption{Measurements of the W-pair production cross-section, compared to the theoretical predictions. 
For explanations, see \cite{Schael:2013ita}.}

\label{F13_4}

\end{figure}

 The triple vertex of the Yang--Mills interaction:
 $$\Gamma_{m_1m_2m_3}^{WW\gamma /Z}(p_1p_2p_3)=g_{\gamma,Z}\left[ (p_1-p_2)_{m_3}g_{m_1m_2}+(p_3-p_1)_{m_2}g_{m_1m_3}+(p_2-p_3)_{m_1}g_{m_2m_3} \right],$$
 where $g_{\gamma}=e$, $g_Z = g_2 c_{\mathrm{W}} = e \frac{c_{\mathrm{W}}}{s_{\mathrm{W}}}$, is confirmed perfectly experimentally, as shown in Fig.~\ref{F13_4}.

The quartic gauge coupling WW$\gamma \gamma$ has been tested recently at the Tevatron and the LHC in W-pair production in association with
two protons or a proton and an antiproton. The quartic couplings WW$\gamma$Z, WWZZ have not been tested yet.
This is a challenging task for the LHC and will require a high-luminosity regime at a linear collider.

\section{The electroweak SM beyond the leading order} 

All the above examples confirming the structure of the SM interactions are leading order processes.
 However, in many cases a high accuracy of experimental measurements requires the SM computations beyond the leading order.
 In the SM, being a  quantum field theory,
computations of higher-order corrections face divergences of
ultraviolet and infrared/collinear nature. In general, the ultraviolet (hard) divergences are
treated with the help of the renormalization procedure while the infrared/collinear (soft) divergences
are cancelled out due to the Kinoshita--Lee--Nauenberg theorem in summing virtual and real contributions to squared matrix elements.

The renormalization procedure is the usual way to deal with the ultraviolet divergences.
We describe briefly only the main ideas of the procedure. In the SM the dimensions of all the coupling constants are
equal to zero. This property has important consequences making the theory renormalizable. In simple words, the renormalizability
 means that all the UV divergences may be incorporated into a redefinition of a few constants such as coupling constants,
masses and field normalization constants. In renormalizable theories only a few diagrams are UV divergent.

As an example, let us consider QED.

The divergency index of a diagram depends only on the number of external legs, and for QED can be expressed in a well-known form
$$ w = 4 - L_{\gamma} - 3/2 L_{\rm e},$$
where $L_{\gamma}$ is the number of external photon lines and  $L_{\rm e}$ is the number of external electron lines.

So, there are only three types of divergent diagrams
 with two external photon lines, the photon self-energy~~~ \includegraphics{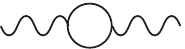}~~~,
with two external electron lines, the electron self-energy~~~\includegraphics{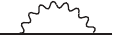}~~~
and, with one photon and two electron external lines, the electron--photon vertex~~~ \includegraphics{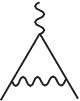}.

It is convenient to consider the renormalization procedure in the functional integral approach already familiar to us.
The generating functional integral in QED in the covariant gauge is given by
\begin{equation}
\begin{array}{rcl}
Z[J, \eta, \bar\eta]& = &\int D(\bar \Psi \Psi A)\exp\left(\mathrm{i}\int {\rm d}^4x \bar\Psi (\mathrm{i}\lefteqn{D}{\;/}-m)\Psi + \mathrm{i}eA+J_{\mu}A^{\mu} \right.\\[10pt]
&&+\left.  \bar \eta \Psi + \bar \Psi \eta -\frac{1}{4}F_{\mu\nu}F^{\mu\nu} +\frac{1}{2\xi}\int {\rm d}^4x(\partial_{\mu}A^{\mu})^2\right).
 \end{array}
\label{14_1}
\end{equation}
The photon propagator is obtained from (\ref{14_1}) by taking two functional derivatives
on $J$ and setting the sources $J$, $\eta$ and $\bar \eta$ to zero:
\begin{equation}
\begin{array}{rcl}
\mathrm{i}D_{\alpha\beta}(x_1,x_2)& = &\int D(\bar \Psi \Psi A)A_{\mu}(x_1)A_{\mu}(x_2)\\[10pt]
&& \cdot\exp\left( \mathrm{i}\int {\rm d}^4x [ -\frac{1}{4}F_{\mu\nu}F^{\mu\nu} +\bar \Psi(i\lefteqn{\partial}{/}-m)\Psi+e\bar \Psi \lefteqn{A}{\;/}\Psi]\right).
 \end{array}
\label{14_2}
\end{equation}

The Dyson--Schwinger equation for the photon propagator is obtained as a consequence
 of the invariance of the measure of the functional integral with respect to the  shift
$A_{\mu}(x) \rightarrow A_{\mu} (x) + \varepsilon_{\mu} (x)$.

The equation for the inverse propagator takes, after Fourier transformation, the form
\begin{equation}
D^{-1}_{\alpha \beta}(k) = (D_0)^{-1}_{\alpha \beta}+\Pi_{\alpha \beta}
\label{14_3}
\end{equation}
or, graphically,\\
$$\includegraphics{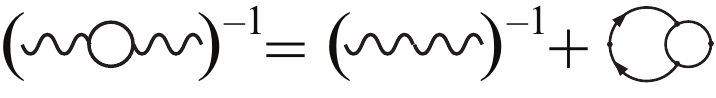},$$
where \includegraphics{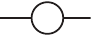} denotes the dressed fermion propagator and  \includegraphics{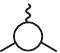} is
the truncated one-particle irreducible vertex function $\Gamma_{\mu}(p_1, p_2, k)$.

At one-loop level the function $\Pi_{\alpha \beta}(k)$ is given by the following Feynman integral:
\begin{equation}
(-{\rm i}e)^2\int \frac{{\rm d}^4 p}{(2\pi)^4}\mbox{Tr}\left[ \frac{\lefteqn{p}{/}+m}{p^2-m^2+{\rm i}0}\gamma_{\alpha} \frac{(\lefteqn{p}{/} - \lefteqn{k}{/})+m}{(p-k)^2-m^2+{\rm i}0}\gamma_{\beta}\right].
\label{14_4}
\end{equation}
The integral (\ref{14_4}) is quadratically divergent from formal power counting.
In order to deal with divergent integrals, we need to introduce some regularization.
We use the dimensional regularization
$${\rm d}^4p \rightarrow \mathrm{d}^D p (\mu ^2)^{2-D/2}.$$
One can see that (\ref{14_4}) gives zero, being convoluted with the external moments $k_{\alpha}$ or $k_{\beta}$. Indeed,
 \begin{equation}
k_{\alpha}\gamma^{\alpha}=( \lefteqn{p}{/}) - (\lefteqn{p}{/}- \lefteqn{k}{/}) = (\lefteqn{p}{/} - m) - [(\lefteqn{p}{/}- \lefteqn{k}{/})-m].
\label{14_5}
\end{equation}
If we substitute (\ref{14_5}) into (\ref{14_4}), we  get
  \begin{equation}
(\mu^{2-D/2})(-{\rm i}e)^2\int \frac{\mathrm{d}^Dp}{(2\pi)^D}
\mbox{Tr}\left[\left( \frac{\lefteqn{p}{/}+m}{p^2-m^2+{\rm i}0}-
\frac{(\lefteqn{p}{/} - \lefteqn{k}{/})+m}{(p-k)^2-m^2+{\rm i}0}\right)\gamma_{\beta}\right]=0.
\label{14_6}
\end{equation}
In fact, this result is valid to all perturbation orders due to the Ward identity
 \begin{equation}
k^{\mu}\Gamma_{\mu}(p_1, p_2, k) = S^{-1}(p_1) - S^{-1}(p_2).
\label{14_61}
\end{equation}
The identity (\ref{14_6}) can be easily derived from $U(1)$ gauge invariance of (\ref{14_1}).

The property means that $\Pi_{\alpha \beta}$ has the following form:
 \begin{equation}
\Pi_{\alpha \beta} (k) = \left( g_{\alpha \beta} k^2 - k_{\alpha}k_{\beta}\right)\Pi(k^2).
\label{14_7}
\end{equation}
Therefore, the dressed photon propagator can be written as
 \begin{equation}
D_{\alpha \beta} (k) =-\frac{\rm i}{k^2}\left[\frac{1}{1+\Pi_{\gamma}(k^2,\varepsilon,\mu^2)}
 \left( g_{\alpha \beta}  - \frac{k_{\alpha}k_{\beta}}{k^2}\right)+\xi\frac{k_{\alpha}k_{\beta}}{k^2}\right].
\label{14_8}
\end{equation}
Of course, in the case where $\Pi_{\gamma} = 0$, we obtain the free photon propagator.

The factor $(1+\Pi_{\gamma}(k^2,\varepsilon,\mu^2))^{-1}$, being taken at zero momentum, should be removed for
the correct normalization
of the kinetic term. It can be done  by rescaling the field $A_{\mu}(x)$ in the following way:
 \begin{equation}
A_{\mu}(x)\rightarrow \frac{1}{\sqrt{Z_3}}A_{\mu}, \mbox{ where } Z_3^{(a)} = (1+\Pi_{\gamma}(0,\varepsilon))^{-1}.
\label{14_9}
\end{equation}
Direct computation of (\ref{14_4}) with well-known Feynman techniques gives
\begin{equation}
\begin{array}{c}
\Pi_{\gamma}(k^2,\varepsilon, \mu^2)=\frac{\displaystyle \alpha}{\displaystyle 3\pi \varepsilon} +\Pi_{\rm finite}\\[10 pt]
\mbox{and therefore}\; Z_3^{-1} = 1+\frac{\displaystyle \alpha}{\displaystyle 3\pi \varepsilon},
\end{array}
\label{14_10}
\end{equation}
where $\varepsilon = (4-D)/2$.

Now let us consider the fermion propagator taking functional derivatives of (\ref{14_1}) on the fermion sources $\bar \eta$ and $\eta$.

Now the expression for the dressed inverse fermion propagator is
\begin{equation}
S^{-1}(p) = S_0^{-1}(p) - \Sigma(p).
\label{14_11}
\end{equation}
The formula (\ref{14_10}) is also the Dyson--Schwinger equation for the dressed fermion propagator, graphically presented as
$$\includegraphics{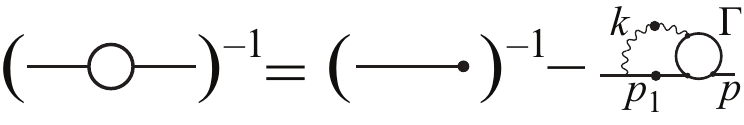}.$$
 The equation involves the same truncated vertex function $\Gamma_{\mu}(p_1 p_2; k)$.
 In the second order of perturbation theory, $\Sigma^{(2)}(p)$ is as follows:
\begin{equation}
-{\rm i}\Sigma^{(2)}(p)=(-{\rm i}e)^2\int \frac{\mathrm{d}^D k}{(2\pi)^D}(\mu^2)^{2-D/2}\cdot \gamma^{\mu}D_{\mu\nu}(k)\frac{\lefteqn{p}{/} - \lefteqn{k}{/} +m}{(p-k)^2-m^2}\gamma_{\nu}.
\label{14_12}
\end{equation}
Direct computation gives the following answer:
\begin{equation}
 \Sigma_2=\frac{\alpha}{8 \pi}(4m-\lefteqn{p}{/})\frac{2}{\varepsilon}+\Sigma_{\mathrm{finite}}.
\label{14_13}
\end{equation}
The generic structure of $\Sigma (p)$ is as follows:
$$
 \Sigma (p)= \lefteqn{p}{/} f_1(p^2)-m f_2 (p^2).
$$
Due to (\ref{14_11}), this means that the fermion propagator has the form
\begin{equation}
S(p) = \frac{1}{\lefteqn{p}{/} (1-f_1(p^2))-m(1- f_2 (p^2))}=-\frac{1}{1-f_1(p^2)}\frac{1}{\lefteqn{p}{/} - m\frac{\displaystyle 1-f_2(p^2)}{\displaystyle 1-f_1(p^2)}}.
\label{14_14}
\end{equation}
Close to physics mass, one should have
$$m_{\rm phys} = m\frac{1-f_2(m_{\rm phys}^2)}{1-f_1(m_{\rm phys}^2)}.$$
So, the fermion propagator has the following form close to physics mass:

\begin{equation}
S(p) = \frac{Z_2(\varepsilon, \mu)}{\hat p -m_{\rm phys}(\varepsilon, \mu)},
\label{14_15}
\end{equation}
where $Z_2 = (1-f_1)^{-1}$, $m_{\rm phys} = m\frac{Z_2}{Z_m}$ and $Z_m = (1-f_2)^{-1}$.
From the one-loop result (\ref{14_13}), we obtain
\begin{equation}
Z^{-1}_2 =1+ \frac{\alpha}{4 \pi \varepsilon} +O(\alpha),
\label{14_16}
\end{equation}
\begin{equation}
Z^{-1}_m =1+ \frac{\alpha}{ \pi \varepsilon} +O(\alpha).
\label{14_17}
\end{equation}

The remaining divergent QED diagram is the vertex function correction given by the integral
\begin{equation}
\includegraphics{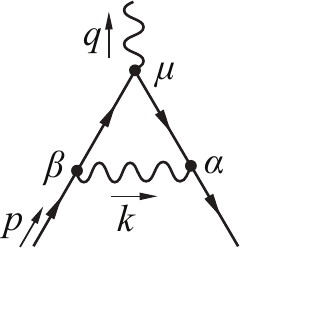}
\begin{array}{rcl}
{\rm i}e\Gamma_{\mu}^{(2)}(p,q)&=&(-{\rm i}e)^3(\mu^2)^{2-D/2}\int \frac{\displaystyle \mathrm{d}^D k}{\displaystyle (2\pi)^D}\\[10pt]
&&\cdot\gamma^{\alpha}(i)\frac{\displaystyle \lefteqn{p}{/}-\lefteqn{q}{/}-\lefteqn{k}{/}+\displaystyle m}{\displaystyle (p-q-k)^2-m^2+{\rm i}0}\\[10pt]
&&\cdot\gamma_{\mu}(i)\frac{\displaystyle \lefteqn{p}{/}-\lefteqn{q}{/}+\displaystyle m}{\displaystyle (p-q)^2-m^2+{\rm i}0}
\cdot\gamma_{\beta}\frac{\displaystyle -{\rm i}}{\displaystyle k^2+{\rm i}0}g^{\alpha \beta}\\
&&\\
&&
\end{array}
\label{14_18}
\end{equation}
In order to compute the divergent part, one can simplify the problem and  compute (\ref{14_18})
in the limit $q\rightarrow 0$. The answer is
\begin{equation}
\Gamma_{\mu}^{(2)}(p,0) = \gamma_{\mu}\left[ \frac{\alpha}{4 \pi}\frac{1}{\varepsilon} + O(\alpha)\right].
\label{14_19}
\end{equation}
Therefore, the vertex function including one-loop correction may be written in the form
\begin{equation}
-{\rm i}e\Gamma_{\mu} = -{\rm i}eZ_1\gamma_{\mu},
\label{14_20}
\end{equation}
where
$$Z_1 = 1-\frac{\alpha}{4\pi}\frac{1}{\varepsilon} + O(\alpha).$$
We can see from (\ref{14_16}) and (\ref{14_20}) that $Z_1=Z_2$ including the one-loop part.
The equality $Z_1=Z_2$ takes place to all orders of perturbation theory due to the Ward identity
\begin{equation}
\Gamma_{\mu} (p,0)= \partial_{\mu}S^{-1}(p),
\label{14_21}
\end{equation}
as follows from (\ref{14_6}) in the limit of the photon momentum $k\rightarrow 0$.

Why did we do all the above computations of divergent
graphs~~~ \includegraphics{14g1-eps-converted-to.pdf},~~~\includegraphics{14g2-eps-converted-to.pdf} and~~~ \includegraphics{14g3-eps-converted-to.pdf}?

Let us rewrite our initial (before renormalization) QED Lagrangian
\begin{equation}
L=-\frac{1}{4} F^0_{\mu\nu}F^{0\mu\nu}+\bar \Psi_0(\mathrm{i}\lefteqn{D}{\,/\,}^0-m_0)\Psi_0,
\label{14_22}
\end{equation}
where $F_{\mu\nu}^0 = \partial_{\mu}A_{\nu}^0-\partial_{\nu}A_{\mu}^0$, $D_{\mu}^0=
\partial_{\mu}-{\rm i}e_0A^0_{\mu}$ (we use the symbol $(0)$ to stress that all the objects are not renormalized,
 or bare, as one usually says) in terms of physical fields and parameters labelled by the symbol ph and
additional terms $\Delta L$:
 \begin{equation}
L=-\frac{1}{4} F^{\rm ph}_{\mu\nu}F^{\rm{ph}\mu\nu}+\bar \Psi_{\rm ph}({\rm i}\lefteqn{D}{\,/}_{\rm ph}-m_{\rm ph})\Psi_{\rm ph}+\Delta L,
\label{14_23}
\end{equation}
where $A^{\rm ph}_{\mu} = Z^{-1/2}_3 A^0_{\mu}$, $\Psi_{\rm ph} = Z^{-1/2}_2 \Psi$, $m_{\rm ph} = ({Z_2}/{Z_m})m_0$,
$e_0 = Z_1 Z_2^{-1} Z_3^{-1/2}(\mu)^{D/2-2}e_{\rm ph}$, $D_{\mu}^{\rm ph} = \partial_{\mu}-{\rm i}e_{\rm ph}A^{\rm ph}_{\mu}$ and
 \begin{equation}
 \begin{array}{rcl}
\Delta L&=&-(Z_3-1)\frac{1}{4} F^{\rm ph}_{\mu\nu}F^{{\rm ph}\mu\nu}+(Z_2-1)\bar \Psi_{\rm ph}({\rm i}\lefteqn{\partial}{/})\Psi_{\rm ph}\\[10pt]
&&+(Z_m-1)m_{\rm ph}\bar \Psi_{\rm ph}\Psi_{\rm ph}+(Z_1-1)e_{\rm ph}\bar \Psi_{\rm ph}(\lefteqn{A}{\,/}_{\rm ph})\Psi_{\rm ph}.
\end{array}
\label{14_24}
\end{equation}
 The Lagrangian $\Delta L$ contains so-called counter-terms.
In the leading order we computed all the coefficients in front of the counter-terms.
Now when one computes some effect using the Lagrangian (\ref{14_23})
all UV divergences are cancelled out order by order in perturbation theory by contributions of the counter-terms.

\begin{figure}[hbt]
\centering

\includegraphics[width=0.6\textwidth]{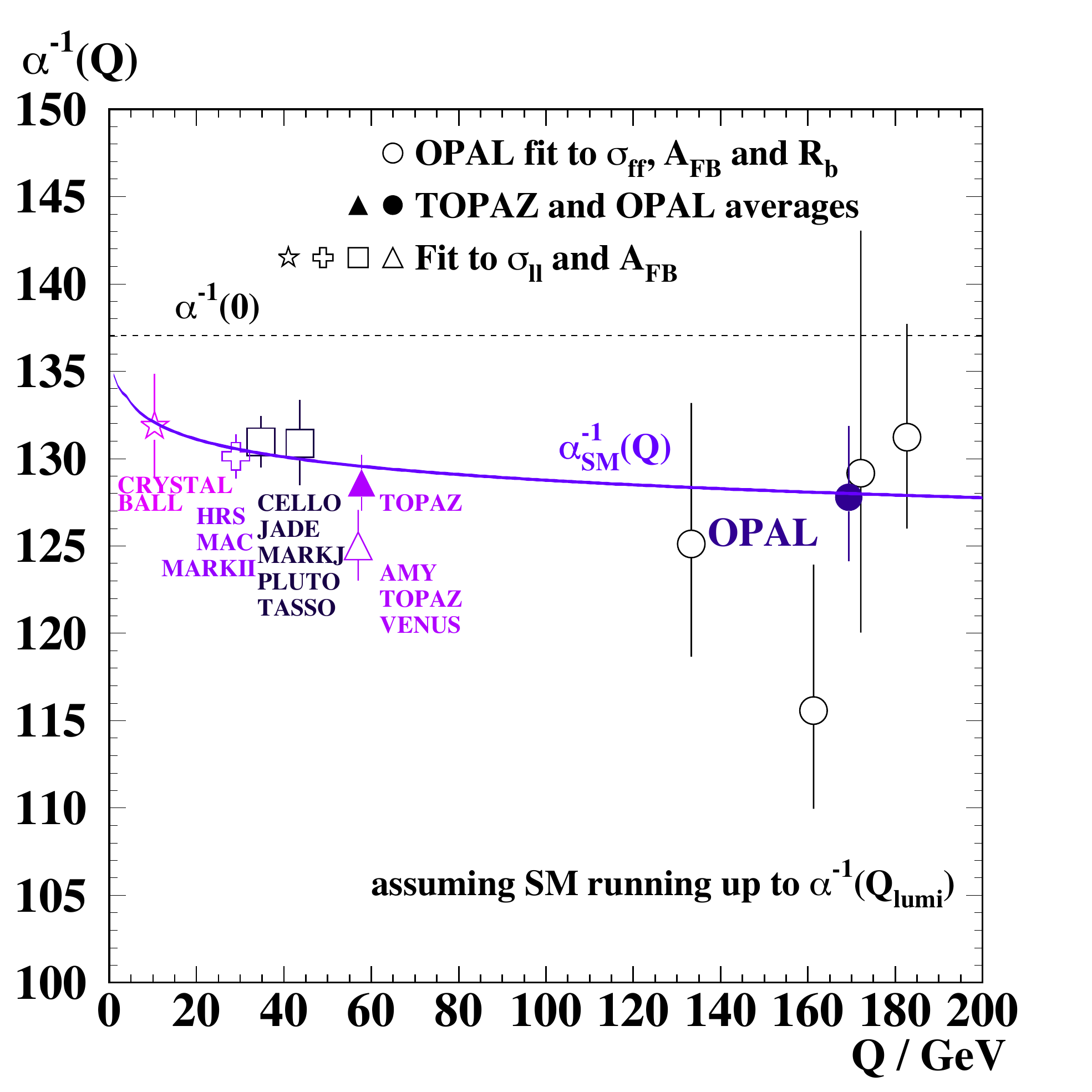}

\caption{Running electromagnetic coupling as a function of collision energy
measured at various energies, in particular, at LEP by the OPAL collaboration  \cite{Abbiendi:1998ea}. }

\label{alpha}

\end{figure}

 Let us look in more detail at the relation for the coupling constant
 $$e_0 = Z_1 Z_2^{-1}Z_3^{-1/2}(\mu)^{D/2-2}e_{\rm ph}(\mu),$$
 where $ (\mu)^{D/2-2}$ is the dimension of the charge. As we discussed above, $Z_1 = Z_2$  due to the Ward identity, and
 \begin{equation}
e_0 = Z_3^{-1/2}(\mu)^{D/2-2}e_{\rm ph}(\mu).
\label{14_25}
\end{equation}
Note that $e_{\rm ph}(\mu)$ is a function of the dimension regularization parameter $\mu$,
 while $e_0$ does not depend on $\mu$. From (\ref{14_25}), one gets the following equality for $\alpha=\frac{{\rm e}^2}{4\pi}$ ($D=4-2\varepsilon$):
  \begin{equation}
\alpha_0 = Z_3^{-1}(\mu^2)^{-\varepsilon}\alpha_{\rm ph}(\mu).
\label{14_26}
\end{equation}
Taking the derivative $\mu\frac{\partial}{\partial \mu}$ on both sides of (\ref{14_26}), we get the following equation:
\begin{equation}
\mu\frac{\partial \alpha}{\partial \mu} = \frac{2\alpha^2}{3\pi} \equiv \beta(\alpha).
\label{14_27}
\end{equation}
The equation (\ref{14_27}) is a particular example of the renormalization group equation, which we do not discuss
in this brief lecture course.
The function on the right-hand side of (\ref{14_27}) is called the $\beta$-function.
 So, at the one-loop level the $\beta$-function in QED is given by the following formula:
 \begin{equation}
\beta(\alpha) = \frac{b_0}{\pi}\alpha^2,\;\;\;b_0=\frac{2}{3}.
\label{14_28}
\end{equation}
One should stress that the coefficient $b_0=\frac{2}{3}$ in QED is positive. The equation (\ref{14_27}) can be easily solved:
 \begin{equation}
\alpha(\mu)= \frac{\alpha(\mu_1)}{1-\frac{\displaystyle \alpha(\mu_1)}{\displaystyle 3\pi}\ln({\mu/\mu_1})^2}.
\label{14_29}
\end{equation}
This is the running coupling constant. If one measures the constant $\alpha$ at some scale $\mu_1$,
 one gets values for the constant at other scales. The coupling constant $\alpha = 1/137$ being measured at very small
scale (small momentum transfer or large distance) in Thompson scattering increases with the scale growing
and becomes $\alpha(M^2_Z)\approx \frac{1}{129}$ at the $Z$ mass. This fact was confirmed nicely in
 LEP experiments. Note that, in order to get $1/129$, one should take into account the contribution of all
SM charged particles to the photon vacuum polarization function $\Pi_{\gamma}$.
This means that the charged particle--antiparticle pairs screen the bare charge at small $\mu^2$ or at large distances.
The QED running coupling constant is illustrated in Fig.\ref{alpha}
\cite{Abbiendi:1998ea}.

Note that in QCD the $\beta$-function is negative, leading to an antiscreening effect;
$\alpha_S$ becomes smaller with increasing of the momentum scale (momentum transfer)
or decreasing distances. This is a famous asymptotic freedom property in QCD.
In QED the situation is the opposite. If the scale $\mu$ increases to very large values the well-known Landau pole
$$\frac{b_0}{\pi} \ln({\mu/\mu_1})^2=1$$
is approached where the perturbation picture in QED breaks down.

One should note that in QED all terms in the four-dimensional Lagrangian (gauge-invariant operators)
have dimension four. As a result, the coupling constant in QED is dimensionless.
This is crucial  to have renormalizable theory, where all UV divergences are cancelled to all
orders with the help of a finite number of counter-terms. This is also the case in the SM.
 All terms of the SM Lagrangian have dimension four and all the coupling constants are dimensionless.
So, the SM is a renormalizable theory in the same manner as QED.

\begin{figure}[hbt]
\centering

\includegraphics{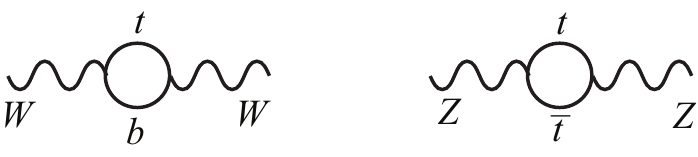}

\caption{Loop corrections}

\label{F14}

\end{figure}

Naively, one may think that the EW higher-order corrections are not that important.
The perturbation theory expansion parameters $\alpha/\pi$ with $\alpha_{\rm em} \sim 1/129$ and
 $\alpha_{\rm weak} \sim 1/30$ are very small. However, the experimental accuracies 
are so high in various cases
that even one-loop EW corrections might not be sufficient. Indeed, selected lists of measured parameters
by LEP1, LEP2, SLD and Tevatron are given below:
$$\begin{array}{cclclcl}
M_Z &=& \;\,91.1875&\pm&0.0021&\mbox{GeV}&0.002\%\\
\Gamma_Z &=&\;\,\;\,2.4952&\pm&0.0023&\mbox{GeV} &0.09\%\\
M_W &=& \;\,80.385&\pm&0.015&\mbox{GeV} &0.02\%\\
M_{\rm top}& = &173.2&\pm&0.9&\mbox{GeV} &0.52\%
\end{array}$$
The most important higher-order corrections come from resummation of the large logarithms, $\log \frac{M_t^2}{m_{e}^2}\approx 24.2$,
as we have seen with running $\alpha$ (1/137 $\rightarrow$ 1/129).
The second class of large corrections comes from contributions of the order of $M^2_{\rm top}/M^2_W$,
which originate from EW Goldstone boson (or longitudinal W/Z polarization state) couplings to the quarks of the third generation.
The later corrections lead to the shifts in W- and Z-boson masses coming from the diagram in Fig. \ref{F14}.

\begin{figure}[hbt]
\centering

\includegraphics[width=0.5\textwidth]{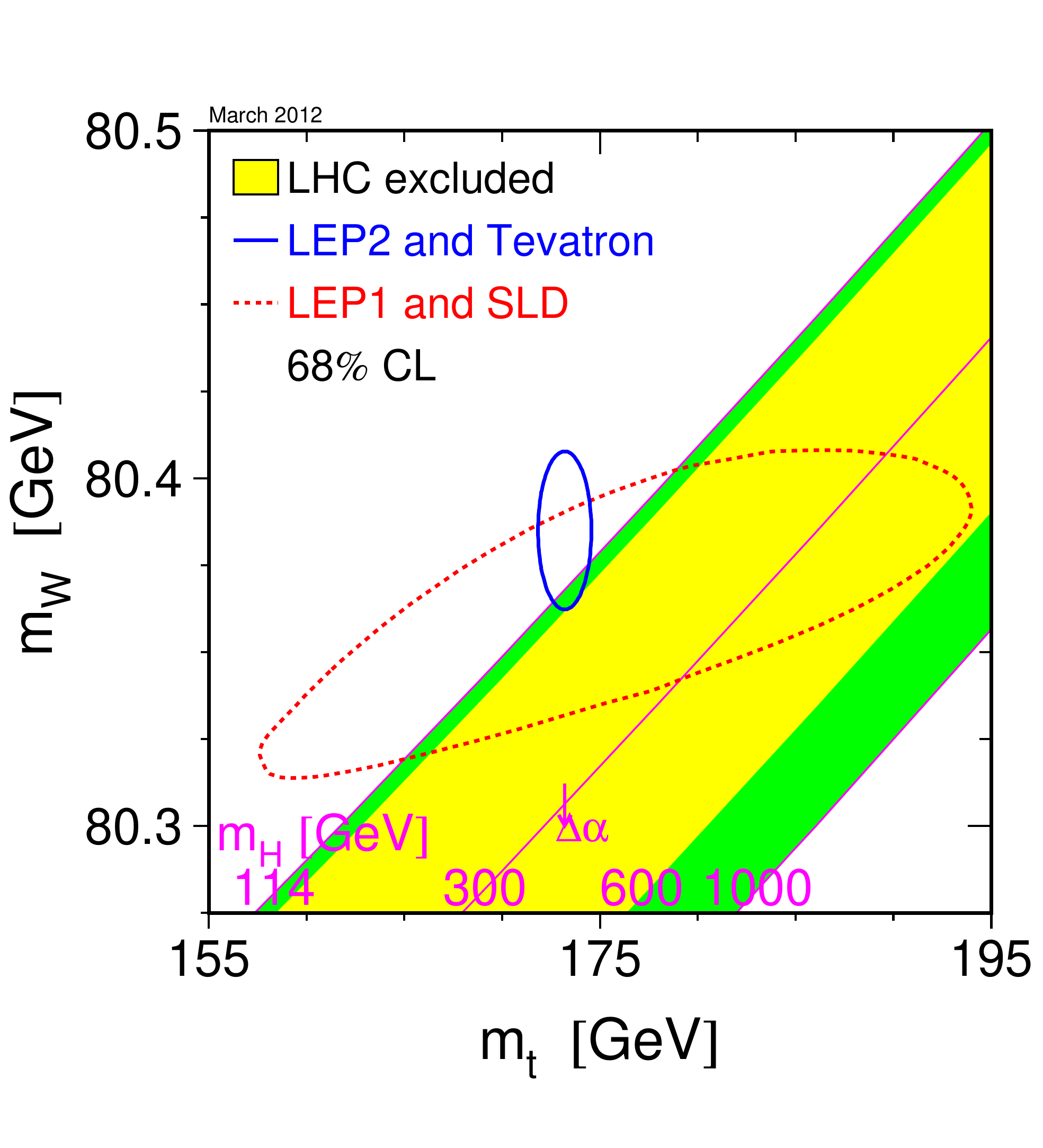}

\caption{W boson mass as a function of the top quark mass at varios masses of the Higgs bosos
(see details in \cite{Schael:2013ita}).}

\label{W-mass-top-mass}

\end{figure}

Loop corrections lead to the fact that SM parameters, coupling constants and masses, 
are running parameters, and they are non-trivial functions of each other.
A famous example is given in Fig.\ref{W-mass-top-mass} \cite{Schael:2013ita} showing the dependence of the W-boson mass as a function of
the top-quark mass at different values of the Higgs boson mass.

One should recall that the top-quark mass has been determined indirectly from the analysis of loop corrections, it being
$$m_t = 178 \pm 8 \begin{array}{l}+17\\-20 \end{array}\;\;\mbox{GeV},$$
which is remarkably close to today's precise measured value $173.2\pm0.9$ GeV.

The low Higgs mass range was preferred by a similar analysis of the Tevatron and LEP data, 
as one can clearly see in Fig.\ref{W-mass-top-mass}.

\begin{figure}[hbt]
\centering

\includegraphics[width=0.7\textwidth]{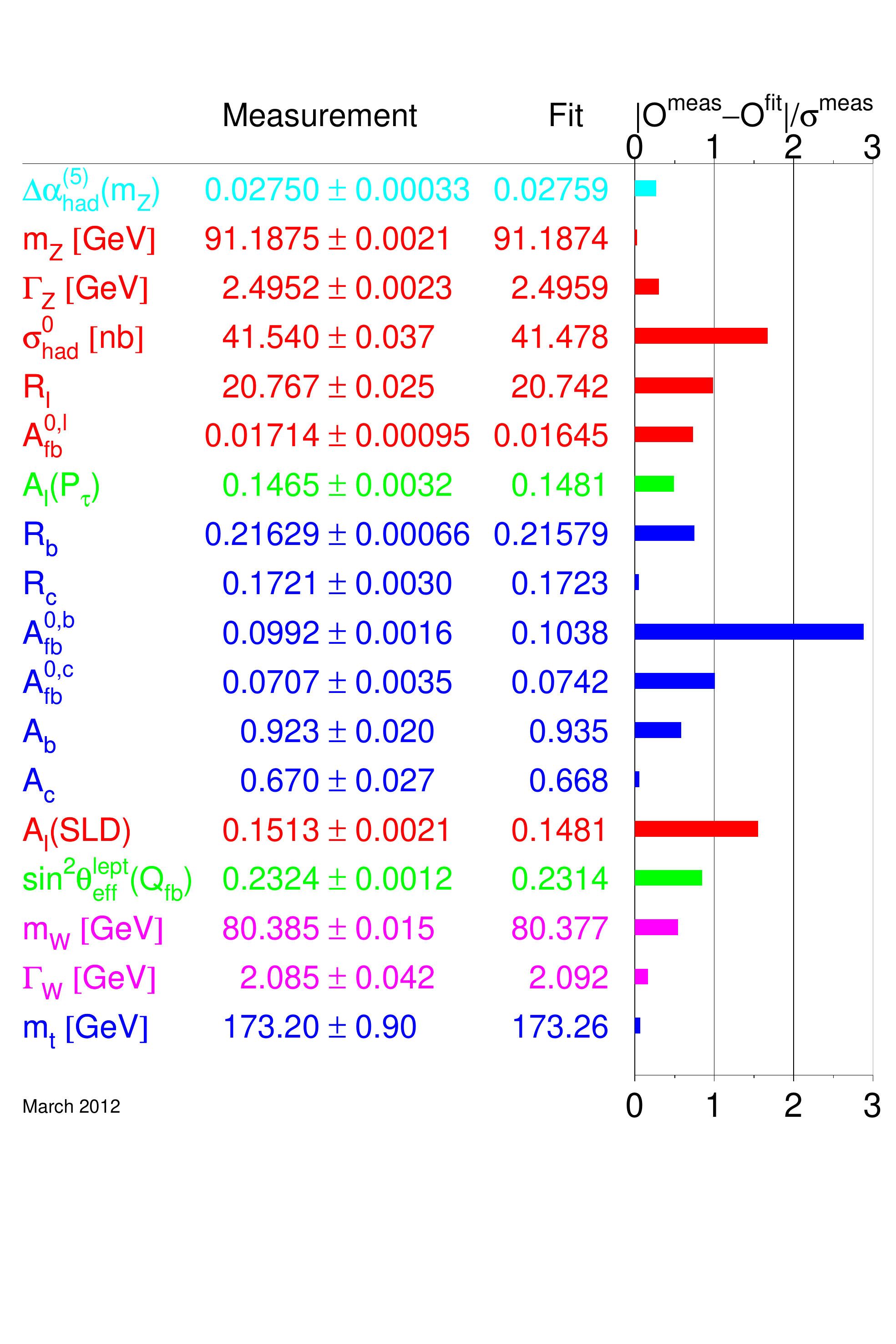}

\caption{Global fit of the EW precision measurements at LEP1, LEP2, SLD and the Tevatron
by the SM computations including loop corrections \cite{ALEPH:2005ab} (the latest update
version of the plot http://lepewwg.web.cern.ch/LEPEWWG/plots/winter2012/) }

\label{EW-global-fit}

\end{figure}

A summary of comparisons of the EW precision measurements at LEP1, LEP2, SLD and the Tevatron and
a global parameter fit is given in the well-known plot shown in Fig.\ref{EW-global-fit} 
\cite{ALEPH:2005ab}.

The only one discrepancy on the level of 3$\sigma$ is observed for $b\bar b$-pair forward--backward asymmetry.

\section{Concluding remarks}
\begin{enumerate}
\item The SM is a renormalizable anomaly-free gauge quantum
field theory with spontaneously broken EW symmetry.
 Remarkable agreement with many experimental measurements is observed.\\
\item All SM leptons, quarks, gauge bosons and, very probably, the Higgs boson have been discovered.\\
\item The SM predicts the structure of all interactions: fermion-gauge, gauge self-interactions,
Higgs-gauge, Higgs-fermion and Higgs self-interactions. As a result, the SM allows us to compute
various cross-sections, distributions and decay rates taking into account higher-order corrections.
However, not yet all of the interactions  were tested experimentally. Parameters of the theory are
not predicted by the theory itself but extracted from the measurements.\\
\item The EW SM has 17 parameters and QCD has one more parameter fixed from experiments:
\begin{itemize}
\item{the gauge-Higgs sector contains four parameters
$g1, g2, \mu^2, \lambda$ or in terms of best measured $\alpha_{\rm em}, G_{\rm F}, M_Z, M_h$;}
\item{six quark masses, three lepton masses;}
\item{three  mixing angles and one phase of the CKM matrix (more parameters come from the
neutrino mixing matrix, which we do not consider here);}
\item{the QCD coupling constant $\alpha_{s}$}.
\end{itemize}
\item There are facts which cannot be explained in the SM:
\begin{itemize}
\item{fermions have very much different masses ($M_{\rm top} = 173$ GeV, $M_{e} = 0.5$ MeV) coming from the same mechanism;}
\item{dark matter exists in the Universe, and there are no dark-matter candidates in the SM;}
\item{the CKM phase as a source of CP violation in the SM is too small to explain particle--antiparticle asymmetry in the Universe;}
\item{neutrino masses, mixing and oscillations cannot be understood in the framework of the SM EW symmetry breaking mechanism;}
\item{there is some tension in explaining the muon anomalous magnetic moment.}
\end{itemize}
\item As is well known, the simplest Higgs mechanism in the SM is not stable with respect to quantum corrections (naturalness problem).
In the SM there is no symmetry which protects a strong (quadratic) dependence of the Higgs mass on a possible new scale.
Something is needed in addition to the SM to stabilize the mass parameter.\\
\item In addition, the SM does not give answers to many questions, such as:
\begin{itemize}
\item{What is a generation? Why are there only three generations?}
\item{How are quarks and leptons related to each other?; what is the nature of the quark--lepton analogy?}
\item{What is responsible for gauge symmetries, why are charges quantized? Are there additional gauge symmetries?}
\item{What is responsible for the formation of the Higgs potential?}
\item{To which accuracy is the CPT (charge, parity, and time) symmetry exact?}
\item{Why is gravity so weak compared to other interactions?}
\end{itemize}
\end{enumerate}

In our lecture we focused mainly on the EW part of the SM and  aspects
of the field theory needed clarifying, and we did not discuss QCD physics, Higgs boson physics, neutrino physics,
flavour physics, problems of the SM models leading to BSM (beyond the Standard Model) scenarios and sequences for cosmology.
These are the subjects of the lectures by Z. Tr\'{o}cs\'{a}nyi, J.~Ellis, B.~Gavelo,
Z.~Ligetti, C.~Csaki and D.~Gorbunov at the 2013 European School.

\section*{Acknowledgements}
I would like to thank the organizers and the participants of the School for creating a very warm and productive
atmosphere. Many thanks to the organizers for their help and support. The work was partially
supported by RFBR grant 12-02-93108 and NSh grant 3042.2014.2.



\begin{thebibliography}{99}
\bibitem{bib:bjorkin65}  J.D. Bjorken and S. Drell, \emph{Relativistic Quantum Mechanics/Fields} (McGraw-Hill, New York, 1965), Vols. I, II.
\bibitem{bib:abers73}  E.S. Abers and B.W. Lee, \emph{Phys. Rep.} \textbf{9} (1973) 1.
\bibitem{bib:itzykson80} C. Itzykson and J. Zuber, \emph{Introduction to Quantum Field Theory} (McGraw-Hill, New York, 1980).
 \bibitem{bib:halzen84} F. Halzen and A.D. Martin, \emph{Quarks and Leptons: An Introductory Course in Modern Particle Physics} (Wiley, New York, 1984).
 \bibitem{bib:cheng84} T.P. Cheng and L.F. Li, \emph{Gauge Theory of Elementary Particle Physics} (Oxford University Press, New York, 1984).
  \bibitem{bib:faddeev91} L.D.~Faddeev and A.A.~Slavnov,  \emph{Gauge Fields, Introduction to Quantum Theory} (Addison-Wesley, Redwood, CA, 1991). Translation of:  A.A.~Slavnov and L.D.~Faddeev, \emph{Vvedenie v Kvantovuiu Teoriiu Kalibrovochnykh Polei} (Nauka, Moscow, 1988), Izd. 3.
 \bibitem{bib:peskin95} M.E. Peskin and D.V. Schroeder, \emph{An Introduction to Quantum Field Theory} (Addison-Wesley, Reading, MA, 1995).
 \bibitem{bib:weinberg96} S. Weinberg, \emph{The Quantum Theory of Fields} (Cambridge University Press, Cambridge, MA, 1996), Vols. I, II.
\bibitem{bib:zee03} A. Zee, \emph{Quantum Field Theory in a Nutshell} (Princeton University Press, Princeton, NJ, 2003).
 \bibitem{bib:buchmuller05} W. Buchm\"uller and C. L\"udeling, Field theory and the Standard Model,
 European School of High-Energy Physics, Kitzb\"uhel, Austria, 2005, Ed. R. Fleischer (CERN, Geneva, 2006),
 CERN-2006-014, p. 1 \url{[arXiv:hep-ph/0609174].}
  \bibitem{bib:pich07} A. Pich, The Standard Model of electroweak interactions, European School of High-Energy Physics, 
Aronsborg, Sweden, 2006, Ed. R. Fleischer (CERN, Geneva, 2007), CERN-2007-005, p. 1 \url{[arXiv:hep-ph/0705.4264]}.
 \bibitem{bib:altarelli08} G. Altarelli, \emph{The Standard Model of Electroweak Interactions}, 
Ed. H. Schopper, Springer Materials -- The Landolt-B\"ornstein Database Groups I, V (Springer, Berlin Heidelberg, 2008),
 Vol. 21A, Chap. 3.
 \bibitem{bib:rubakov09}  V. Rubakov, Field theory and the Standard Model, European School of High-Energy Physics, Herbeumont-sur-Semois, Belgium, 2008, Eds. N. Ellis and R. Fleischer (CERN, Geneva, 2009), CERN-2009-002, p. 1.
 \bibitem{bib:hollik09}  W. Hollik, Quantum field theory and the Standard Model, 
European School of High-Energy Physics, Bautzen, Germany, 2009, Eds. 
C. Grojean and M. Spiropulu (CERN, Geneva, 2010), CERN-2010-002, p. 1 \url{[arXiv:hep-ph/1012.3883v1]}.
 \bibitem{bib:alterelli13}  
G. Altarelli, Collider physics within the Standard Model: a primer, 
CERN-PH-TH-2013-020, 2013, \url{[arXiv:hep-ph/1303.2842]}.

\bibitem{ALEPH:2005ab} 
  S.~Schael {\it et al.}  [ALEPH and DELPHI and L3 and OPAL and SLD and LEP Electroweak Working Group and SLD Electroweak Group and SLD Heavy Flavour Group Collaborations],
 ``Precision electroweak measurements on the $Z$ resonance,''
  Phys.\ Rept.\  {\bf 427}, 257 (2006)
  [hep-ex/0509008].

\bibitem{Schael:2013ita}
  S.~Schael {\it et al.} [ALEPH and DELPHI and L3 and OPAL and LEP Electroweak Collaborations],
  ``Electroweak Measurements in Electron-Positron Collisions at W-Boson-Pair Energies at LEP,''
  Phys.\ Rept.\  {\bf 532} (2013) 119
  [arXiv:1302.3415 [hep-ex]].

\bibitem{Abbiendi:1998ea} 
G.~Abbiendi {\it et al.} [OPAL Collaboration],
 Eur.\ Phys.\ J.\ C {\bf 6} (1999) 1
 [hep-ex/9808023].
\end{thebibliography}
\end{document}